%% file: DmesonInpPbAllInOne.tex
\documentclass[ALICE,manyauthors]{cernphprep}

\pdfoutput=1
\usepackage[english]{babel}
\usepackage{graphicx}
\usepackage{verbatim}
\usepackage{amsfonts}
\usepackage{makeidx}
\usepackage{color}
\usepackage{amsmath}
\usepackage{lineno}
\usepackage{epsfig}
\usepackage{epstopdf}
\usepackage{hyperref}
\usepackage{cite}
\newcommand{\sqrts}{\sqrt{s}}
\newcommand{\sqrtsNN}{\sqrt{s_{\rm \scriptscriptstyle NN}}}

\newcommand{\GeV}{\mathrm{GeV}}
\newcommand{\TeV}{\mathrm{TeV}}

\newcommand{\mev}{\mathrm{MeV}}
\newcommand{\gev}{\mathrm{GeV}}
\newcommand{\gevc}{\mathrm{GeV}/c}
\newcommand{\tev}{\mathrm{TeV}}

\newcommand{\RpPb}{R_{\rm pPb}}

\newcommand{\pt}{p_{\rm T}}

\newcommand{\DtoKpi}{{\rm D}^0 \to {\rm K}^-\pi^+}
\newcommand{\DtoKpipi}{{\rm D}^+\to {\rm K}^-\pi^+\pi^+}
\newcommand{\DstartoDpi}{{\rm D}^{*+} \to {\rm D}^0 \pi^+}

\newcommand{\DstophipitoKKpi}{{\rm D_s^{+}\to \phi\pi^+\to K^-K^+\pi^+}}

\newcommand{\phitoKK}{{\rm \phi\to  K^-K^+}}
\newcommand{\DstoKzerostarK}{{\rm D_s^{+}\to \overline{K}^{*0} K^+}}
\newcommand{\Dstofzeropi}{{\rm D_s^{+}\to f_{0}(980) \pi^+}}

\newcommand{\Dzero}{{\rm D^0}}
\newcommand{\Dzerobar}{{\overline{{\rm D}}\,^0}}
\newcommand{\Dstar}{{\rm D^{*+}}}

\newcommand{\Dplus}{{\rm D^+}}

\newcommand{\Ds}{{\rm D_s^+}}

\newcommand{\KKpi}{{\rm K^-K^+\pi^+}}

\newcommand{\dEdx}{{\rm d}E/{\rm d}x}

\newcommand{\meanpt}{{\langle p_{ {\mathrm T} } \rangle}}

\begin{document}

\begin{titlepage}
\PHyear{2016}
\PHnumber{127}      
\PHdate{20 May}  
%

\title{D-meson production in p--Pb collisions at $\mathbf{\sqrtsNN=5.02}$~TeV \\ and in pp collisions at $\mathbf{\sqrts=7}$~TeV}
\ShortTitle{D-meson production in p--Pb and pp collisions}

\Collaboration{ALICE Collaboration\thanks{See Appendix~\ref{app:collab} for the list of collaboration members}}
\ShortAuthor{ALICE Collaboration} 

\begin{abstract} 
The production cross sections of the prompt charmed 
mesons $\Dzero$, $\Dplus$, $\Dstar$ and $\Ds$ 
were measured at mid-rapidity
in p--Pb collisions at a centre-of-mass energy per nucleon pair $\sqrtsNN=5.02~\tev$ with the ALICE detector at the LHC.
D mesons were reconstructed from their decays $\DtoKpi$, $\DtoKpipi$, 
$\DstartoDpi$, $\DstophipitoKKpi$, and their charge conjugates.
The $\pt$-differential production cross sections  were measured
at mid-rapidity in the transverse momentum interval $1<\pt<24~\GeV/c$ 
for $\Dzero$, $\Dplus$ and $\Dstar$ mesons and in $2<\pt<12~\gev/c$ for 
$\Ds$ mesons,
using an analysis method based on the selection of decay topologies
displaced from the interaction vertex.
The production cross sections of the $\Dzero$, $\Dplus$ and $\Dstar$ mesons 
were also measured in three $\pt$ intervals
as a function of the rapidity $y_{\rm cms}$  in the centre-of-mass system in $-1.26<y_{\rm cms}<0.34$.
In addition, the prompt $\Dzero$ production cross section was measured in
pp collisions at $\sqrt{s}=7~\TeV$ and p--Pb collisions at
$\sqrtsNN=5.02~\tev$ down to $\pt=0$ using an analysis technique
that is based on the estimation and subtraction of the combinatorial background, without reconstruction of the $\Dzero$ decay vertex.
The nuclear modification factor $\RpPb(\pt)$, defined as the ratio of 
the $\pt$-differential D-meson cross section in p--Pb collisions and that in 
pp collisions scaled by the mass number of the Pb nucleus, was calculated for
the four D-meson species and found to be compatible with unity within experimental uncertainties.
The results are compared to theoretical calculations 
that include cold-nuclear-matter effects and to transport model 
calculations incorporating the interactions of charm quarks with an 
expanding deconfined medium.

\end{abstract}

\end{titlepage}

\setcounter{page}{2}

\section{Introduction}
\label{sec:intro}

The measurement of the production cross section of hadrons containing heavy 
quarks, charm and beauty, in proton--proton (pp) collisions is a
sensitive test of perturbative Quantum Chromodynamics (pQCD) calculations.
The inclusive transverse momentum ($\pt$) and rapidity ($y$) differential 
cross sections can be calculated in the collinear factorisation 
approach as a convolution of three terms: i)
the parton distribution functions (PDF) of the incoming protons; ii) the 
partonic hard scattering cross section; and iii) the fragmentation function, 
which models the non-perturbative transition of a heavy quark to 
a given heavy-flavour hadron species~\cite{Andronic:2015wma}.
At LHC energies, implementations of these calculations are available
at next-to-leading order (NLO) accuracy in the general-mass 
variable-flavour-number scheme, 
GM-VFNS~\cite{Kniehl:2004fy,Kniehl:2005mk,Kniehl:2012ti}, and
at fixed order with next-to-leading-log resummation, 
FONLL~\cite{Cacciari:1998it,Cacciari:2012ny}.
Calculations of heavy-flavour production cross sections
in hadronic collisions also exist within the framework of $k_{\rm T}$-factorisation, at leading order 
(LO) approximation, with unintegrated gluon distributions (UGDFs) to account for 
the transverse momenta of the initial 
partons~\cite{Luszczak:2008je,Maciula:2013wg,Catani:1990eg}.
At LHC energies, the measurement of charm production at low $\pt$ probes 
the parton distribution functions of the proton at small values of parton 
fractional momentum $x$ and squared momentum transfer $Q^2$.
For illustration, in the simplified scenario of a $2\to2$ process at leading 
order, charm quarks ($m_{\rm c}\approx 1.5~\gev/c^2$) with $\pt = 0.5~\gev/c$ and 
rapidity $y = 0$ probe the parton distribution functions at 
$x\approx 4\times 10^{-4}$ and $Q^2\approx 10~\gev^2$.
Perturbative QCD calculations have substantial uncertainties at low $\pt$, 
owing both to the large effect of the choice of the factorisation and 
renormalisation scales at low $Q^2$ and to the sizeable uncertainties on the 
gluon PDFs at small $x$~\cite{Cacciari:2015fta}.
Therefore, a precise measurement of the D-meson production cross section
down to $\pt=0$ could provide an important constraint to pQCD calculations
and to low-$x$ gluon PDFs.
This is also relevant for cosmic-ray and neutrino astrophysics, where 
high-energy neutrinos from the decay of charmed hadrons produced in particle showers
in the atmosphere constitute
an important background for neutrinos from astrophysical 
sources~\cite{Bhattacharya:2015jpa,Bhattacharya:2016jce,Gauld:2015yia,Garzelli:2015psa}. 
Furthermore, the measurement in pp collisions provides the reference
for results in heavy-ion collisions, where heavy quarks are sensitive probes 
of the properties of the hot and dense medium with
partonic degrees of freedom formed in the collision ---the Quark-Gluon Plasma.
In this context, the measurement of D-meson production down to $\pt=0$ in pp collisions also
allows the precise determination of the total charm-production cross section, which is
a crucial ingredient for the models of charmonium regeneration in the Quark-Gluon 
Plasma~\cite{Andronic:2011yq,Zhao:2011cv,Liu:2009nb}.

Measurements in proton--nucleus collisions allow an assessment of the various
effects related to the presence of nuclei in the colliding system 
and denoted as cold-nuclear-matter (CNM) effects.
In the initial state, the PDFs are modified in bound nucleons as compared to free nucleons, depending on $x$ and $Q^2$~\cite{Arneodo:1992wf,Malace:2014uea}.
At LHC energies, the most relevant effect is shadowing: a reduction of the
parton densities at low $x$, which becomes stronger when $Q^2$ decreases and the nucleus mass number $A$ increases.
This effect, induced by the high phase-space density of small-$x$ partons, can be described, 
within the collinear factorisation framework,
by means of phenomenological parametrisations of the 
modification of the PDFs (denoted as 
nPDFs)~\cite{Eskola:2009uj,Hirai:2007sx,deFlorian:2003qf}.
If the parton phase-space reaches saturation, PDF evolution equations are not applicable 
and the most appropriate theoretical description is
the Colour 
Glass 
Condensate effective theory  (CGC)~\cite{Gelis:2010nm,Tribedy:2011aa,Albacete:2012xq,Rezaeian:2012ye,Fujii:2013yja}.
The modification of the small-$x$ parton dynamics can significantly reduce D-meson production at low $\pt$. 
Furthermore, the multiple scattering of partons in the nucleus before and/or
after the hard scattering can modify the kinematic distribution of the
produced hadrons: partons can lose energy in the initial stages of the 
collision via initial-state radiation~\cite{Vitev:2007ve}, or experience 
transverse momentum broadening due to multiple soft collisions before the 
heavy-quark pair is produced \cite{Lev:1983hh, Wang:1998ww, Kopeliovich:2002yh}.
These initial-state effects are expected to have a small impact 
on D-meson production at high $\pt$ ($\pt>3$--$4~\GeV/c$), but they
can induce a significant modification of the D-meson cross section and
momentum distribution at lower momenta.
For this reason, a measurement of the D-meson production cross section and its 
nuclear modification factor $\RpPb$ (the ratio of the cross section in 
p--Pb collisions to that in pp interactions scaled by the mass 
number of the Pb nucleus) down to $\pt=0$ could provide important 
information.
In addition to the initial-state effects discussed above, also final-state 
effects may be responsible for a modification of heavy-flavour hadron 
yields and momentum distributions.
The presence of significant final-state effects in high-multiplicity p--Pb 
collisions is suggested by different observations, e.g.\ the 
presence of long-range correlations of charged 
hadrons~\cite{CMS:2012qk, Abelev:2012ola, ABELEV:2013wsa, Aad:2012gla,Adam:2015bka}, the evolution with multiplicity of the identified-hadron 
transverse-momentum distributions~\cite{Abelev:2013haa,Chatrchyan:2013eya},
and the suppression of the $\psi{\rm (2S)}$ production with respect to 
the ${\rm J}/\psi$ one~\cite{Abelev:2014zpa,Aaij:2016eyl,Adam:2016ohd}.
The correlation measurements can be described by hydrodynamic 
calculations assuming the formation of a medium with some degree of 
collectivity (see e.g.\,\cite{Bozek:2012gr, Bozek:2013uha}),
even though alternative explanations exist, based on the CGC effective theory 
(see e.g.\,\cite{Dusling:2012cg}) or on the anisotropic escape 
probability of partons from the collision zone~\cite{He:2015hfa}.
If a collective expansion in the final state were present, 
the medium could also impart a flow to heavy-flavour hadrons.
The possible effect on the
D-meson transverse momentum distributions was first estimated in 
Ref.~\cite{Sickles:2013yna} by employing an approach based on a blast-wave 
function with parameters extracted from fits to the light-hadron 
spectra.
More detailed calculations were subsequently carried out in the framework of 
transport models assuming that also in p--Pb collisions at LHC energies a hot and deconfined 
medium is formed, which modifies the propagation and hadronisation of heavy 
quarks~\cite{Xu:2015iha,Beraudo:2015wsd}.
The results of these calculations show a modification of the 
D-meson $\pt$ distributions at $\pt<4~\GeV/c$ by 
radial flow, possibly accompanied by a moderate 
($<20$--30\%) suppression at higher $\pt$, caused by in-medium energy loss.

In this article, we report on the measurements of production cross sections 
and nuclear modification factors of D mesons performed in minimum-bias p--Pb 
collisions at $\sqrtsNN=5.02~\TeV$ with the ALICE detector.
In Ref.~\cite{Abelev:2014hha}, the results of $\pt$-differential
cross sections and $\RpPb$ of $\Dzero$, $\Dplus$ and $\Dstar$ mesons for 
$\pt>1~\gev/c$, and of $\Ds$ mesons for $\pt>2~\gev/c$, at mid-rapidity were 
reported.
We complement them in this article with measurements of 
production cross sections of $\Dzero$, $\Dplus$ and $\Dstar$ mesons
as a function of rapidity in three $\pt$ intervals.
For the $\Dzero$ meson, we also report an extension down to 
$\pt=0$ of the measurements of the $\pt$-differential production cross sections 
in p--Pb collisions at $\sqrtsNN=5.02~\TeV$ and in pp collisions at 
$\sqrt{s}=7~\TeV$ published in Refs.~\cite{Abelev:2014hha} 
and~\cite{ALICE:2011aa}, respectively.
This allowed a determination of the $\pt$-integrated $\Dzero$ cross section
at mid-rapidity, which for pp collisions at $\sqrt{s}=7~\TeV$ is more precise
than the previous result~\cite{ALICE:2011aa}.

The paper is organized as follows.
In Section~\ref{sec:sample}, the ALICE apparatus, its performance and the 
data samples used for the measurement are briefly described.
The analysis technique utilized for a first set of measurements of
$\Dzero$, $\Dplus$, $\Dstar$ and $\Ds$ production is presented in
Section~\ref{sec:topol} together with the corrections
and the systematic uncertainties.
This analysis technique is based on the reconstruction of the D-meson displaced
decay vertex and will be, for brevity, indicated as the analysis
`with decay-vertex reconstruction' in this article.
With this technique the $\pt$-differential production cross section 
was measured down to $\pt=1~\GeV/c$ both in pp collisions at 
$\sqrts=7~\TeV$~\cite{ALICE:2011aa} and in p--Pb collisions at 
$\sqrtsNN=5.02~\TeV$~\cite{Abelev:2014hha}, as well as in pp and Pb--Pb
collisions at $\sqrtsNN=2.76~\TeV$~\cite{Abelev:2012vra,Adam:2015sza}.
In order to extend the measurement down to $\pt=0$, 
where the decay-vertex selection becomes very inefficient,
a different analysis
technique, which does not exploit the displaced decay-vertex topology, 
was developed for the $\Dzero$-meson reconstruction
in pp collisions at $\sqrts=7~\TeV$ and in p--Pb collisions at 
$\sqrtsNN=5.02~\TeV$.
This analysis technique, denoted as `without decay-vertex 
reconstruction' throughout this article,  is described in 
Section~\ref{sec:lowpt}.
The results are presented and discussed in Section~\ref{sec:resul}.
The cross sections measured in pp collisions are compared to the results of 
pQCD calculations, while the measurements of the D-meson nuclear modification 
factor in p--Pb collisions are compared to models including cold and hot
nuclear matter effects.

\section{Apparatus and data samples}
\label{sec:sample}

The ALICE apparatus~\cite{Aamodt:2008zz,Abelev:2014ffa} consists of a central 
barrel detector covering the pseudo-rapidity range $|\eta|<0.9$, a forward 
muon spectrometer covering the pseudo-rapidity range $-4.0<\eta<-2.5$ and a 
set of detectors at forward and backward rapidities used for 
triggering and event characterization. In the following, the detectors used 
for the D-meson analysis are described. 

The D mesons are reconstructed in the mid-rapidity region using the tracking and particle identification 
capabilities of the central barrel detectors, which are located in a large solenoidal magnet that produces 
a magnetic field of 0.5~T along the beam direction ($z$ axis). The innermost detector of the central barrel is the 
Inner Tracking System (ITS), which is comprised of six cylindrical layers of silicon detectors with radii between 3.9 and 43.0~cm. 
The two innermost layers, with average radii of 3.9~cm and 7.6~cm, are equipped with Silicon Pixel Detectors (SPD); the two 
intermediate layers, with average radii of 15.0~cm and 23.9~cm, are equipped with Silicon Drift Detectors (SDD) and the two outermost 
layers, with average radii of 38.0 cm and 43.0~cm, are equipped with double-sided Silicon Strip Detectors (SSD). 
The low material budget (on average 7.7\% of a radiation length for tracks crossing the ITS at $\eta=0$), the high spatial resolution, 
and the small distance of the innermost layer from the beam vacuum tube, allow 
the measurement of the track impact parameter
in the transverse plane ($d_{0}$), i.e.\ the distance of closest approach of the track to the interaction vertex in the plane 
transverse to the beam direction, with a resolution better than 75~$\mu$m for $\pt> 1$ GeV/$c$~\cite{Aamodt:2010aa}.

The ITS is surrounded by a large cylindrical Time Projection Chamber 
(TPC)~\cite{Alme:2010ke} with an active radial range from about 85 to 250 cm 
and an overall length along the beam direction of 500 cm. 
It covers the full azimuth in the pseudo-rapidity range $|\eta| < 0.9$ and 
provides track reconstruction with up to 159 points along the trajectory of a 
charged particle as well as particle identification via the measurement of
specific energy loss $\dEdx$. 
The charged particle identification capability of the TPC is supplemented by 
the Time-Of-Flight detector (TOF)~\cite{Akindinov:2013tea}, which is based 
on Multi-gap Resistive Plate Chambers (MRPCs) and is positioned at radial 
distances between 377 and 399 cm from the beam axis. 
The TOF detector measures the flight time of the particles from the 
interaction point. 
The start time of the event can be determined either from the information
provided by the T0 detector~\cite{Bondila:2005xy} or via a combinatorial 
analysis of the particle arrival times at the TOF 
detector~\cite{Akindinov:2013tea}.
The T0 detector is composed of two arrays of Cherenkov counters located on 
either side of the interaction point at $+350$~cm and $-70$~cm from the 
nominal vertex position along the beam line. The T0 time resolution is about 40~ps for pp
collisions.
The overall TOF resolution, including the uncertainty on the start time of 
the event, and the tracking and momentum resolution contributions, is 
about 150~ps in pp collisions and 85~ps for high-multiplicity 
p--Pb collisions~\cite{Abelev:2014ffa}.

Triggering and event selection are based on the V0 and SPD detectors
and on the Zero Degree Calorimeters (ZDC).
The V0 detector consists of two scintillator arrays, denoted V0A and V0C,  
covering the pseudo-rapidity ranges $2.8<\eta<5.1$ and $-3.7<\eta<-1.7$, 
respectively~\cite{Abbas:2013taa}. 
The ZDCs are two sets of neutron and 
proton calorimeters positioned along the beam axis on both sides of the ALICE 
apparatus at about 110 m from the interaction point.

The data samples used for the analyses presented here include p--Pb collisions at $\sqrtsNN=5.02~\rm{TeV}$ and 
pp collisions at $\sqrt{s}=7~\rm{TeV}$, collected in 2013 and 2010, respectively. 
During the p--Pb run, the beam energies were 4~TeV for protons and 1.58~TeV per nucleon for lead nuclei. 
With this beam configuration, 
the nucleon--nucleon centre-of-mass system moves in rapidity by $\Delta y_{\mathrm{cms}}=0.465$ in the direction 
of the proton beam. The D-meson analyses were performed in the laboratory-frame interval $|y_{\mathrm{lab}}|<0.5$, 
which leads to a shifted centre-of-mass rapidity coverage of $-0.96 < y_{\mathrm{cms}} < 0.04$.
In p--Pb collisions, minimum-bias events were selected requiring at least one hit in both of the V0A and V0C
scintillator arrays.
In pp collisions, minimum-bias events were triggered by requiring at least one hit in either of the V0 counters or 
in the SPD. The minimum-bias (MB) trigger 
was estimated to be sensitive to about 96.4\% and 87\% of the p--Pb and pp inelastic cross sections, respectively~\cite{Abelev:2014epa,Abelev:2012sea}.
Beam-gas and other machine-induced background collisions were removed via offline selections based on the timing 
information provided by the V0 and the ZDCs, and the correlation between 
the number of hits and track segments (tracklets) in the SPD detector.
For the data samples considered in this paper, the probability of collision 
pile-up was below 4\% per triggered pp event and below the per-cent level 
per triggered p--Pb event.
An algorithm to detect multiple interaction vertices was used to reduce the pile-up contribution. 
An event was rejected if a second interaction vertex was found.
The remaining undetected pile-up was negligible in the present analysis. 
Only events with a primary vertex reconstructed within $\pm 10$~cm from the 
centre of the detector along the beam line were considered. 
The number of events passing these selection criteria was about $10^8$ for p--Pb collisions
and about $3.1 \cdot 10^8$ for pp collisions.
The corresponding integrated luminosities, $L_{\rm int}=N_{\rm MB}/\sigma_{\rm MB}$, 
where $\sigma_{\rm MB}$ is the MB trigger cross section measured with
van der Meer scans, are $48.6~\rm{\mu b^{-1}}$, with an uncertainty of 3.7\%, 
for the p--Pb sample~\cite{Abelev:2014epa}, and 
$5.0~{\rm nb}^{-1}$ ($\pm3.5\%$) for the pp sample~\cite{Abelev:2012sea}.

\section{Analysis with decay-vertex reconstruction in p-Pb collisions}
\label{sec:topol}

\subsection{$\Dzero$, $\Dplus$, $\Dstar$ and $\Ds$ meson reconstruction and selection}
\label{sec:selection}

$\Dzero$, $\Dplus$, $\Dstar$ and $\Ds$ mesons, and their charge conjugates, were reconstructed via their hadronic decay channels $\DtoKpi$ (with a branching ratio, BR, of $3.88\pm 0.05\%$), $\DtoKpipi$ ($\mathrm{BR}=9.13\pm0.19\%$), $\DstartoDpi$ ($\mathrm{BR}=67.7\pm 0.5\%$) followed by $\DtoKpi$, and $\DstophipitoKKpi$ ($\mathrm{BR}=2.24 \pm 0.10\%$)~\cite{Agashe:2014kda}. 
The $\Dzero$, $\Dplus$, and $\Ds$ mesons decay weakly with mean proper decay lengths ($c\tau$) of about 123, 312 and 150~$\mu$m~\cite{Agashe:2014kda}, respectively. The analysis strategy was based on the reconstruction of secondary vertices separated by a few hundred $\mu$m from the interaction point. The $\Dstar$ meson decays strongly at the primary vertex, and the decay topology of the produced $\Dzero$ was reconstructed along with a soft pion originating from the primary vertex. The transverse momentum of the soft pion produced in the $\Dstar$ decays typically ranges from 0.1 to 1.5~$\gevc$, depending on the $\Dstar$ $\pt$.

$\Dzero$, $\Dplus$ and $\Ds$ candidates were formed using pairs and triplets of tracks with the correct charge-sign combination. Tracks were selected by requiring $|\eta| < 0.8$, $\pt > 0.3~\gevc$, at least 70 (out of a maximum of 159) associated space points and a fit quality $\chi^2/\mathrm{ndf} < 2$ in the TPC, and at least two (out of six) hits in the ITS, out of which at least one had to be in either of the two SPD layers. $\Dstar$ candidates were formed by combining $\Dzero$ candidates with tracks with $\pt>0.1~\gevc$ and at least three hits in the ITS, out of which at least one had to be in the SPD. The track selection criteria reduce the D-meson acceptance, which drops steeply to zero for $|y_{\rm lab}|>0.5$ at low $\pt$ and for $|y_{\rm lab}|>0.8$ at $\pt > 5~\gevc$. A $\pt$-dependent fiducial acceptance region was therefore defined as $|y_{\rm lab}|<y_{\mathrm{fid}}(\pt)$, with $y_{\mathrm{fid}}(\pt)$ increasing from 0.5 to 0.8 in the transverse momentum range $0 < \pt < 5~\gevc$ according to a second-order polynomial function, and $y_{\mathrm{fid}}=0.8$ for $\pt > 5~\gevc$.

The selection of the D-meson decay topology was mainly based on the displacement of the tracks from the interaction vertex, the separation of the primary and secondary vertices, and the pointing of the reconstructed D-meson momentum to the primary vertex. 
A detailed description of the variables used to select the D-meson candidates
can be found in Refs.~\cite{ALICE:2011aa,Abelev:2012tca}.
The actual cut values were optimized for the signal and background levels of 
the p--Pb sample; they depend on the D-meson species and $\pt$, but they are 
the same in all the considered rapidity intervals. 

Further reduction of the combinatorial background was obtained by applying particle identification (PID) to the decay tracks. A $3 \sigma$ compatibility cut was applied to the difference between the measured and expected signals for pions and kaons for the TPC $\mathrm{d}E/\mathrm{d}x$ and the time-of-flight measured with the TOF detector. Tracks without hits in the TOF detector were identified using only the TPC information. PID selections were not applied to the pion track from the $\Dstar$ strong decay. A tighter PID selection was applied to the $\Ds$ candidates: tracks without a TOF signal (mostly at low momentum) were identified using only the TPC information and requiring a $2\sigma$ compatibility with the expected $\dEdx$. 
This stricter PID selection strategy was needed in the $\Ds$ case
due to the large background of track triplets and the
short $\Ds$ lifetime, which limits the effectiveness of the geometrical 
selections on the displaced decay-vertex topology.
In addition, in the cases of $\Dplus \to \rm K^- \pi^+ \pi^+$ and
$\Ds \to \rm K^- \rm K^+ \pi^+$ decays, the charge signs of the decay
particles were exploited in combination with the pion and kaon identification. 
Since in both these decay modes, the decay particle with the opposite charge 
sign with respect to the D meson has to be a kaon, a candidate was rejected if 
the opposite-sign track was not compatible with the kaon hypothesis. 
The applied PID strategy provides a reduction of the combinatorial background by a factor of about three at low $\pt$ while preserving an efficiency of 95\% for the $\Dzero$, $\Dplus$ and $\Dstar$ signals and of 85$\%$ for the $\Ds$ signal.
The fraction of signal candidates passing the PID selections is lower than 
that expected from a perfectly Gaussian response due to the non-Gaussian tail 
of the TOF signal and the non-negligible contamination originating from wrong 
associations between reconstructed tracks and TOF hits~\cite{Adam:2016acv}.

In the $\Ds$ case, in order to select $\Ds \rightarrow \phi \pi^+$ decays with 
$\phitoKK$, candidates were rejected if none of the two pairs of 
opposite-charge tracks (required to be compatible with the kaon hypothesis) 
had an invariant mass compatible with the PDG world average for the $\phi$ 
meson mass (1.0195 GeV/$c^2$)~\cite{Agashe:2014kda}.
The difference between the reconstructed K$^+$K$^-$ invariant mass and 
world-average $\phi$ mass was required to be less than 5--$10~\mev/c^2$ 
depending on the $\Ds$ $\pt$ interval. This selection preserves 70--85\% of
the $\Ds$ signal.
\begin{figure}[!tb]
\begin{center}
\includegraphics[width=\textwidth]{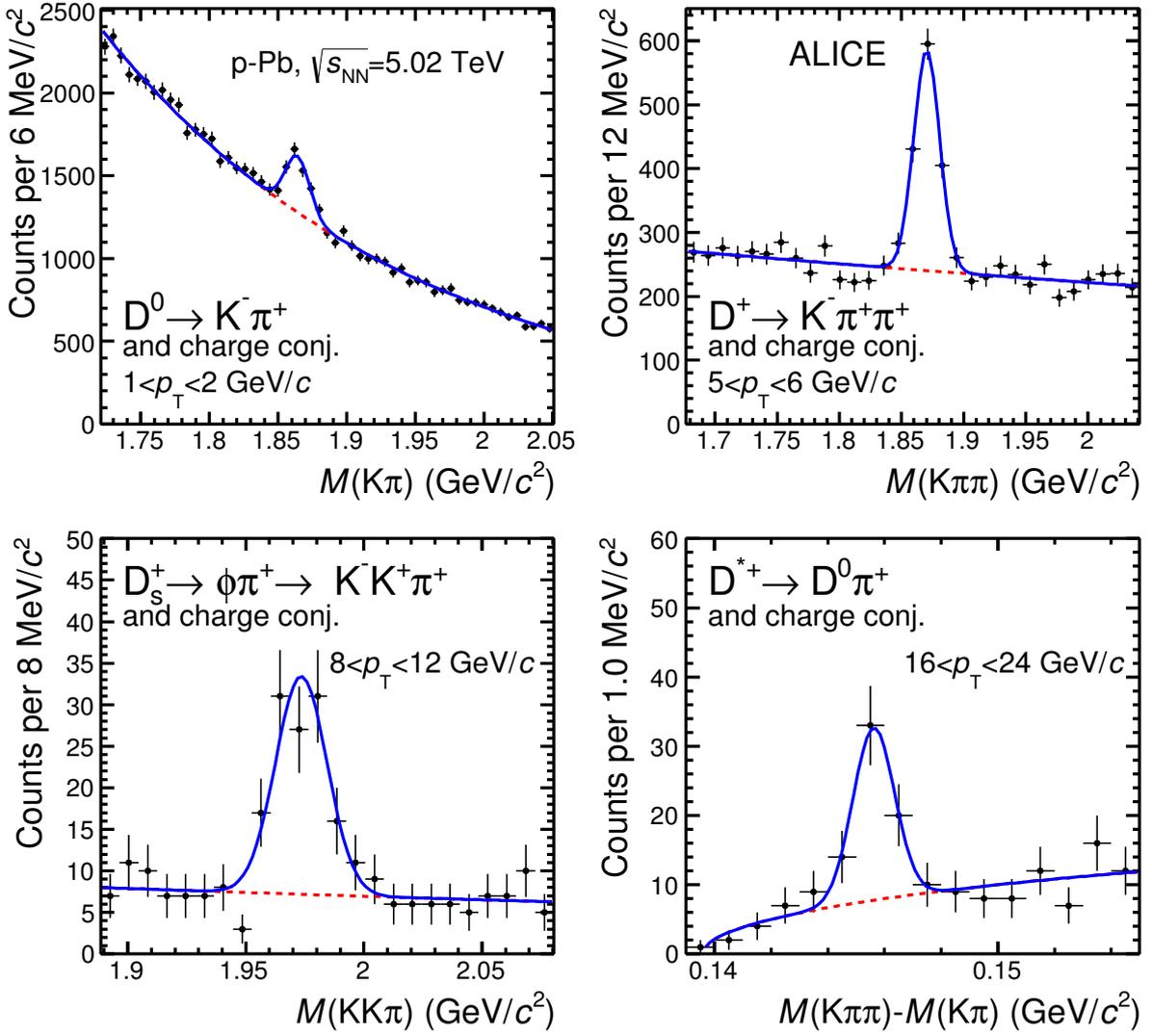}
\caption{Distributions of the invariant mass for $\Dzero$ (top left), $\Dplus$ (top right), $\Ds$ (bottom left) candidates and their charge conjugates and of the mass difference for $\Dstar$ (bottom right) candidates (and charge conjugates) in the rapidity interval $|y_{\rm lab}|<y_{\mathrm{fid}}(\pt)$ in p--Pb collisions. The dashed lines represent the fit to the background while the solid lines represent the total fit function. One $\pt$ interval is shown for each species: $1<\pt<2~\gevc$ for $\Dzero$, $5<\pt<6~\gevc$ for $\Dplus$, $8<\pt<12~\gevc$ for $\Ds$ and $16<\pt<24~\gevc$ for $\Dstar$. }  
\label{fig:invmass4specpPb} 
\end{center}
\end{figure}

The D-meson raw yields were extracted from fits to the $\Dzero$, $\Dplus$ and $\Ds$ candidate invariant-mass distributions and to the mass difference $\Delta M = M (\mathrm{K} \pi \pi) - M(\mathrm{K} \pi)$ distributions for $\Dstar$ candidates. In the fit function, the signal is modeled with a Gaussian and the background is described by an exponential term for $\Dzero$, $\Dplus$ and $\Ds$ candidates and by a threshold function multiplied by an exponential ($a \sqrt{\Delta M - m_{\pi}} \cdot {\rm e}^{b(\Delta M - m_{\pi})}$) for the $\Dstar$ case. 
For all four D-meson species, the mean values of the Gaussian functions in all 
transverse momentum and rapidity intervals were found to be compatible within  
uncertainties with the PDG world average values~\cite{Agashe:2014kda}.
The Gaussian widths are consistent with the simulation results with deviations
of at most 15\%.

With the analysis based on the decay-vertex reconstruction, D-meson yields were extracted as a function of the transverse momentum in the range $1<\pt<24~\gevc$ for $\Dzero$, $\Dplus$ and $\Dstar$ ($2<\pt<12~\gevc$ for $\Ds$) in a rapidity interval $|y_{\rm lab}|<y_{\mathrm{fid}}(\pt)$. The yield of $\Dzero$, $\Dplus$ and $\Dstar$ mesons was measured also as a function of rapidity in three $\pt$ intervals: $2<\pt<5~\gevc$, $5<\pt<8~\gevc$ and $8<\pt<16~\gevc$. The rapidity interval of the measurement was $|y_{\mathrm{lab}}|< 0.7$ for the lowest $\pt$ interval and  $|y_{\mathrm{lab}}|<0.8$ for the other two $\pt$ intervals.

\begin{figure}[!tb]
\begin{center}
\includegraphics[width=\textwidth]{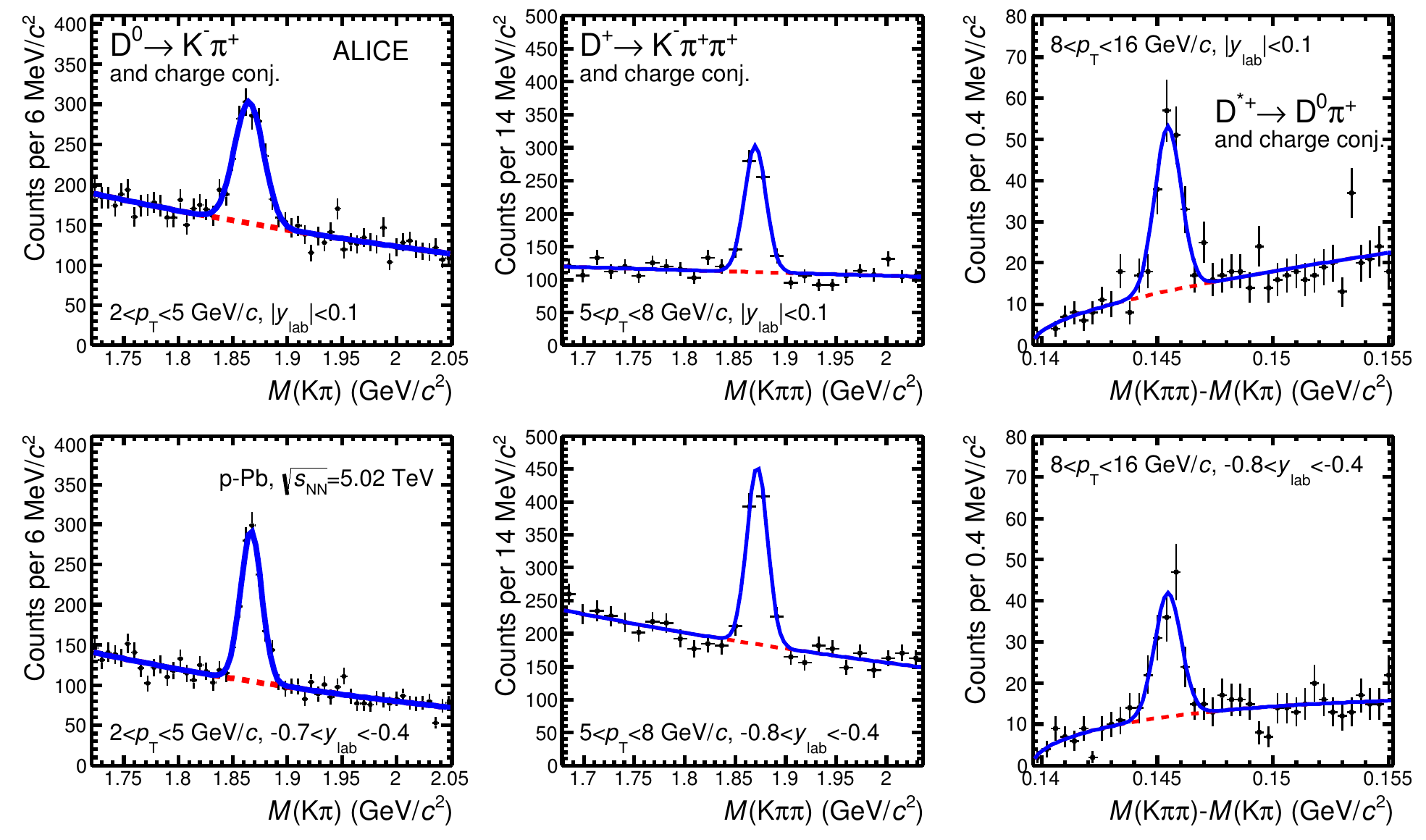}
\caption{Distributions of the invariant mass for $\Dzero$ (left column), $\Dplus$ (middle column) candidates and their charge conjugates and of the mass difference for $\Dstar$ (right column) candidates (and charge conjugates) in p--Pb collisions in the rapidity intervals $|y_{\rm lab}| <0.1 $ (top row) and $-0.8 < y_{\mathrm{lab}} < -0.4$ ($-0.7 < y_{\mathrm{lab}} < -0.4$ for $\pt<5~\gev/c$) (bottom row). The dashed lines represent the fit to the background while the solid lines represent the total fit function. One $\pt$ interval is shown for each species: $2<\pt<5~\gevc$ for $\Dzero$, $5<\pt<8~\gevc$ for $\Dplus$ and $8<\pt<16~\gevc$ for $\Dstar$.}
\label{fig:invmassybins} 
\end{center}
\end{figure}

Figure~\ref{fig:invmass4specpPb} shows the $\Dzero$, $\Dplus$ and $\Ds$ 
candidate invariant-mass distributions and the $\Dstar$ mass-difference 
distribution in four $\pt$ intervals in the fiducial acceptance region 
$|y_{\rm lab}|<y_{\mathrm{fid}}(\pt)$.
In addition, the invariant-mass (mass-difference) distributions of $\Dzero$, 
$\Dplus$ and $\Dstar$ candidates in two rapidity intervals, namely 
$|y_{\mathrm{lab}}| < 0.1$ and $-0.8 < y_{\mathrm{lab}} < -0.4$ 
($-0.7 < y_{\mathrm{lab}} < -0.4$ for $\pt<5~\gev/c$), are shown in the 
upper and lower panels of Fig.~\ref{fig:invmassybins} for three $\pt$ 
intervals.
%
%

\subsection{Acceptance, efficiency and subtraction of beauty feed-down contribution}
\label{sec:topcorrections}

The D-meson raw yields extracted in each $\pt$ and $y$ interval were corrected to obtain the prompt D-meson cross sections
\begin{equation}
  \label{eq:topolcrosssectionPromptD}
  \frac{{\rm d^2}\sigma^{\rm D}}{{\rm d}\pt {\rm d} y}=
  \frac{1}{\Delta\pt} \cdot \frac{f_{\rm prompt} \cdot \frac{1}{2} \cdot N^{\rm D+\overline{D},raw}(\pt)}{\Delta y} \cdot \frac{1}{({\rm Acc}\times\epsilon)_{\rm prompt}(\pt)} \cdot \frac{1}{{\rm BR} \cdot L_{\rm int}}\,.
\end{equation}

In the formula, $N^{\rm D+\overline{D},raw}$ is the raw yield (sum of particles and 
antiparticles). 
It includes contributions from both prompt (i.e. produced in the charm quark 
fragmentation, either directly or through decays of excited open charm 
and charmonium states) and from feed-down D mesons (i.e. originating from beauty-hadron 
decays). 
The factor $1/2$ accounts for the fact that the measured yields include 
particles and antiparticles while the cross sections are given for particles 
only;
$f_{\rm prompt}$ is the fraction of prompt D mesons in the raw yield;
$({\rm Acc}\times\epsilon)_{\rm prompt}$ is the product of acceptance and 
efficiency for prompt D mesons, where $\epsilon$ accounts for primary vertex reconstruction,
D-meson decay track reconstruction and selection, and for D-meson candidate 
selection with secondary vertex and PID cuts; 
$\Delta \pt$ and $\Delta y$ are the widths of the transverse momentum and 
rapidity intervals; 
BR is the branching ratio of the considered decay channel, 
and $L_{\rm int}$ is the integrated luminosity.
  
\begin{figure}[!tb]
\begin{center}
\includegraphics[width=\textwidth]{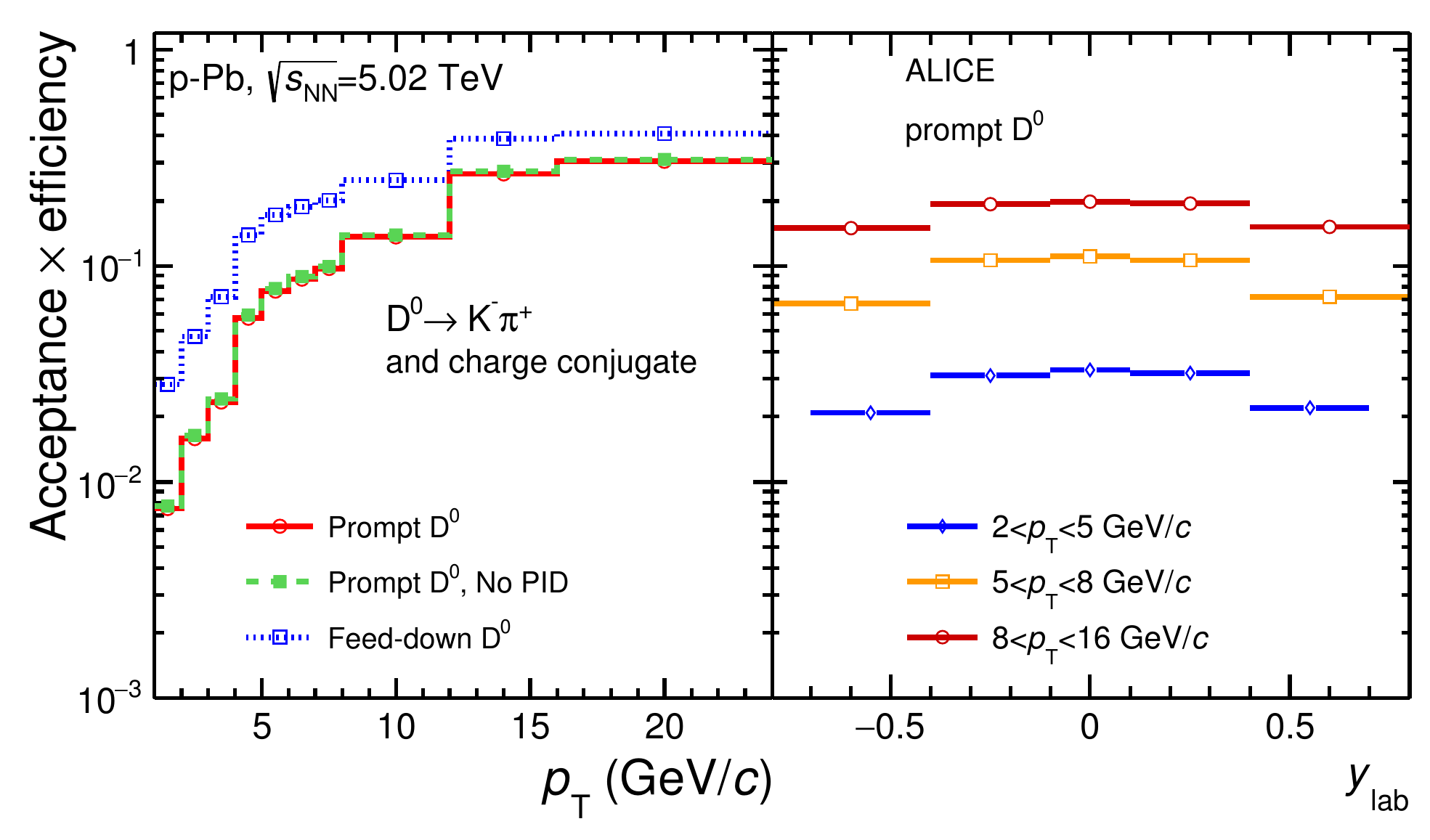}
\caption{Product of acceptance and efficiency for $\Dzero$ mesons as a function of $\pt$ (left) and as a function of $y_{\rm lab}$ (right). In the left panel, efficiencies are shown for prompt $\Dzero$ with (solid line) and without (dashed line) PID selection applied, and feed-down $\Dzero$ (dotted line). In the right panel, the ${\rm Acc}\times\epsilon$ values are shown for prompt $\Dzero$ mesons for the three $\pt$ intervals considered in the analysis as a function of rapidity.}
\label{fig:accefftopol} 
\end{center}
\end{figure}

The acceptance and efficiency correction factors were obtained from Monte Carlo simulations including detailed descriptions of the geometry of the apparatus and of the detector response. 
Proton-proton collisions were generated by using the PYTHIA v6.4.21 event 
generator~\cite{Sjostrand:2006za} with the Perugia-0 tune~\cite{Skands:2010ak}.
Events containing a $\rm{c\overline{c}}$ or $\rm{b\overline{b}}$ pair were
selected and an underlying p--Pb collision generated with 
HIJING 1.36~\cite{Wang:1991hta} was added to each of them in order to obtain a 
better description of the multiplicity distributions observed in data.
The generated D-meson $\pt$ distribution was weighted in order to match the
shape predicted by FONLL calculations~\cite{Cacciari:1998it}
at $\sqrt{s} = 5.02$~TeV, based on the observation that FONLL provides
a good description of the measured D-meson $\pt$-differential cross sections at 
$\sqrt{s}=2.76$ and 7~TeV~\cite{Cacciari:2012ny,ALICE:2011aa,Abelev:2012vra,Aaij:2013mga}.

The efficiency depends on the multiplicity of charged particles produced in the collision, since the primary vertex resolution, thus the resolution for the topological selection variables, improves at high multiplicity. Therefore, the generated events were weighted on the basis of their charged-particle multiplicity in order to match the multiplicity distribution observed in data. 
The weight function was defined as the ratio between the distribution of the 
number of tracklets (segments of tracks connecting two hits in the SPD layers 
and aligned with the primary vertex) measured in data and that obtained in the 
Monte Carlo simulation. 
The efficiency varies from about 1\% to 30\% depending on D-meson $\pt$ and species. As an example, the product of acceptance and efficiency  $\mathrm{Acc} \times \epsilon$ for prompt $\Dzero$ mesons is shown in Fig.~\ref{fig:accefftopol} (left panel) as a function of $\pt$ in the rapidity range $|y_{\rm lab}|<y_{\mathrm{fid}}(\pt)$. In the same figure, the efficiencies when the PID selection is not applied (about 5\% higher as expected from the PID strategy utilized) and  efficiencies for $\Dzero$ mesons from B decays are also shown (about a factor of two higher because the decay vertices of feed-down D mesons are more displaced from the primary vertex and they are more efficiently selected by the topological selections). The figures of $\mathrm{Acc} \times \epsilon$ as a function
of $\pt$ for $\Dplus$, $\Dstar$ and $\Ds$ mesons can be found in 
Ref.~\cite{ALICE-PUBLIC-2015-001}.
The right-hand panel of Fig.~\ref{fig:accefftopol} shows the prompt $\Dzero$ $\mathrm{Acc} \times \epsilon$ as a function of $y_{\mathrm{lab}}$ for the three momentum intervals considered in this analysis. The small decrease at $|y_{\rm lab}|>0.4$ is due to the detector acceptance. 

The correction factor $f_{\mathrm{prompt}}$ was calculated with a FONLL-based method as 
\begin{equation}
\label{eq:fpr}
f_{\mathrm{prompt}} =1- \frac{N^{\text{D~feed-down}}_{\mathrm{raw}}}{N^{\mathrm{D}}_{\mathrm{raw}}}= 1- A \cdot \left (\frac{\rm d^2 \sigma}{\mathrm d\pt \mathrm d y} \right)^{\rm FONLL}_{\text{feed-down}} \cdot R_{\rm pPb}^\text{feed-down} \cdot \frac{(\mathrm{Acc} \times \epsilon)_\text{feed-down} \cdot \Delta y \Delta \pt \cdot \mathrm{BR} \cdot L_{\rm int}}{N^{\rm D +\overline{D},raw}/2}\,,
\end{equation}

where $A$ is the mass number of the Pb nucleus.
The procedure uses the B-meson production cross section in pp collisions at $\sqrts = 5.02~\TeV$ estimated with FONLL calculations, the $\mathrm{B} \rightarrow \mathrm{D} + X$ decay kinematics from the EvtGen package~\cite{Lange2001152}, the efficiencies for D mesons from beauty-hadron decays and a hypothesis on the nuclear modification factor $R_{\mathrm{pPb}}^\text{feed-down}$ of D mesons from B decays. On the basis of calculations including initial state effects through the EPS09 nuclear PDF parametrisations~\cite{Eskola:2009uj} or the Color Glass Condensate formalism~\cite{Fujii:2013yja}, 
 it was assumed that the $R_{\mathrm{pPb}}$ of prompt and feed-down D mesons were equal and their ratio was varied in the range $0.9 < R^{\text{feed-down}}_{\mathrm{pPb}}/R^{\mathrm{prompt}}_{\mathrm{pPb}} < 1.3$ to evaluate the systematic uncertainties.  
The resulting $f_{\rm prompt}$ values and their uncertainties are shown 
in the right-hand panels of Fig.~\ref{fig:fprompt} for $\Dzero$, $\Dplus$ and 
$\Dstar$ mesons in the $|y_{\rm lab}|<y_{\mathrm{fid}}(\pt)$ interval.
The central values of $f_{\rm prompt}$ range between 0.81 and 0.96 depending on 
D-meson species and $\pt$ with no significant rapidity dependence.

\subsection{Systematic uncertainties}
\label{sec:topsystematics}

The systematic uncertainties on the raw yield values were determined for each $\pt$ and $y$ interval by repeating the fit in a different mass range, by varying the background fit function and by counting the candidates in the invariant-mass 
region of the signal peak after subtracting the background estimated from the 
side bands. 
The alternative background fit functions considered were a linear or a second order polynomial function for $\Dzero$, $\Dplus$ and $\Ds$ and $a \cdot (\Delta M - m_{\pi})^b$  for the $\Dstar$.  
For the $\Dzero$ meson, the systematic uncertainty on the raw yield extraction also includes a contribution due to signal candidates reconstructed when swapping the masses of the final state kaon and pion ({\it reflections}). 
This contribution, which is strongly reduced by the PID selection, was  estimated to be 3\% (4\%) at low (high) $\pt$ based on the invariant-mass distribution of these candidates in the simulation.   

For $\Ds$ mesons, it was also verified that the contribution to the measured 
yield due to other decay channels giving rise to the same $\KKpi$ final state,
in particular $\DstoKzerostarK$ and $\Dstofzeropi$, is completely negligible 
due to the much lower efficiency for the selection of these decays induced 
by the cut on the KK invariant mass in combination with the kaon 
and pion identification~\cite{Abelev:2012tca}.

The systematic uncertainty on the tracking efficiency was estimated by comparing the probability to match the TPC tracks to the ITS hits in data and 
simulation, and by varying the track quality selection criteria. 
It amounts to 3\% for each track, which results in a 6\% uncertainty for the two-body decay of $\Dzero$ mesons and 9\% for $\Dplus$, $\Dstar$, and $\Ds$ 
mesons, which are reconstructed from three-body final states. 

The systematic uncertainty on the D-meson selection efficiency reflects 
residual discrepancies between data and simulations on the variables used in
the displaced decay-vertex topology selection criteria.
This effect was estimated by repeating the analysis with different values of 
the selection cuts, which significantly vary the signal-to-background ratio 
and efficiencies. 
The value of the uncertainty was estimated from the variation of the corrected yields. The systematic uncertainties are largest at low $\pt$, where the efficiencies are lowest, and decrease with increasing $\pt$, with no dependence on rapidity.

The systematic uncertainty associated with particle identification was estimated for $\Dzero$, $\Dplus$ and $\Dstar$ mesons by comparing the corrected yields with and without applying PID to select pions and kaons. 
The results for the two cases were found to be compatible; therefore no 
systematic uncertainty was assigned.  
In the $\Ds$ case, due to the tighter kaon and pion identification criteria, 
a PID systematic uncertainty of 10\% in the interval $2<\pt<4~\gevc$ and 5\% 
at $\pt>4~\gevc$ was estimated by varying the PID selection criteria with 
the procedure described in Ref.~\cite{Abelev:2012tca}.

The effect on the efficiencies due to the shape of the simulated D-meson $\pt$ distribution was evaluated by considering different shapes (PYTHIA, FONLL) and was found to range from 0 to 4\% depending on $\pt$. No significant systematic effect is induced by the rapidity distribution of the generated D mesons because the efficiency does not have a pronounced rapidity dependence.
The effect of possible differences between the charged-multiplicity distributions in data and simulations was found to be negligible.

The systematic uncertainty due to the subtraction of feed-down D mesons from B decays was estimated as in previous measurements~\cite{ALICE:2011aa} by varying the FONLL parameters (b-quark mass, factorisation and renormalisation scales) as prescribed in~\cite{Cacciari:2012ny} and by varying the hypothesis on the $R_{\rm pPb}^\text{feed-down}$ as described in Section~\ref{sec:topcorrections}. An alternative method based on the ratio of FONLL predictions for D and B meson cross sections was also used~\cite{ALICE:2011aa}. 

The cross sections have a systematic uncertainty on the normalisation induced by the uncertainties on the integrated luminosity (3.7\%~\cite{Abelev:2014epa}) and on the branching ratios of the considered D-meson decays. 

A summary of the systematic uncertainties is reported in Tables~\ref{tab:topolsystVsPt} and~\ref{tab:topolsystVsY}. 
The systematic uncertainties on PID, tracking and selection 
efficiencies are mostly correlated among the different $\pt$ and rapidity 
intervals, while the raw-yield extraction uncertainty is mostly uncorrelated.

\begin{table}[htp]
\begin{center}
\begin{tabular}{l|cc|cc|cc|cc}
\hline
					&\multicolumn{2}{c}{$\Dzero$}	&\multicolumn{2}{c}{$\Dplus$}	&\multicolumn{2}{c}{$\Dstar$}	&\multicolumn{2}{c}{$\Ds$}	\\
\hline
$\pt$ interval ($\gevc$)	&1--2	& 5--6 &1--2	&5--6&1--2&12--16&2--4	& 6--8\\
\hline
Raw yield extraction                    & 8\%	& 4\%	&10\%                   &5\%	&\phantom{0}8\%	&\phantom{0}2\%	&10\%	        &\phantom{0}5\%\\
Correction factor & & & & \\
$\qquad$ Tracking efficiency 		& 6\%	& 6\%	&\phantom{0}9\% 	&9\%	&\phantom{0}9\%	&\phantom{0}9\%	&\phantom{0}9\%	&\phantom{0}9\%\\
$\qquad$ Selection efficiency 		& 8\%	& 5\%	&10\%	                &6\%	&10\%	        &\phantom{0}5\%	&15\%	        &15\%\\
$\qquad$ PID efficiency			& negl.	& negl.	& negl.	                &negl.	&negl.	        &negl.	        &10\%	        &\phantom{0}5\%\\
$\qquad$ MC $\pt$ shape		 	& 2\%	& negl.	& 2\%	& negl.	&3\%	&1\%	&\phantom{0}4\%	&\phantom{0}4\%\\
$\qquad$ MC $N_{\rm ch}$ shape		& negl.	& negl.	& negl.	                &negl.	&negl.	        &negl.	&negl.	&negl.\\[1ex]	
Feed-down from B	& $^{+\phantom{0}5}_{-47}\%$	& $^{+\phantom{0}5}_{-12}\%$	& $^{+\phantom{0}1}_{-22}\%$	& $^{+3}_{-7}\%$	
          &$^{+\phantom{0}2}_{-30}\%$	& $^{+2}_{-5}\%$	& $^{+\phantom{0}4}_{-24}\%$ & $^{+\phantom{0}7}_{-14}\%$\\[1ex]
\hline
Luminosity  
  & \multicolumn{2}{c|}{3.7\%} & \multicolumn{2}{c|}{3.7\%} & \multicolumn{2}{c|}{3.7\%} & \multicolumn{2}{c}{3.7\%}\\
Branching ratio		
  & \multicolumn{2}{c|}{1.3\%}	&\multicolumn{2}{c|}{2.1\%}&\multicolumn{2}{c|}{1.5\%}&\multicolumn{2}{c}{4.5\%}\\
\hline
\end{tabular}
\caption{Relative systematic uncertainties on prompt D-meson production cross sections in p--Pb collisions in two $\pt$ intervals and the rapidity range $|y|<y_{\rm fid}(\pt)$. 
\label{tab:topolsystVsPt}
}
\end{center}
 \end{table}
\begin{table}[htp]
\begin{center}
\begin{tabular}{l|cc|cc|cc}
\hline
&\multicolumn{2}{c}{$\Dzero$}	&\multicolumn{2}{c}{$\Dplus$}	&\multicolumn{2}{c}{$\Dstar$}\\
\hline
$y_{\rm lab}$ interval 	&$-$0.1,0.1&0.4, 0.8 &$-$0.1,0.1&0.4, 0.8  &$-$0.1,0.1&0.4, 0.8 \\
\hline
Raw yield extraction  		&10\%	&6\%	&5\%	&5\%	&3\%	&6\%	\\
Correction factor & & & & \\
$\qquad$ Tracking efficiency 		&\phantom{0}6\%	&6\%	&9\%	&9\%	&9\%	&9\%	\\
$\qquad$ Selection efficiency		&\phantom{0}5\%	&5\%	&8\%	&8\%	&5\% 	&5\%	\\
$\qquad$ PID efficiency			&negl.	&negl.	&negl.	&negl.	&negl.	&negl.	\\
$\qquad$ MC $\pt$ shape 			&\phantom{0}3\%	&3\%	&5\%	&5\%	&5\%	&5\%	\\
$\qquad$ MC $N_{\rm ch}$ shape		& negl.	& negl.	& negl.	&negl.	&negl.	&negl.	\\[1ex]	
Feed-down from B	&$^{+\phantom{0}5}_{-11}\%$	& $^{+\phantom{0}5}_{-11}\%$	& $^{+3}_{-7}\%$	& $^{+3}_{-7}\%$	& $^{+2}_{-5}\%$	& $^{+2}_{-4}\%$	\\[1ex]
\hline
Luminosity  
  & \multicolumn{2}{c|}{3.7\%} & \multicolumn{2}{c|}{3.7\%} & \multicolumn{2}{c}{3.7\%} \\
Branching ratio			& \multicolumn{2}{c|}{1.3\%}	&\multicolumn{2}{c|}{2.1\%}&\multicolumn{2}{c}{4.5\%}\\
\hline
\end{tabular}
\caption{Relative systematic uncertainties on prompt D-meson production cross sections in p--Pb collisions in the $\pt$ interval $5<\pt<8~\gevc$ and two rapidity intervals.
\label{tab:topolsystVsY}
}
\end{center}
 \end{table}

\subsection{Prompt fraction with a data-driven approach}
\label{sec:topfprompt}

\begin{figure}[!tb]
\begin{center}
\includegraphics[width=\textwidth]{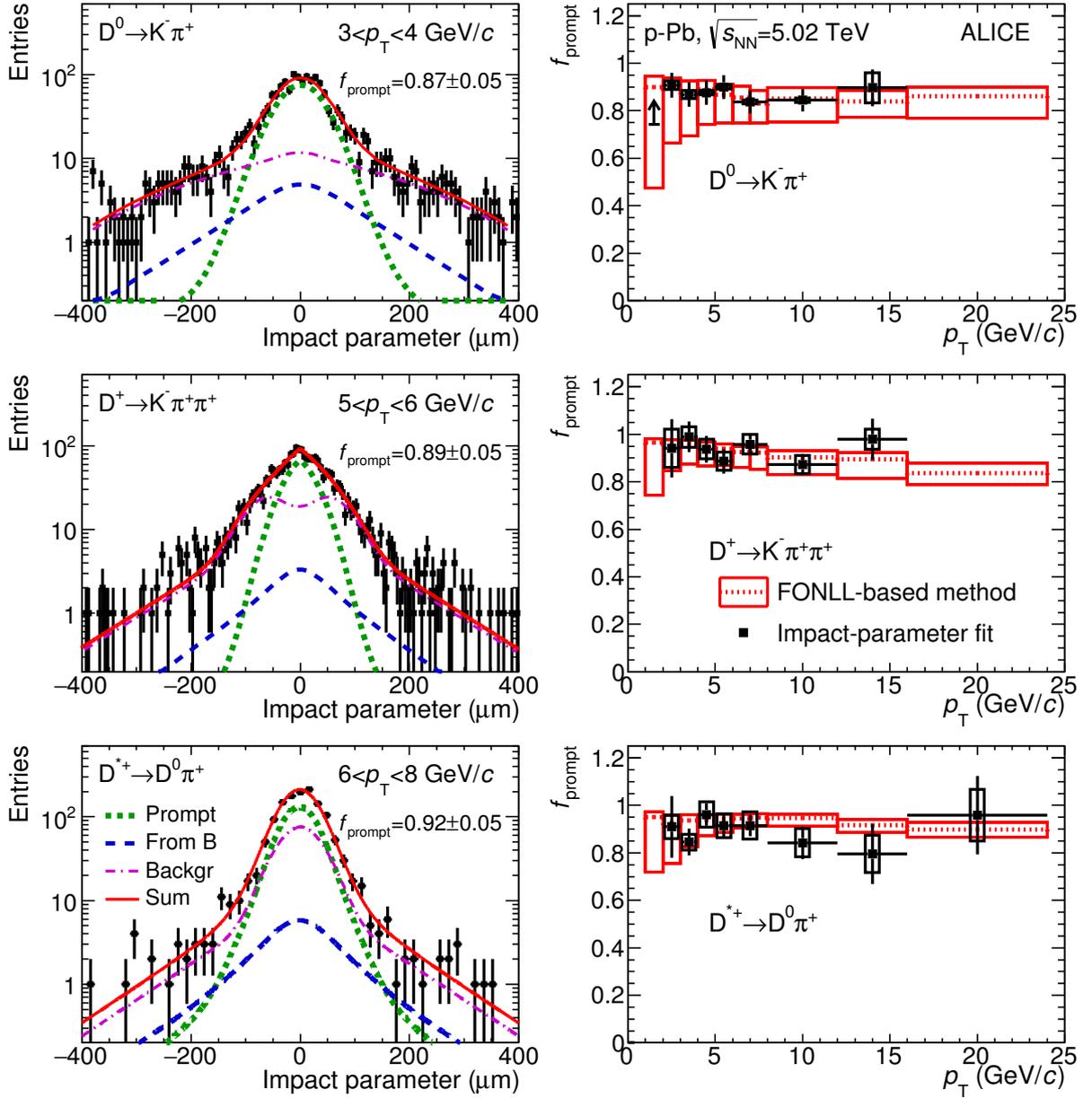}
\caption{Left: Examples of fits to $\Dzero$ (top), $\Dplus$ (middle) and $\Dstar$ (bottom) impact-parameter distributions in the $\pt$ intervals 
$3<\pt<4~\gevc$, $5<\pt<6~\gevc$ and $6<\pt<8~\gevc$, respectively. 
The curves show the fit functions describing the prompt, feed-down and background contributions, as well as their sum, as described in the text.
Right: fraction of prompt $\Dzero$ (top), $\Dplus$ (middle) and $\Dstar$ (bottom) raw yield as a function of $\pt$ compared to the FONLL-based approach. 
The results from the data-driven method are shown as square markers with the 
error bars (boxes) representing the statistical (systematic) uncertainty.  
The arrow in the interval $1<\pt<2~\gev/c$ represents the minimum value within a 95\% confidence level.
The central values of $f_{\rm prompt}$ from the FONLL-based approach are shown
by the dashed line and their uncertainty by the red boxes.}
\label{fig:fprompt} 
\end{center}
\end{figure}

The prompt fractions in the raw yields of $\Dzero$, $\Dplus$ and $\Dstar$ 
mesons, $f_{\rm prompt}$, calculated with the FONLL-based method of 
Eq.~(\ref{eq:fpr}) were cross-checked with a data-driven method that exploits 
the different shapes of the distributions 
of the transverse-plane impact parameter to the primary vertex ($d_0$) of 
prompt and feed-down D mesons. 
The prompt fraction was estimated via an unbinned likelihood fit of the $d_0$ distribution of $\Dzero$($\Dplus$)-meson candidates 
with invariant mass $|M-M_{\mathrm{D}}|<1.5(2) \sigma$ (where $\sigma$ is the
width of the Gaussian function describing the D-meson signal in the 
invariant-mass fits)
and of $\Dstar$-meson candidates with a mass difference $|\Delta M-\Delta M_{\mathrm{D^{*+}}}|<2.5 \sigma$, 
using the fit function 
\begin{equation}
 F(d_0) = S\cdot \left[ (1-f_{\rm prompt}) F^{\rm feed\mbox{-}down} (d_0) + f_{\rm prompt} F^{\rm prompt} (d_0) \right] + B\cdot F^{\rm backgr}(d_0)\,.
  \label{eq:ImpParFit}
\end{equation} 
In this function, $S$ and $B$ are the signal raw yield and background 
in the selected invariant-mass range; $F^{\rm prompt} (d_0)$, 
$F^{\rm feed\mbox{-}down} (d_0)$  and $F^{\rm backgr}(d_0)$ 
are functions describing the impact parameter distributions of prompt D mesons, feed-down D mesons, and background, respectively. 
The function $F^{\rm prompt}$ is a detector resolution term modelled with a Gaussian and a symmetric exponential term, 
$\frac{1}{2\lambda}\exp{\left( -\frac{\lvert d_0 \rvert}{\lambda} \right) }$, 
describing the tails of the impact-parameter distribution of prompt 
D mesons. $F^{\rm feed\mbox{-}down}$ is the convolution of the detector resolution term with a symmetric double-exponential
function ($F^{\rm feed\mbox{-}down}_{\rm true}$) describing the intrinsic impact parameter distribution of secondary 
D mesons from B-meson decays, which is determined by the decay length and decay kinematics of B mesons. 
The parameters of the $F^{\rm prompt}$ and $F^{\rm feed\mbox{-}down}_{\rm true}$ functions were fixed to the values obtained by fitting the distributions from Monte Carlo simulations,
except for the Gaussian width of the detector-resolution term, which was kept free in the data fit to compensate for a possible imperfect description of the impact-parameter resolution in the simulation. 
The widths recovered from the fit to the data were found to be in agreement 
with the simulation for $\pt>3~\GeV/c$ and slightly larger at lower $\pt$.
For $\Dzero$ and $\Dstar$ mesons, the background fit function, $F^{\rm backgr}$, 
is the sum of a Gaussian and a symmetric exponential term centred at 
zero. For $\Dplus$ mesons, the background impact-parameter distribution has a double-peak structure with a depletion around zero 
induced by the selections applied. The shape was thus modelled with two Gaussians and two symmetric exponential terms. The parameters of 
$F^{\rm backgr}$ were fixed by fitting the impact parameter distribution of background candidates in the side-bands of the signal peak in the invariant-mass 
distributions (mass difference for $\Dstar$ mesons), namely in the interval $4\sigma < |M-M_{\mathrm{D^{0},D^{+}}}| < 15\sigma$ ($6\sigma < \Delta M-\Delta M_{\mathrm{D^{*+}}} < 15\sigma$). Figure~\ref{fig:fprompt} (left) 
shows examples of fits to the impact-parameter distributions of $\Dzero$, $\Dplus$ and $\Dstar$ 
mesons in the transverse-momentum intervals $3<\pt<4~\gevc$, $5<\pt<6~\gevc$ and $6<\pt<8~\gevc$, respectively. 

The prompt fraction estimated with the data-driven approach has systematic uncertainties due to 
i) the shape assumed for prompt D-meson, feed-down D-meson, 
and background impact-parameter distributions, 
ii) the uncertainty on the signal and background yields, and
iii) the consistency of the procedure, evaluated with a Monte Carlo closure test.
Several checks were carried out to estimate 
the systematic uncertainty from the shape assumed for the 
impact-parameter distributions of the prompt and feed-down components.
The fit was repeated fixing the Gaussian width in the $F^{\rm prompt}$ 
functions to the values expected from the simulation and using 
template distributions from the simulation in place of the 
$F^{\rm feed\mbox{-}down}$ and $F^{\rm prompt}$ functional forms.
Furthermore, the stability of the results against a possible imperfect 
description of the impact parameter resolution in the simulation was verified 
with a dedicated ``fast'' simulation in which the reconstructed track 
properties were modified to match the impact parameter resolution measured in 
data, following the procedure described in~\cite{Musa:1475244}. 
In addition, the fit procedure was also repeated after tuning the $\pt$ 
distributions of prompt and feed-down D mesons in the simulation to match those
predicted by FONLL calculations. 
The uncertainty deriving from the parametrisation of $F^{\rm backgr}$ was 
estimated by extracting the background impact-parameter distribution from 
different invariant-mass intervals. Overall, the systematic uncertainty arising from the shape assumed for prompt D-meson, 
feed-down D-meson, and background impact parameter distributions is typically smaller than $4\%$. 
The systematic effect due to the uncertainty on the signal and background yields was determined by repeating the fit with $S$ and $B$ varied
according to the quadratic sum of the statistical and systematic uncertainties
on the raw yield described in Section~\ref{sec:topsystematics}. The resulting deviation of the prompt D-meson 
raw yield, $f_{\rm prompt}\cdot S$, was used to define the related systematic uncertainty, which ranges
from $0$ to $10\%$ depending on the meson species and $\pt$, with typical values around $2\%$ at intermediate $\pt$. It 
was also checked that the variation of the width of the invariant-mass (mass-difference for $\Dstar$ mesons) interval around the D-meson peak
in which $f_{\rm prompt}$ is evaluated yields a sizable effect ($3\%$) only for $\Dstar$ mesons. 
Finally, a Monte Carlo closure test was carried out to verify the 
consistency of the procedure with simulated data by comparing the 
$f_{\rm prompt}$ values recovered with the impact-parameter fit and the input 
ones: the difference,
typically about $1\%$, was considered as a systematic uncertainty.
The total systematic uncertainty on $f_{\rm prompt}$ with the data-driven 
approach is about $2\%$ for $\Dzero$ mesons and $5\%$ for $\Dplus$ and 
$\Dstar$ mesons for $\pt<12~\gev/c$, and increases at higher $\pt$ up to $11\%$ 
for $\Dstar$ mesons in the interval $16<\pt<24~\gev/c$. 

The prompt fraction of $\Dzero$, $\Dplus$ and $\Dstar$ mesons 
measured with this method is shown in Fig.~\ref{fig:fprompt} (right). 
For the interval $1<\pt<2~\gev/c$, given the poor precision of 
the impact-parameter fit, a lower limit could be estimated only for 
$\Dzero$ mesons at a $95\%$ confidence level on the basis of statistical 
and systematic uncertainties.
For the same reason, in the highest $\pt$ interval, $16<\pt<24~\gev/c$, 
the prompt fraction could be determined with the data-driven 
method only for $\Dstar$ mesons.
The prompt fraction measured with the impact-parameter fits is found to be 
compatible with the FONLL-based estimation within uncertainties.
For $\Dzero$ mesons, the data-driven approach provides a more precise
determination of the prompt fraction, while for $\Dstar$ and $\Dplus$ mesons
smaller uncertainties are obtained with the FONLL-based method. 
In addition, the data-driven results are not available at low 
$\pt$ ($\pt<2~\gev/c$) and, for $\Dzero$ and $\Dplus$ mesons, at high $\pt$ 
($\pt>16~\gev/c$).
Finally, it should also be considered that the systematic uncertainty on 
the FONLL-based $f_{\rm prompt}$ calculation partially cancels in the 
computation of the nuclear modification factor, because it is correlated between
the p--Pb cross-section and the pp reference.
Note that for the data sample of pp collisions at $\sqrt{s}=7~\TeV$ used
to compute the reference for the nuclear modification factor, 
$f_{\rm prompt}$ could be measured with the data-driven method only for $\Dzero$ 
mesons with poor statistical precision in a limited $\pt$ interval 
($2<\pt<12~\GeV/c$)~\cite{ALICE:2011aa}.
For these reasons, the FONLL-based method was used in the calculation 
of the production cross sections and nuclear modification factors  with 
the current data samples. 
The analysis presented here demonstrates that the data-driven method will 
become fully applicable on the upcoming larger data samples.

\section{$\mathbf{\Dzero}$ analysis in pp and p--Pb collisions without decay-vertex reconstruction}
\label{sec:lowpt}

\subsection{Analysis method}
\label{sec:lowmeth}

In order to extend the measurement of D-meson production to
$\pt<1~\GeV/c$, a different analysis method, not based on 
geometrical selections on the displaced decay-vertex topology, was developed
for the two-body decay $\DtoKpi$ (and its charge conjugate).
Indeed, at very low $\pt$, the D-meson decay topology can not be efficiently 
resolved because of the insufficient resolution of the track impact parameter 
and the small Lorentz boost.
Furthermore, selection criteria based on secondary-vertex displacement 
tend to select with higher efficiency non-prompt D mesons from beauty-hadron
decays, thus increasing the systematic uncertainty on 
the subtraction of the beauty feed-down contribution.
Using an analysis technique mainly based on particle identification and 
on the estimation and subtraction of the combinatorial background,
it was possible to measure the $\Dzero$-meson yield down to $\pt=0$
in pp and p--Pb collisions.

\begin{figure}[!tb]
\begin{center}
\includegraphics[width=\textwidth]{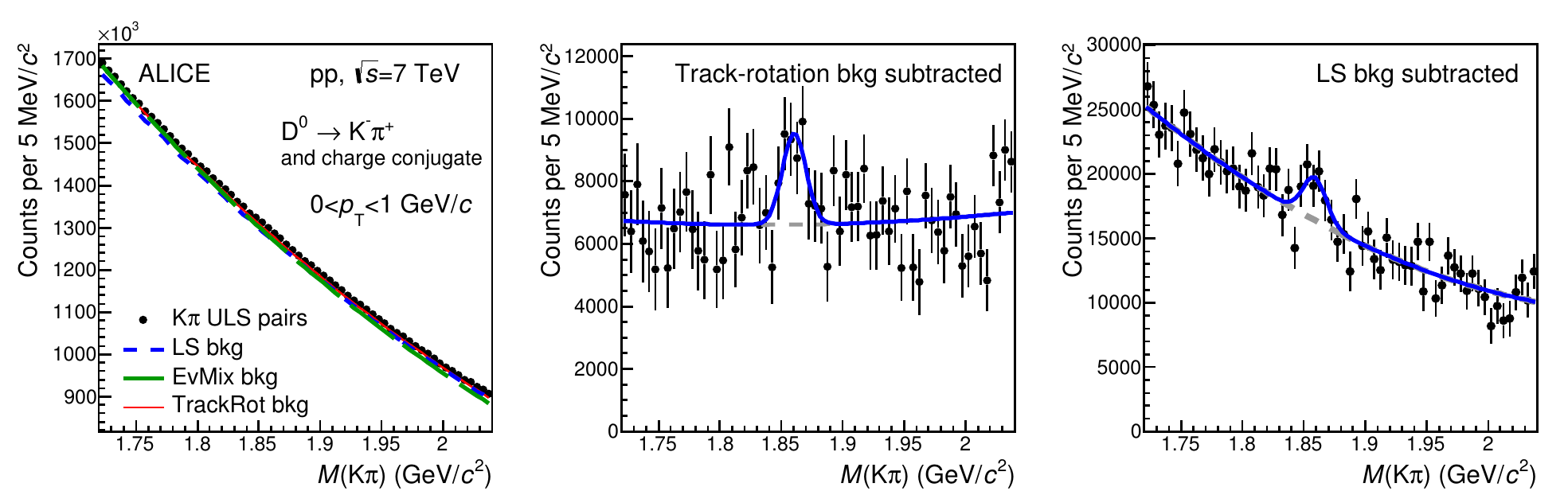}
\includegraphics[width=\textwidth]{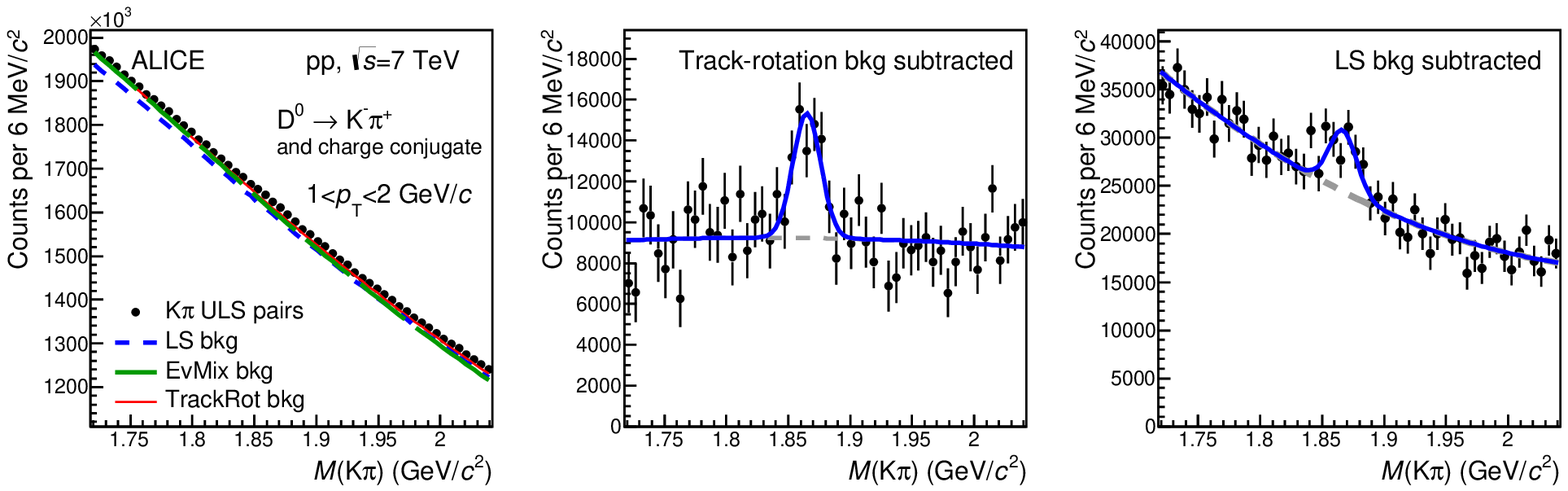}
\caption{Invariant-mass distributions of $\DtoKpi$ candidates (and
charge conjugates) in pp collisions at $\sqrts=7~\TeV$ for
two $\pt$ intervals: $0<\pt<1~\gev/c$ (top panels) and $1<\pt<2~\gev/c$ 
(bottom panels). For both $\pt$ intervals, the left panels display
the invariant-mass distribution of all ULS K$\pi$ pairs together with the
background distributions estimated with the LS, event-mixing and 
track-rotation techniques. The middle and right panels show the invariant-mass 
distributions after subtraction of the background from the track-rotation 
and LS techniques. Fit functions are superimposed.}
\label{fig:invmasssplowptpp} 
\end{center}
\end{figure}

\begin{figure}[!tb]
\begin{center}
\includegraphics[width=\textwidth]{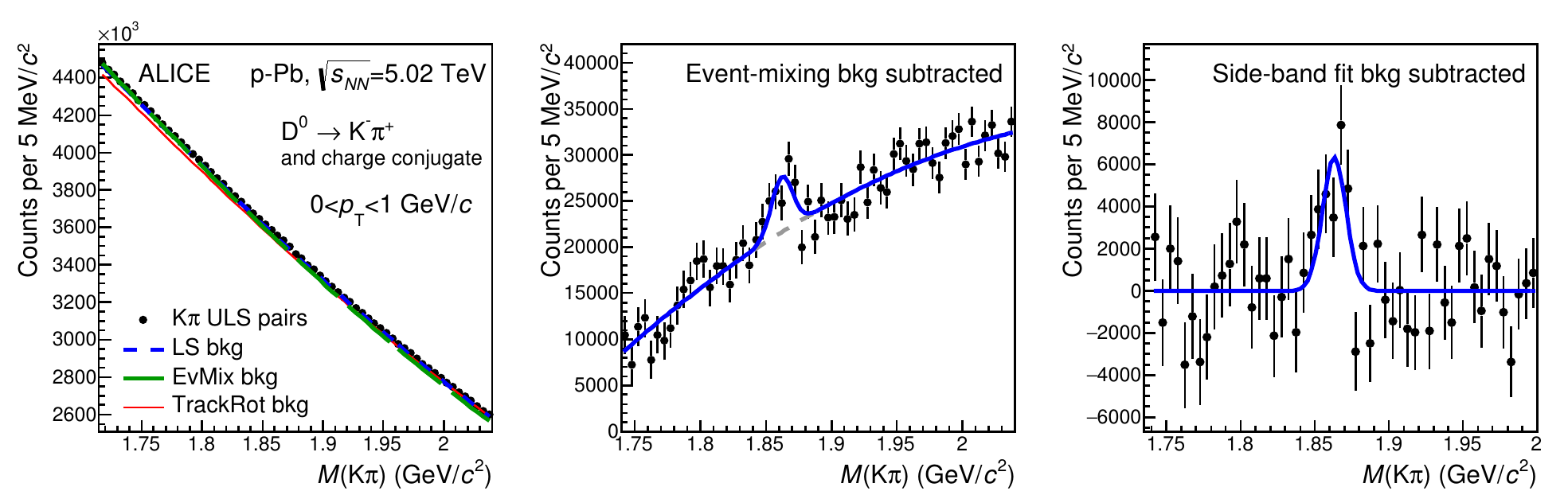}
\includegraphics[width=\textwidth]{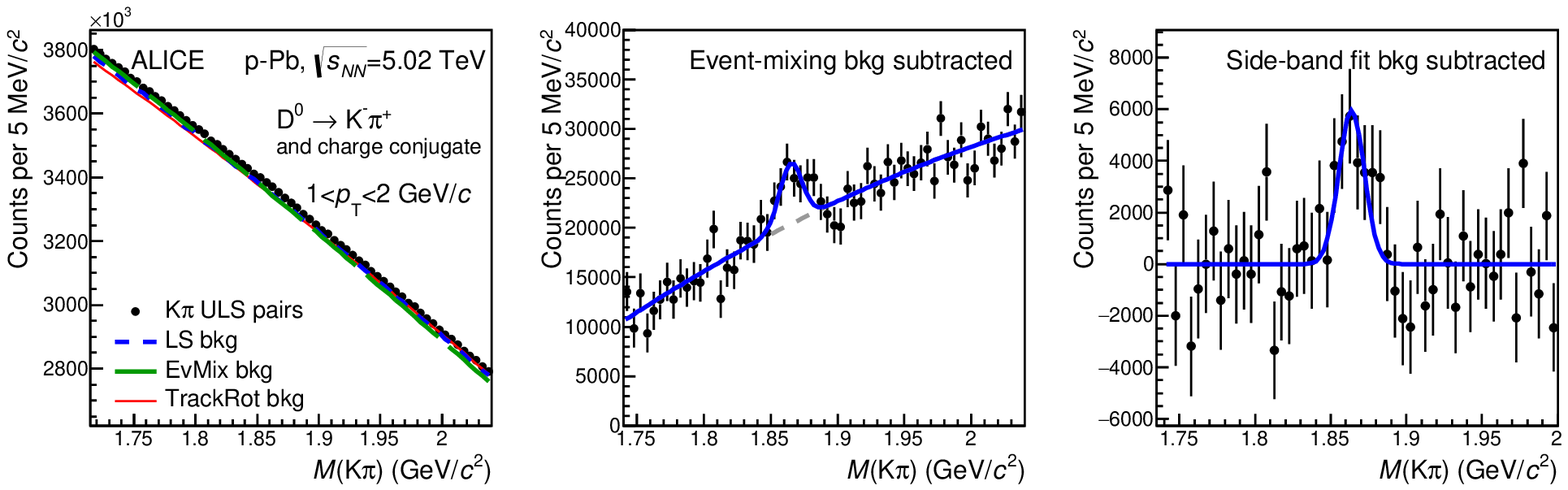}
\caption{Invariant-mass distributions of $\DtoKpi$ candidates (and
charge conjugates) in p--Pb collisions at $\sqrtsNN=5.02~\TeV$ for
two $\pt$ intervals: $0<\pt<1~\gev/c$ (top panels) and $1<\pt<2~\gev/c$ 
(bottom panels). For both $\pt$ intervals, the left panels display
the invariant-mass distribution of all ULS K$\pi$ pairs together with the
background distributions estimated with the LS, event-mixing and 
track-rotation techniques. The middle and right panels show the invariant-mass 
distributions after subtraction of the background from the event-mixing
and side-band fit techniques. Fit functions are superimposed.}
\label{fig:invmasssplowptpPb} 
\end{center}
\end{figure}

The $\Dzero$ yield was extracted in eight $\pt$ intervals in the range 
$0<\pt<12~\GeV/c$ from an invariant-mass analysis of 
pairs of kaons and pions with opposite charge sign (UnLike Sign, ULS).
$\Dzero$ candidates were defined from tracks with $|\eta|<0.8$ and 
$\pt>0.3~\gev/c$ ($0.4~\gev/c$ in the p--Pb analysis).
Tracks were selected with the same criteria described in 
Section~\ref{sec:selection} for the analysis with decay-vertex reconstruction,
with the only difference that the request of at least one 
hit in either of the two layers of the SPD was not applied for pp collisions.
Pion and kaon identification was based on the same strategy used 
in the analysis with decay-vertex reconstruction, i.e.\ based on
compatibility selections at $3\,\sigma$ level between the measured and 
expected d$E$/d$x$ in the TPC and time-of-flight from the interaction 
vertex to the TOF detector.
Tracks without TOF information were identified based only on the TPC
d$E$/d$x$ signal.
The resulting $\Dzero$ and $\Dzerobar$ candidates were selected
by applying a fiducial acceptance cut $|y_{\rm lab}|<0.8$ on their rapidity.
As compared to the analysis with decay-vertex reconstruction described in 
Section~\ref{sec:selection}, a wider fiducial acceptance region was used 
in this analysis to preserve more candidates at low $\pt$.
The resulting invariant-mass distributions of K$\pi$ pairs in the transverse
momentum intervals $0<\pt<1~\GeV/c$ and $1<\pt<2~\GeV/c$ are shown 
in the left-hand panels of 
Figs.~\ref{fig:invmasssplowptpp} and~\ref{fig:invmasssplowptpPb} for pp and 
p--Pb collisions, respectively.

Four different techniques were used to estimate the background distribution: 
(i) like-sign pairs; (ii) event mixing; (iii) track rotation; and 
(iv) side-band fit.
The like-sign (LS) method is based on K$\pi$ combinations with same charge 
sign.
In each $\pt$ interval, the ULS background invariant-mass 
distribution was estimated from the LS ones as 
$N_{\rm K^+\pi^-}~=~2\cdot\sqrt{N_{\rm K^+\pi^+} \cdot N_{\rm K^-\pi^-}}$,
where $N_{\rm K^+\pi^+}$ and $N_{\rm K^-\pi^-}$ are the number of like-sign K$\pi$ 
pairs in a given invariant-mass interval.
The event-mixing method estimates the uncorrelated background by pairing
each kaon of a given event with all pions of other events
having similar multiplicity and vertex position along the beam axis.
In the track-rotation technique, for each $\Dzero$ (and $\Dzerobar$) 
candidate, up to nine combinatorial-background-like candidates were created 
by rotating the kaon track by different angles in the range between
$\frac{5\pi}{6}$ and $\frac{7\pi}{6}$ radians in azimuth.
In the case of the event-mixing and track-rotation methods, the background is 
normalized to match the yield of K$\pi$ pairs at one edge of the invariant-mass 
range considered for the extraction of the $\Dzero$ raw yield.

The invariant-mass distributions of background candidates estimated with
these three methods (i--iii) are shown as lines in the left panels of 
Figs.~\ref{fig:invmasssplowptpp} and~\ref{fig:invmasssplowptpPb} for the 
pp and p--Pb cases, respectively.
The background distribution is subtracted from the ULS K$\pi$ invariant-mass 
distribution. 
Some examples of the resulting distributions, which contain the $\Dzero$ signal and 
the remaining background, are shown in Fig.~\ref{fig:invmasssplowptpp}
for the track-rotation (middle panels) and LS (right-hand panels) methods in 
pp interactions and in the middle panels of Fig.~\ref{fig:invmasssplowptpPb} 
for the event-mixing method in p--Pb collisions.
The $\Dzero$ raw yield (sum of particle and antiparticle contributions)
was extracted via a fit to the background-subtracted 
invariant-mass distribution.
The fit function is composed of a Gaussian term to describe the signal
and a second-order polynomial function to model the remaining background.

The fourth approach to the background treatment consists of a two-step fit 
to the ULS K$\pi$ invariant-mass distribution.
In the first step, the side bands of the $\Dzero$ peak 
($|M({\rm K\pi})-M(\Dzero)|>2.5\,\sigma$, 
where $\sigma$ is the Gaussian width of the $\Dzero$ peak from the simulation), 
were used to evaluate the background shape, which was modeled 
with a fourth order polynomial for $\pt<2~\GeV/c$ and with a second-order polynomial for $\pt>2~\GeV/c$.
In the second step, the invariant-mass distribution was fitted in the whole
range, using a Gaussian function to model the signal and the 
polynomial function from the previous step to describe the background.
In the right-hand panels of Fig.~\ref{fig:invmasssplowptpPb} the 
invariant-mass distribution of $\Dzero$ candidates after subtracting the 
background estimated from the side bands is shown, together with the Gaussian 
function that describes the signal peak.

In the fits for all four methods, the width of the Gaussian was fixed to 
the value from the simulation, while the centroid was left as a free 
parameter of the fit and was found to be compatible, within uncertainties, 
with the PDG world-average value of the $\Dzero$ mass~\cite{Agashe:2014kda}.

The raw-yield values from the four methods for the background subtraction
were found to be consistent within 10\% in all
$\pt$ intervals of the pp and p--Pb data samples.
The arithmetic average of the four values was, therefore, computed and used in the calculation of the cross sections.
The statistical uncertainties on these average raw-yield values were defined 
as the arithmetic average of the uncertainties from the four
background-subtraction methods.
In the case of the pp sample, the signal-to-background ratio ranges from 
$10^{-3}$ (at low $\pt$) to $2\cdot10^{-2}$ (at high $\pt$), while
the statistical significance is about 4 in the bin $0<\pt<1~\GeV/c$ and 
larger than 6 up to $\pt=4~\GeV/c$.
For the p--Pb sample, the signal-to-background ratio increases from
$7\cdot 10^{-4}$ to $4\cdot10^{-2}$ with increasing $\pt$ and the
statistical significance is about 4 in the two lowest $\pt$
intervals and larger than 7 at higher $\pt$.
The statistical uncertainties on the raw yield are larger than those
obtained in the analysis with decay-vertex reconstruction, except
for the interval $1<\pt<2~\GeV/c$ in the case of pp collisions.
In both pp and p--Pb collisions, this strategy  
allowed the measurement of the $\Dzero$ signal in the 
interval $0<\pt<1~\GeV/c$, which was not accessible with the displaced-vertex 
selection technique.

\subsection{Corrections}
\label{sec:lowcorr}

\begin{figure}[!tb]
\begin{center}
\includegraphics[width=0.48\textwidth]{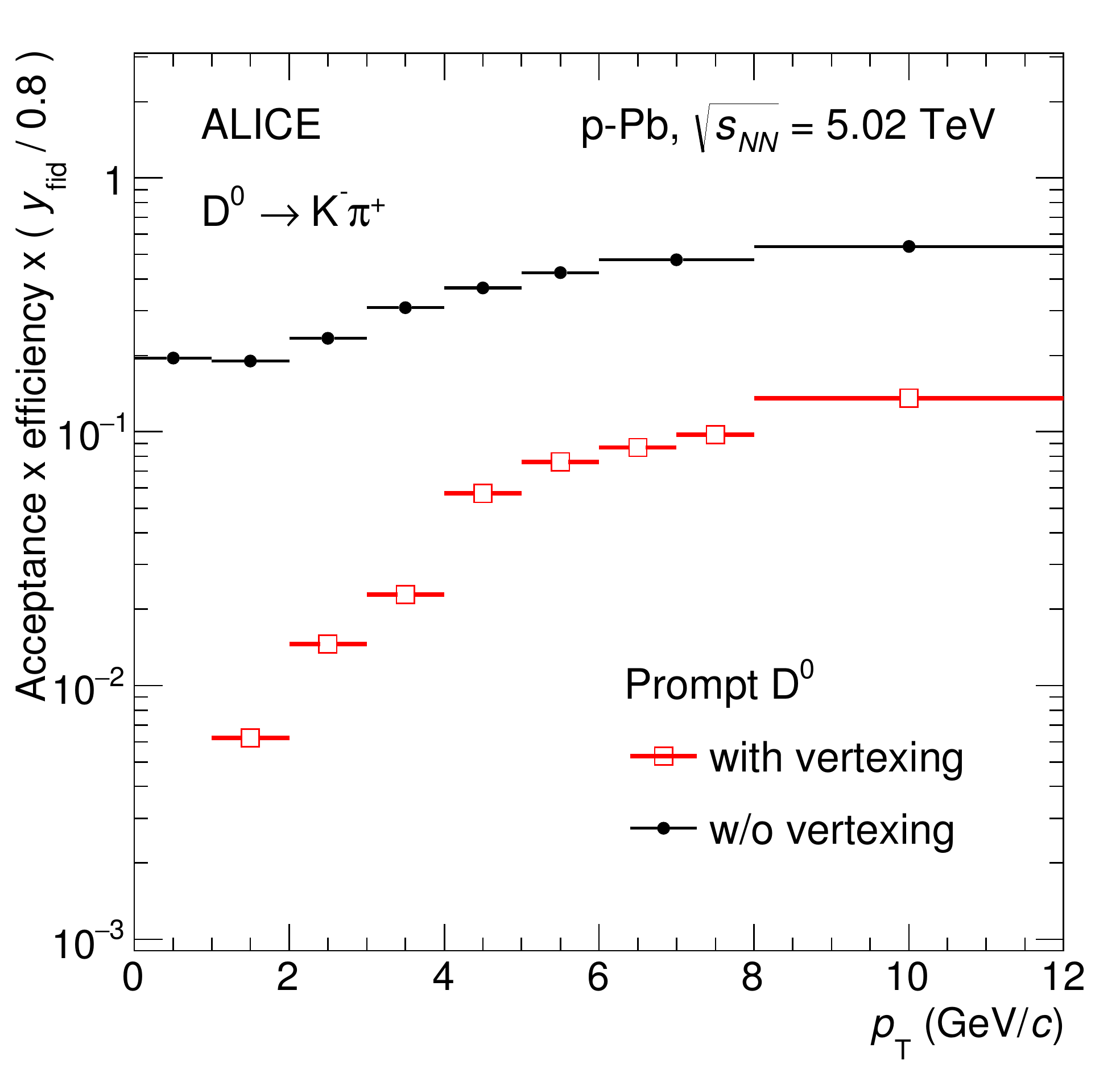}
\caption{Product of acceptance and efficiency in p--Pb collisions.
The ${\rm Acc}\times\epsilon$ values from the analysis with decay-vertex 
reconstruction were rescaled to account for the different fiducial acceptance, 
$y_{\rm fid}(\pt)$, selection on the $\Dzero$ rapidity.}
\label{fig:accefflowpt} 
\end{center}
\end{figure}

The product of the acceptance and the efficiency, ${\rm Acc}\times\epsilon$, 
for $\Dzero$-meson reconstruction and selection with the approach described in
the previous subsection was determined using Monte Carlo simulations.
Events containing prompt and feed-down D-meson signals were simulated using
the PYTHIA~v6.4.21 event generator~\cite{Sjostrand:2006za} with the Perugia-0 tune~\cite{Skands:2010ak}.
In the case of p--Pb collisions, an underlying event generated with HIJING
 1.36~\cite{Wang:1991hta} was added to obtain a realistic multiplicity 
distribution.
The calculation of the $\Dzero$ efficiency was performed utilizing $\pt$ and
event-multiplicity dependent weights, so as to match the D-meson
$\pt$ spectra predicted by FONLL calculations and the measured charged-particle
multiplicity distributions at mid-rapidity.
The resulting ${\rm Acc}\times\epsilon$ of prompt 
$\Dzero$ mesons for the p--Pb sample is shown as a function of $\pt$ 
in Fig.~\ref{fig:accefflowpt} and compared to that for the analysis
with decay-vertex reconstruction.
The efficiency is higher by a factor of about 20 at low $\pt$ (3 at high $\pt$)
in the case of the analysis that does not make use 
of selections on the displacement of the $\Dzero$ decay point.
The $\pt$ dependence of the  ${\rm Acc}\times\epsilon$ is less steep as 
compared to the analysis with decay-vertex reconstruction.
Note that for the analysis without decay-vertex reconstruction the efficiency
is almost independent of $\pt$ and the increase of the 
${\rm Acc}\times\epsilon$ with increasing $\pt$ is mainly determined by the 
geometrical acceptance of the apparatus, i.e.\ by the fraction of $\Dzero$ 
mesons with $|y_{\rm lab}|<0.8$ having the two decay tracks in $|\eta|<0.8$.
Unlike in the analysis with decay-vertex reconstruction,
the efficiency is the same for prompt $\Dzero$
and for $\Dzero$ from beauty-hadron decays, as expected when
no selection is made on the displacement of the $\Dzero$ decay 
vertex from the interaction point.

Since the acceptance and the efficiency are the same for prompt and feed-down
$\Dzero$ mesons, the production cross section for `inclusive' $\Dzero$
mesons (i.e.\ sum of the prompt and feed-down contributions) in the rapidity 
range $|y_{\rm lab}|<0.5$ can be calculated as
\begin{equation}
  \label{eq:crosssectionIncD}
  \frac{{\rm d^2}\sigma^{\rm D^0,incl.}}{{\rm d}\pt {\rm d} y}=
  \frac{1}{\Delta\pt} \cdot \frac{\frac{1}{2} \cdot \left.N^{\rm D^0+\overline{D^0},raw}(\pt)\right|_{|y|<0.8}}{\Delta y} \cdot \frac{1}{({\rm Acc}\times\epsilon)(\pt)} \cdot \frac{1}{{\rm BR} \cdot L_{\rm int}}\,.
\end{equation}
where $N^{\rm D^0+\overline{D^0},raw}(\pt)$ are the $\Dzero$ raw yields.

The production cross section of prompt $\Dzero$ mesons was obtained as
\begin{equation}
  \label{eq:crosssectionPromptD}
  \frac{{\rm d^2}\sigma^{\rm D^0,prompt}}{{\rm d}\pt {\rm d} y}=
  f_{\rm prompt}(\pt)\cdot 
  \frac{{\rm d^2}\sigma^{\rm D^0,incl.}}{{\rm d}\pt {\rm d} y}\,.
\end{equation}

The values of $f_{\rm prompt}$ were estimated 
with the same pQCD-based method used for the analysis with decay-vertex 
reconstruction as described in Section~\ref{sec:topcorrections}.
The resulting $f_{\rm prompt}$ values are similar for pp and p--Pb collisions:
they decrease with increasing $\pt$, from a value of 
about 0.96 at low $\pt$ ($\pt<4~\GeV/c$) to about 0.89 in the interval 
$8<\pt<12~\gev/c$.
The prompt contribution to the $\Dzero$-meson raw yield is larger than 
in the analysis with decay-vertex reconstruction, since the feed-down
component is not enhanced by the selection criteria.

\subsection{Systematic uncertainties}
\label{sec:lowsyst}

The following sources of systematic uncertainty were considered for the
prompt $\Dzero$ cross section: 
(i) systematic uncertainty due to the signal extraction from the invariant-mass
distributions;
(ii) systematic uncertainty affecting the ${\rm Acc}\times\epsilon$
correction factor; and 
(iii) systematic uncertainty due to the beauty feed-down subtraction.
In addition, the cross sections are affected by (iv) a global normalisation 
uncertainty, due to the
determination of the integrated luminosity (3.5\% in pp and 3.7\% in p--Pb)
and the $\DtoKpi$ branching ratio (1.3\%).

The systematic uncertainty on the raw yield extraction was estimated in 
each $\pt$ interval and for each of the four background-subtraction
techniques from the distribution of the results obtained by repeating the fit 
to the invariant-mass distributions varying i) the fit range and ii) 
the functions used to model the signal and background contributions.
In particular, an exponential and a third-order polynomial function were 
used as alternative functional forms to describe the background in the LS, event-mixing and track-rotation analyses, while in the analysis with the
side-band technique polynomials of second, third and fourth order were used.
The signal line shape was varied by using Gaussian functions with the
mean fixed to the PDG world-average $\Dzero$ mass and varying the
widths by $\pm 15\%$ with respect to the value expected from Monte Carlo 
simulations, based on the deviations between the Gaussian width values
observed in data and simulations for the analysis with decay-vertex 
reconstruction.
The effect of the signal line shape was also tested by comparing
the raw yields extracted through the fits with those obtained with a method 
based on the counting of the entries in the invariant-mass distributions after 
subtraction of the (residual) background estimated from a fit to the side 
bands of the $\Dzero$ peak.
The r.m.s.\ of the distribution of the raw yield values obtained from the 
fit variations was assigned as the systematic uncertainty.
A possible additional systematic effect could arise from signal candidates 
that pass the selection criteria also when the ${\rm (K,\pi)}$ mass hypothesis 
for the decay tracks is swapped.
A large fraction of these `reflections' is rejected by the applied PID 
selections.
The effect of the remaining contribution was estimated by repeating
the fits including an additional term to describe this `reflected-signal'
based on its invariant-mass shape in Monte Carlo simulations and was
found to be negligible.
The reflection contribution induces a smaller systematic effect than
in the analysis with decay-vertex reconstruction due to the smaller 
signal-to-background ratio.
In the case of background estimation with the event-mixing technique,
the result was found to be stable against variations of the criteria
on vertex position and event multiplicity used to define the samples of
collisions to be mixed. 
The systematic uncertainty was found to be similar for the four different
techniques for the background treatment and dominated,
in all $\pt$ intervals, by the contribution of the signal line shape, 
which is common to all the background-subtraction approaches.
Therefore, when computing the average of LS, event-mixing, 
track-rotation and side-band results, it was propagated as a fully correlated 
uncertainty.

\begin{table}[!tb]
\centering
\begin{tabular}{l|cccc|cccc} 
\hline 
  & \multicolumn{4}{c|}{pp} 
  & \multicolumn{4}{c}{p--Pb}\\
\hline
  & \multicolumn{4}{c|}{$\pt$ interval ($\gev/c$)}
  & \multicolumn{4}{c}{$\pt$ interval ($\gev/c$)}\\
 & 0--1 & 1--2 & 2--3 & 5--6 
 & 0--1 & 1--2 & 2--3 & 5--6\\
\hline
Raw yield extraction 
  & 14\% & 14\% & 10\% & 14\% 
  & 15\% & 15\% & 10\% & 10\% \\  
Correction factor & & & & \\
$\qquad$ Tracking efficiency 
  & \phantom{0}8\% & \phantom{0}8\% & \phantom{0}8\% & \phantom{0}8\%
  & \phantom{0}6\% & \phantom{0}6\% & \phantom{0}6\% & \phantom{0}6\%\\
$\qquad$ Selection efficiency 
  & negl. & negl. & negl. & negl.
  & negl. & negl. & negl. & negl.\\
$\qquad$ PID efficiency 
  & \phantom{0}5\% & \phantom{0}5\% & \phantom{0}3\% & \phantom{0}3\%
  & negl. & negl. & negl. & negl.\\
$\qquad$ MC $\pt$ shape   
  & negl. & negl. & negl. & negl.
  & negl. & negl. & negl. & negl.\\
$\qquad$ MC $N_{\rm ch}$ shape 
  & negl. & negl. & negl. & negl.
  & negl. & negl. & negl. & negl.\\[1ex]
Feed-down from B   
  & $^{+\phantom{0}2}_{-12}\%$ & $^{+\phantom{0}2}_{-23}\%$ & $^{+\phantom{0}2}_{-10}\%$ & $^{+2}_{-4}\%$
  & $^{+2}_{-9}\%$ & $^{+\phantom{0}2}_{-17}\%$ & $^{+2}_{-9}\%$ & $^{+2}_{-4}\%$\\[1ex]
\hline
Luminosity  
  & \multicolumn{4}{c|}{3.5\%} 
  & \multicolumn{4}{c}{3.7\%} \\
Branching ratio       
  & \multicolumn{4}{c|}{1.3\%} 
  & \multicolumn{4}{c}{1.3\%} \\
\hline 
\end{tabular}
\caption{Relative systematic uncertainties on the $\pt$-differential production
cross section of prompt $\Dzero$ mesons in p--Pb and pp collisions
for the analysis without decay-vertex reconstruction.}
\label{tab:SystLowPt}
\end{table}

The uncertainty on the ${\rm Acc}\times\epsilon$ correction factor originates 
from imperfections in the detector description in the Monte Carlo simulations,
which could affect the particle reconstruction, the $\Dzero$-candidate 
selection efficiency, and the kaon and pion identification.
In addition, the correction factor could also be sensitive to the generated 
shapes of the $\Dzero$-meson $\pt$ distribution and of the multiplicity of 
particles produced in the collision. 
The systematic uncertainty on the tracking efficiency, which includes the 
effects of track reconstruction and selection, was estimated by comparing  
the efficiency of track prolongation from the TPC to the ITS between data 
and simulation, and by varying the track quality selections.
It amounts to 4\% per track in the pp sample and 3\% per track in the p--Pb sample.
The stability of the corrected yield was tested against variations of the 
single-track $\pt$ selection and K/$\pi$ identification criteria used to 
form the $\Dzero$ candidates.
No systematic effect was found to be induced by the single-track $\pt$ cut.
In the case of the particle-identification criteria, different selections
were tested, and the corrected yields were found to be compatible with those
from the standard $3\,\sigma$ cut.
Nevertheless, an analysis without applying PID selections could not be 
performed due to the insufficient statistical significance of the signal.
This test was carried out in the analysis with decay-vertex reconstruction,
resulting for the pp sample in an estimated uncertainty of 5\% for 
$\pt<2~\GeV/c$ and 3\% at higher $\pt$, while no systematic uncertainty due 
to the PID was observed in the p--Pb case.
The same uncertainties were therefore assigned to the cross sections obtained
with the analysis without decay-vertex reconstruction.
The effect on the efficiency due to possible differences between the real 
and simulated $\Dzero$ momentum and charged-multiplicity distributions was
studied by varying the input distributions (using the D-meson $\pt$
shapes predicted by FONLL and PYTHIA and the charged-multiplicity distributions
from HIJING and from data) and was found to be negligible.
 
The systematic uncertainty due to the subtraction of the beauty-feed-down
contribution was estimated following the same procedure of the
analysis with decay-vertex reconstruction as described 
in Section~\ref{sec:topsystematics}.
As compared to the analysis with decay-vertex reconstruction,
the smaller contribution of $\Dzero$ from beauty-hadron decays, due to the 
absence of a selection on the decay-vertex topology, results in a smaller 
systematic uncertainty on the feed-down subtraction.

The assigned uncertainties, estimated with the methods described above, 
are reported in Table~\ref{tab:SystLowPt} for four $\pt$ intervals 
and for pp and p--Pb collisions.


\section{Results}
\label{sec:resul}

\subsection{$\Dzero$-meson and $\rm c\overline c$ production cross section in pp collisions at $\sqrt s=7~\tev$}
\label{sec:resultspp}

\begin{figure}[!t]
\begin{center}
\includegraphics[width=0.48\textwidth]{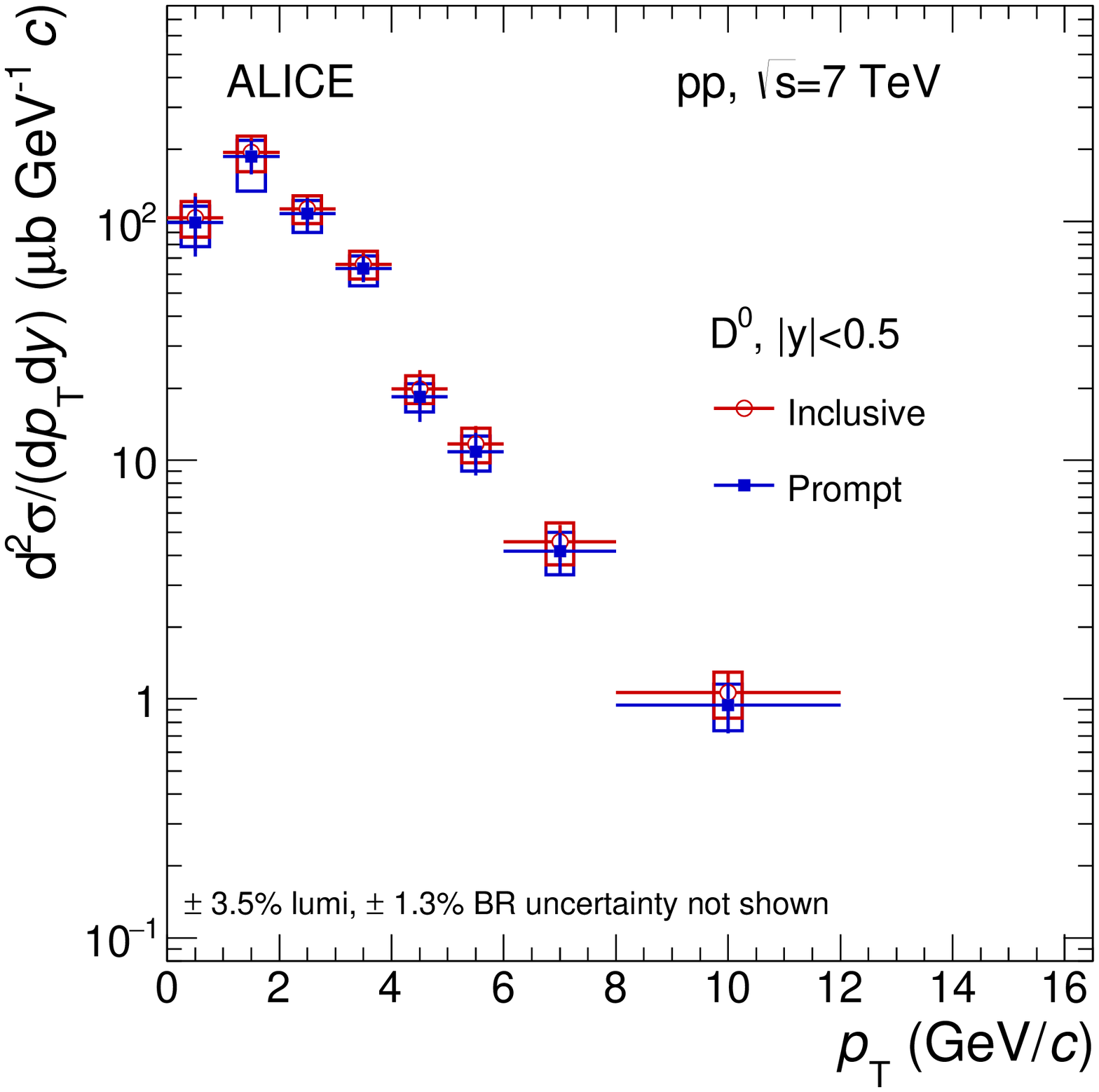}
\includegraphics[width=0.48\textwidth]{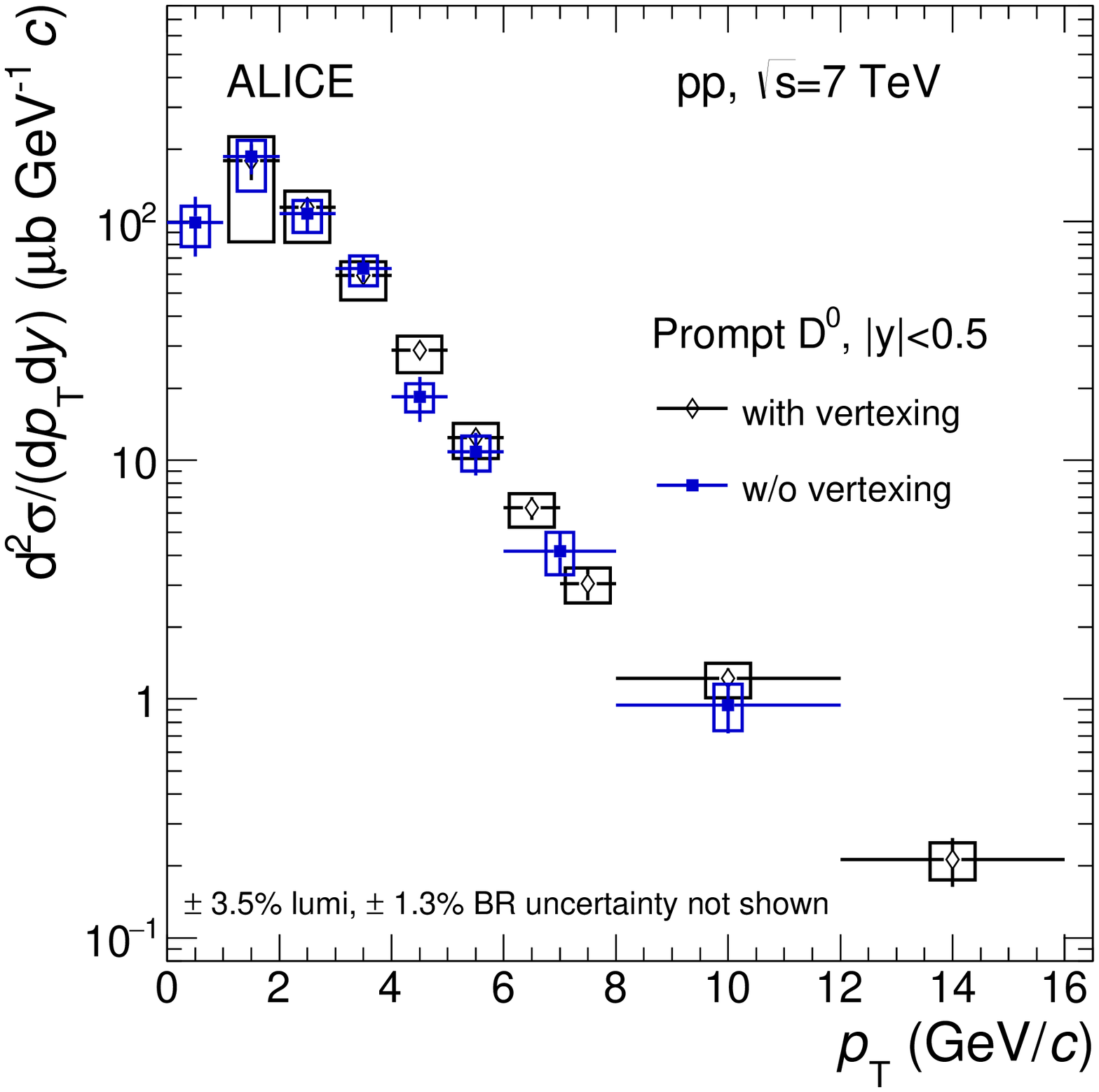}
\caption{$\pt$-differential production cross section of $\Dzero$ mesons with $|y|<0.5$ in pp collisions at $\sqrt s =7~\tev$. Left: comparison of prompt and inclusive $\Dzero$ mesons (the latter including also $\Dzero$ mesons from beauty-hadron decays) from the analysis without decay-vertex reconstruction. Right: comparison between the prompt $\Dzero$ cross sections measured with~\cite{ALICE:2011aa} and without decay-vertex reconstruction. Here and in all the following figures the symbols are plotted at the centre of the $\pt$ intervals (shown by the horizontal lines), the vertical lines represent the statistical uncertainties and the vertical size of the boxes corresponds to the systematic uncertainties.}
\label{fig:CrossSecPP}
\end{center}
\end{figure}

Figure~\ref{fig:CrossSecPP} shows the $\pt$-differential cross section for $\Dzero$ mesons with $|y|<0.5$ in pp collisions at $\sqrt s=7~\tev$. In the left-hand panel of the figure, the cross section obtained from the analysis without decay-vertex reconstruction is shown for 
inclusive and for prompt $\Dzero$ mesons, i.e.\,before and after the subtraction of the cross section of $\Dzero$ mesons from beauty-hadron decays. The subtraction of the feed-down contribution increases the systematic uncertainties at low $\pt$, where
the uncertainty of the correction is largest, and at high $\pt$, because the correction increases 
($f_{\rm prompt}$ decreases) with $\pt$. 
In the right-hand panel of Fig.~\ref{fig:CrossSecPP} the cross section for prompt $\Dzero$ mesons 
is compared with that obtained with decay-vertex reconstruction as published in Ref.~\cite{ALICE:2011aa}. The results are consistent for most of the $\pt$ intervals within one $\sigma$ of the statistical uncertainties,
which are independent for the two measurements because of their very different signal-to-background ratios and efficiencies.

\begin{figure}[!t]
\begin{center}
\includegraphics[width=0.48\textwidth]{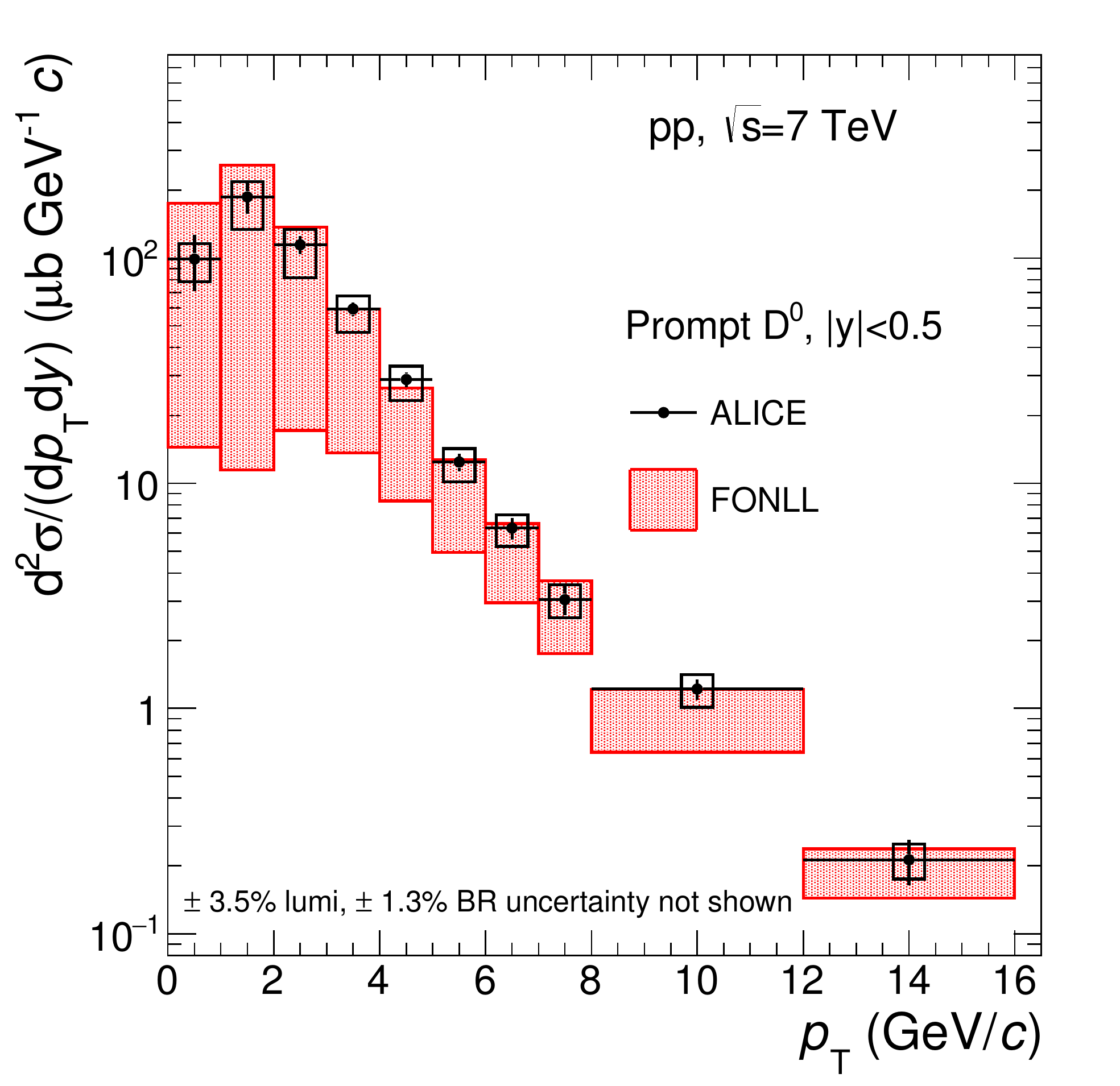}
\includegraphics[width=0.48\textwidth]{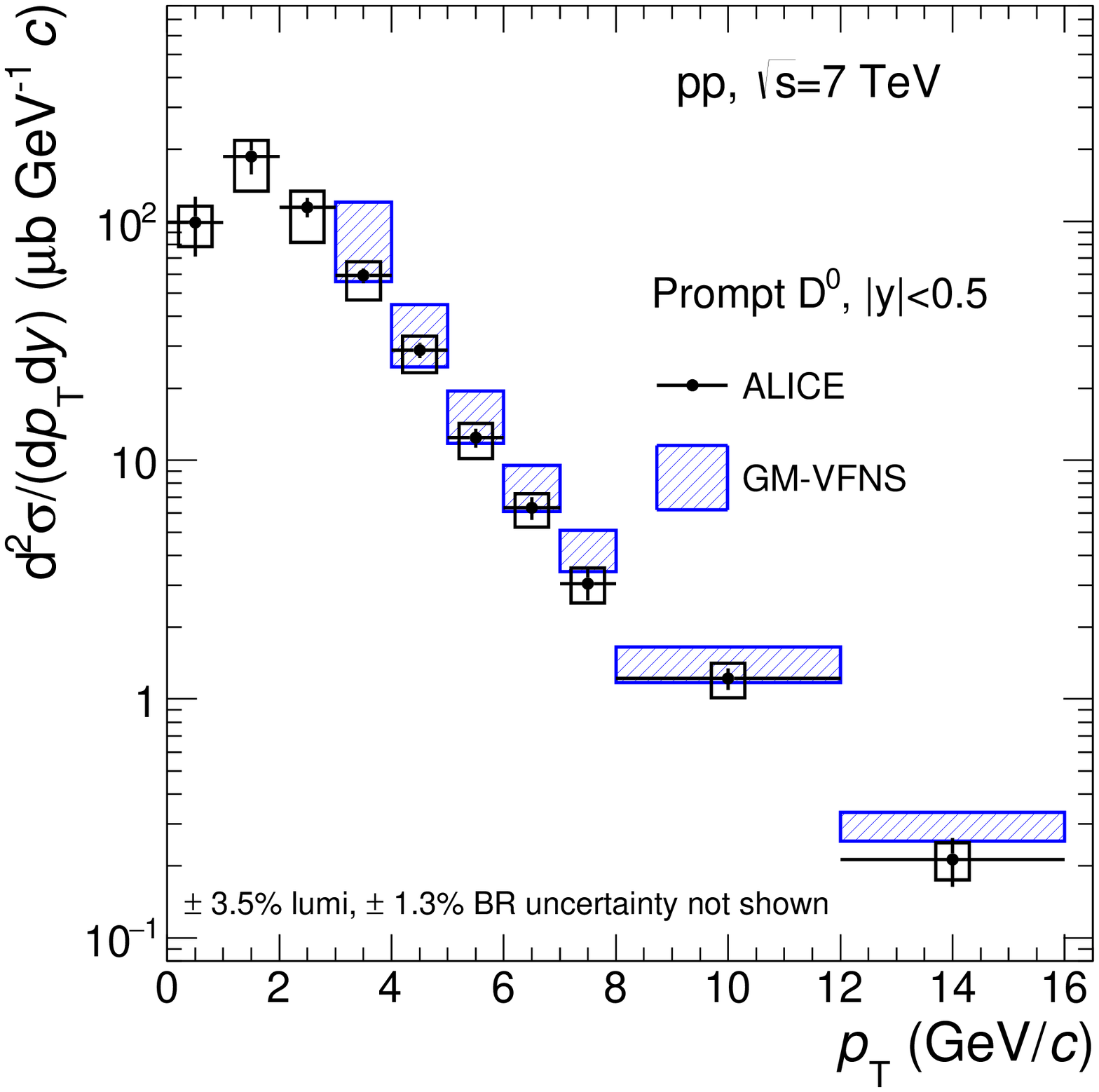}
\includegraphics[width=0.48\textwidth]{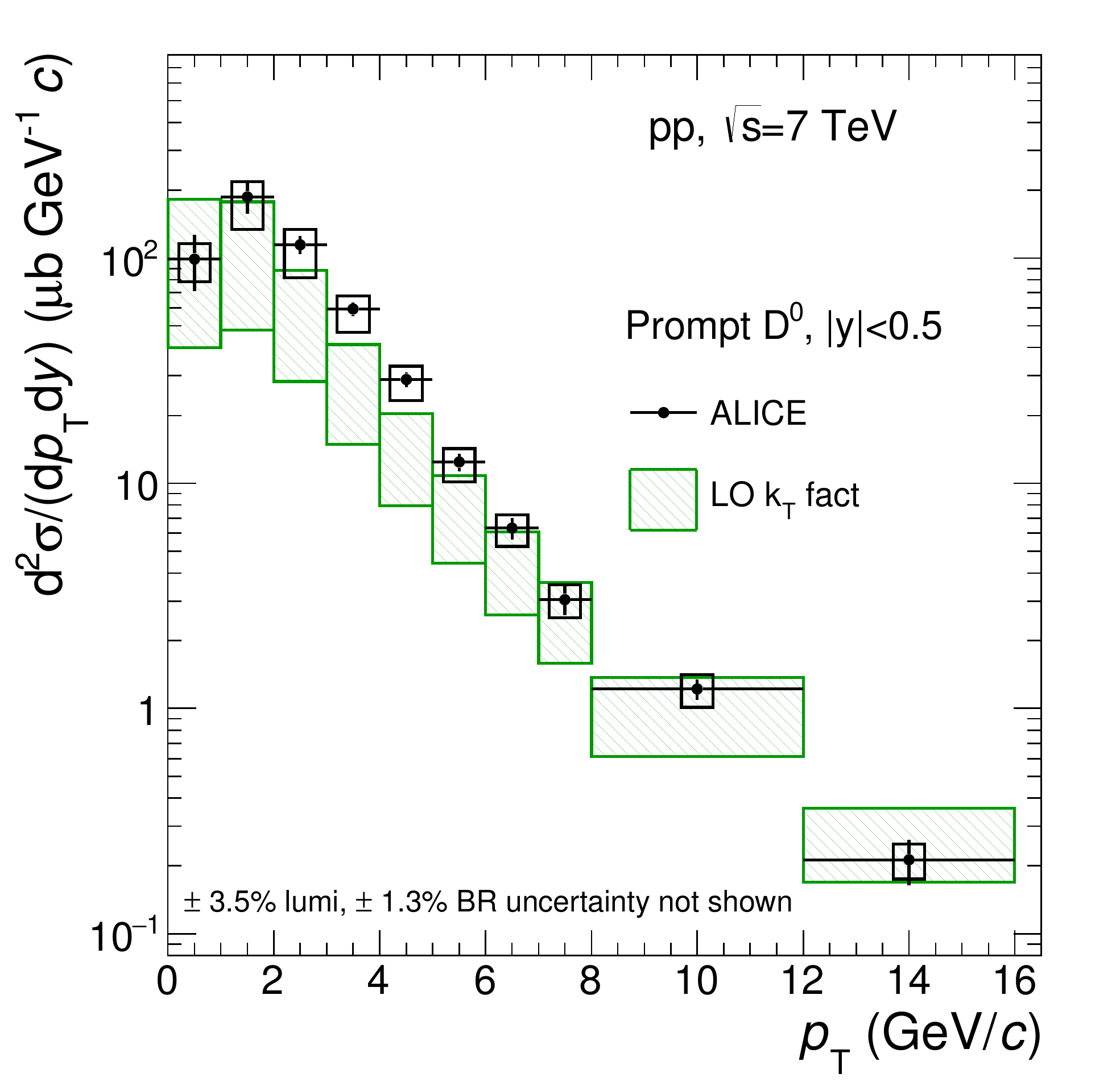}
\includegraphics[width=0.48\textwidth]{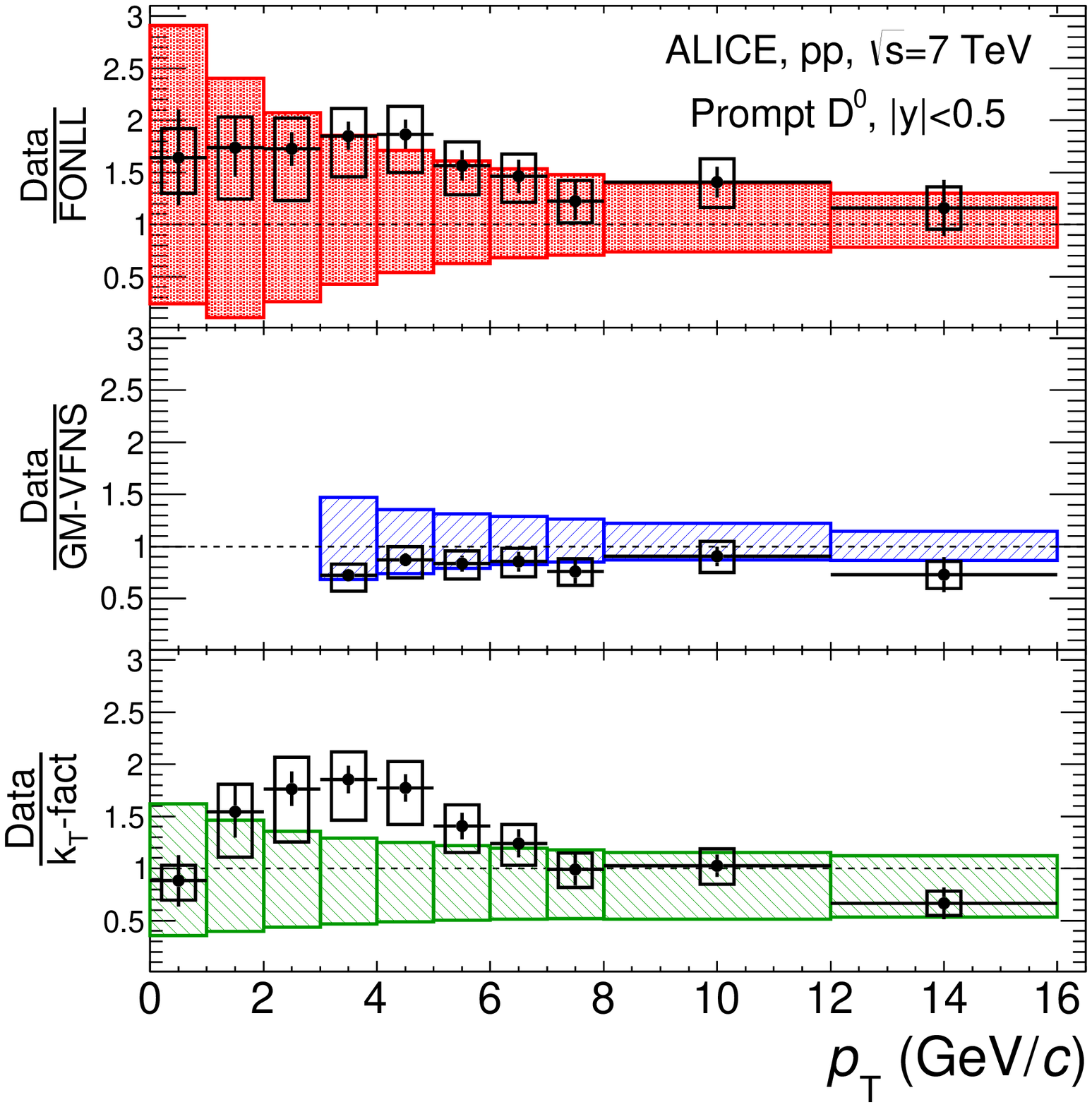}
\caption{$\pt$-differential production cross section of prompt $\Dzero$ mesons with $|y|<0.5$ in the interval $0<\pt<16~\gev/c$, in pp collisions at $\sqrt s =7~\tev$. The data points in $0<\pt<2~\gev/c$ are obtained from the analysis described in this article, while the data points in $2<\pt<16~\gev/c$ are taken from Ref~\cite{ALICE:2011aa}. The cross section is compared to three pQCD calculations: FONLL~\cite{Cacciari:2012ny} (top-left panel), GM-VFNS~\cite{Kniehl:2012ti} (top-right panel) and a leading order (LO) calculation based on $k_{\rm T}$-factorisation~\cite{Maciula:2013wg} (bottom-left panel). 
In the bottom-right panel, the ratios of the data to the three calculated 
cross sections are reported.}
\label{fig:CrossSecPPFONLL}
\end{center}
\end{figure}

Figure~\ref{fig:CrossSecPPFONLL} compiles the most precise ALICE measurement of the 
$\pt$-differential cross section of prompt $\Dzero$ mesons, which uses in each $\pt$ interval the data point with the smallest total uncertainty, namely the results from the analysis without decay-vertex reconstruction in 
$0<\pt<2~\gev/c$ and those from the analysis with decay-vertex reconstruction in $2<\pt<16~\gev/c$. The cross section is compared with results from perturbative QCD calculations, two of which are based on collinear factorisation (FONLL~\cite{Cacciari:1998it,Cacciari:2012ny}\footnote{In the FONLL calculation the $\rm c\to D^0$ fragmentation fraction was updated to the value reported in Ref.~\cite{Gladilin:2014tba}.} and GM-VFNS~\cite{Kniehl:2004fy,Kniehl:2005mk,Kniehl:2012ti}) and one is a leading order (LO) calculation based on $k_{\rm T}$-factorisation~\cite{Maciula:2013wg}.
The ratios of the data to the three calculated cross sections are shown in the bottom-right panel of Fig.~\ref{fig:CrossSecPPFONLL}. The ratio to FONLL is approximately constant at 1.5, but consistent with unity within the theoretical and experimental uncertainties. 
A ratio data/FONLL larger than unity was observed also at other values of $\sqrt s$, from 0.2 to 
13~TeV~\cite{Adamczyk:2012af,Acosta:2003ax,Abelev:2012vra,Aaij:2013mga,Aaij:2015bpa}.
 The ratio to GM-VFNS is approximately constant at 0.75.
 The ratio to the LO $k_{\rm T}$-factorisation calculation is consistent with unity for $\pt<2~\gev/c$ and $\pt>5~\gev/c$, while it is larger than unity for $2<\pt<5~\gev/c$. 

The average transverse momentum $\meanpt$ of prompt $\Dzero$ mesons 
was measured by fitting the cross section
reported in Fig.~\ref{fig:CrossSecPPFONLL} with a power-law function:
\begin{equation}
f(\pt)=C \frac{\pt}{(1+(\pt/p_0)^2)^n}\,,
\label{powlaw}
\end{equation}
where $C$, $p_0$ and $n$ are the free parameters.
The result is:
\begin{equation}
\meanpt^{\rm prompt\,D^0}_{\rm pp,\,7\,TeV} = 2.18 \pm 0.08\,({\rm stat.})\, \pm 0.07\,({\rm syst.})~\gev/c\,.
\end{equation}
The systematic uncertainty has three contributions.
The first accounts for the uncorrelated systematic uncertainties on the 
$\pt$-differential cross section and it was obtained by repeating the fit 
using the uncorrelated systematic uncertainties as errors on the data points.
The second contribution accounts for the uncertainties that are 
correlated among the $\pt$ intervals and it was computed from the variation
of $\meanpt$ observed when repeating the fit by moving all data points 
to the upper (lower) edge of the correlated uncertainties.
The third source of systematic uncertainty is due to the fit function and 
it was estimated using different functions and using an alternative method,
which is not based on fits to the spectrum, but on direct calculations 
of $\meanpt$ from the data points with different assignments of the average
transverse momentum of $\Dzero$ mesons in the intervals of the 
$\pt$-differential measurement.

The production cross section of prompt $\Dzero$ mesons per unit of rapidity at mid-rapidity was obtained by integrating the
$\pt$-differential cross section shown in Fig.~\ref{fig:CrossSecPPFONLL}.
The systematic uncertainty was defined by propagating the yield extraction uncertainties as uncorrelated among $\pt$ intervals 
(quadratic sum) and all the other uncertainties as correlated (linear sum). The resulting cross section is:
\begin{equation}
{\rm d}\sigma^{\rm prompt\,D^0}_{\rm pp,\,7\,TeV}/{\rm d}y=518\pm 43\,({\rm stat.}) ^{+\phantom{1}57}_{-102}\,({\rm syst.})\pm 18\,({\rm lumi.})\pm 7\,({\rm BR})~\mu{\rm b}\,.
\label{eq:sigD0pp}
\end{equation}
This measurement is consistent within statistical uncertainties with the value obtained in the analysis 
with decay-vertex reconstruction~\cite{ALICE:2011aa} ($516\pm 41\,({\rm stat.}) ^{+138}_{-179}\,({\rm syst.})\pm 18\,({\rm lumi.})\pm 7\,({\rm BR})~\mu{\rm b}$),
but it has a total systematic uncertainty reduced by a factor of about two on the low side and almost three on the high side, where the 
earlier measurement was affected by large uncertainties on the feed-down correction and on the extrapolation to $\pt=0$ (a factor $1.25^{+0.29}_{-0.09}$~\cite{ALICE:2011aa}), respectively.
For completeness, we also report the inclusive cross section of $\Dzero$ mesons, without feed-down subtraction, as obtained by integrating the inclusive cross section shown in Fig.~\ref{fig:CrossSecPP} (left):
\begin{equation}
{\rm d}\sigma^{\rm inclusive\,D^0}_{\rm pp,\,7\,TeV}/{\rm d}y=522\pm 45\,({\rm stat.}) \pm 55\,({\rm syst.})\pm 18\,({\rm lumi.})\pm 7\,({\rm BR})~\mu{\rm b}\,.
\end{equation}
The central values of the prompt and inclusive 
${\rm d}\sigma/{\rm d}y$ are numerically very similar. However, this should not lead to a conclusion that the prompt fraction is essentially unity, because
the two cross section determinations are to a large extent independent.
Indeed,  the contribution of
$\Dzero$ mesons with $\pt>2~\GeV/c$ is taken from the results obtained with 
different analysis techniques in the two cases: the analysis `with decay-vertex 
reconstruction' is used for the prompt cross section and the analysis 
`without decay-vertex reconstruction' for the inclusive one.
The uncertainties on the results from these two analyses are to a large extent 
independent, having in common only the 8.5\% contribution 
due to the tracking and PID efficiency correction, and the
contributions from the luminosity and the BR.

The ${\rm c\overline{c}}$ production cross section per unit of rapidity at mid-rapidity ($|y|<0.5$)
was calculated by dividing the prompt $\Dzero$-meson cross section
by the fraction of charm quarks hadronising into $\Dzero$ mesons 
(fragmentation fraction, FF), $0.542 \pm 0.024$~\cite{Gladilin:2014tba} and
correcting for the different shapes of the distributions of $y_{\Dzero}$ and $y_{\rm c\overline{c}}$ (${\rm c\overline{c}}$ pair rapidity).
This correction is composed of two factors.
The first factor accounts for the different rapidity shapes
of $\Dzero$ mesons and single charm quarks and it was
evaluated to be unity based on FONLL calculations.
A 3\% uncertainty on this factor was evaluated from the difference 
between values from  FONLL and the PYTHIA\,6~\cite{Sjostrand:2006za} 
event generator.
The second factor is the ratio ${\rm d}\sigma/{\rm d}y_{\rm c\overline c} \big/  {\rm d}\sigma/{\rm d}y_{\rm c}$, which 
was estimated from NLO pQCD calculations (MNR~\cite{Mangano:1991jk} and 
POWHEG~\cite{Frixione:2007nw}) 
as $\sigma^{\rm c\overline{c}}_{|y|<0.5} / \sigma^{\rm c}_{|y|<0.5}=1.034$.
A 1.5\% uncertainty on this factor was estimated from the difference among 
the values obtained varying the factorisation and renormalisation scales 
in the MNR calculation and interfacing, via the POWHEG-BOX 
package~\cite{Alioli:2010xd}, the NLO calculations with a parton shower 
simulation with PYTHIA.
The resulting  ${\rm c\overline{c}}$ cross section per unit of rapidity at 
mid-rapidity is:
\begin{equation}
{\rm d}\sigma^{\rm c\overline{c}}_{\rm pp,\,7\,TeV}/{\rm d}y =  988\pm 81\,({\rm stat.}) ^{+108}_{-195}\,({\rm syst.})\pm 35\,({\rm lumi.})\pm 44\,({\rm FF})\pm 33\,({\rm rap.\,shape})~\mu{\rm b}\,.
\end{equation}

\begin{figure}[!htb]
\begin{center}
\includegraphics[width=0.55\textwidth]{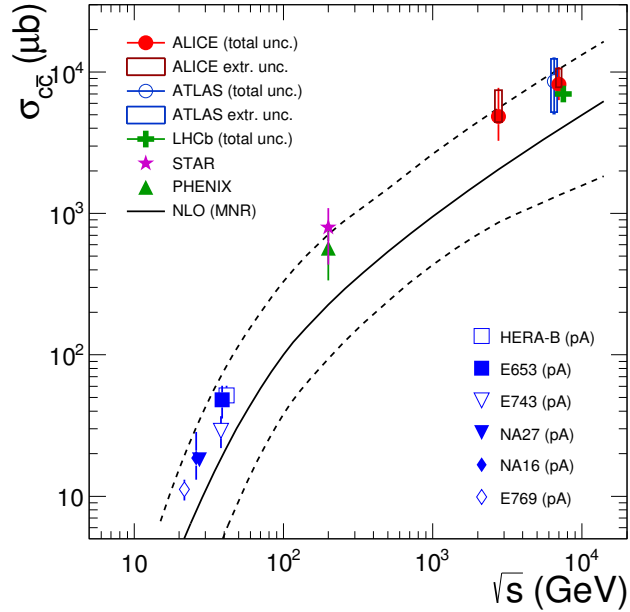}
\caption{
Total inclusive charm production cross section in nucleon--nucleon collisions 
as a function of 
$\sqrt{s}$~\cite{Lourenco:2006vw,Adare:2010de,Adamczyk:2012af,Aaij:2013mga,Abelev:2012vra,Aad:2015zix}.
Data are from pA collisions for $\sqrt s<100$~GeV and from pp collisions for $\sqrt s>100$~GeV.  
Data from pA collisions 
were scaled by $1/A$.
Results from NLO pQCD calculations (MNR~\cite{Mangano:1991jk}) and their 
uncertainties are shown as solid and dashed lines.
}
\label{fig:totcs}
\end{center}
\end{figure}

The total production cross section of prompt $\Dzero$ mesons 
(average of particles and antiparticles) was calculated
by extrapolating to full phase space the cross section measured at mid-rapidity.
The extrapolation factor was defined as the ratio of the $\Dzero$ 
production cross sections in full rapidity and in $|y|<0.5$ calculated with 
the FONLL central parameters: $8.57 ^{+2.52}_{-0.38}$.
The systematic uncertainty on the extrapolation factor was estimated by
considering the contributions due to i) the uncertainties on the 
CTEQ6.6 PDFs~\cite{Pumplin:2002vw} and ii) the variation of the charm-quark 
mass and the renormalisation and factorisation scales in the FONLL 
calculation, as proposed in~\cite{Cacciari:2012ny}.
The resulting cross section is:
\begin{equation}
{\rm \sigma^{\rm prompt\,D^0}_{\rm pp,\,7\,TeV}= 4.43 \pm 0.36 \,({\rm stat.})  \,^{+0.49}_{-0.88}\,({\rm syst.})  \,^{+1.30}_{-0.19}\ ({\rm extr.}) \pm 0.16 \,({\rm lumi.})\pm 0.06 \,({\rm BR})~{\rm mb}\,.}
\end{equation}
The total charm production cross section was calculated by dividing the 
total prompt $\Dzero$-meson production cross section by 
the fragmentation fraction reported above.
The resulting ${\rm c\overline{c}}$ production cross section in pp collisions at $\sqrt{s}=7~\TeV$ is:
\begin{equation}
{\rm \sigma^{\rm c\overline{c}}_{\rm pp,\,7\,TeV}= 8.18 \pm 0.67 \,({\rm stat.}) \,^{+0.90}_{-1.62}\,({\rm syst.})  \,^{+2.40}_{-0.36} ({\rm extr.}) \pm 0.29\,({\rm lumi.})\, \pm 0.36\,({\rm FF})~{\rm mb}\,,}
\end{equation}
which has smaller systematic and extrapolation uncertainties
as compared to the value of Ref.~\cite{Abelev:2012vra}.
We verified that the precision of the ${\rm c\overline{c}}$ production 
cross-section determination does not improve if the results calculated from
$\Dplus$ and $\Dstar$ mesons, which have significantly larger extrapolation 
uncertainties as compared to the $\Dzero$ one, are included via a weighted 
average procedure, as done in Ref.~\cite{Abelev:2012vra}.
In Fig.~\ref{fig:totcs}, the total charm production cross section is shown
as a function of the centre-of-mass energy of the collision together with
other measurements~\cite{Lourenco:2006vw,Adare:2010de,Adamczyk:2012af,Aaij:2013mga,Abelev:2012vra,Aad:2015zix}.
The LHCb value was computed by multiplying the 
$\pt$-integrated charm cross section at forward rapidity~\cite{Aaij:2013mga} by the 
rapidity extrapolation factor given in Ref.~\cite{LHCb:2010lga}.
The proton--nucleus (pA) measurements were scaled by $1/A$, assuming no nuclear effects.
The curves show the results of next-to-leading-order pQCD calculations  
(MNR~\cite{Mangano:1991jk}) together with their uncertainties obtained
varying the calculation parameters as suggested in~\cite{Cacciari:2012ny}.
The dependence of the charm production cross section on the collision energy 
is described by the pQCD calculation, with all the
data points lying close to the upper edge of the uncertainty band.

\subsection{$\rm D$-meson production cross section in p--Pb collisions at $\sqrtsNN=5.02~\tev$}
\label{sec:resultspPb}

\begin{figure}[!t]
\begin{center}
\includegraphics[width=0.48\textwidth]{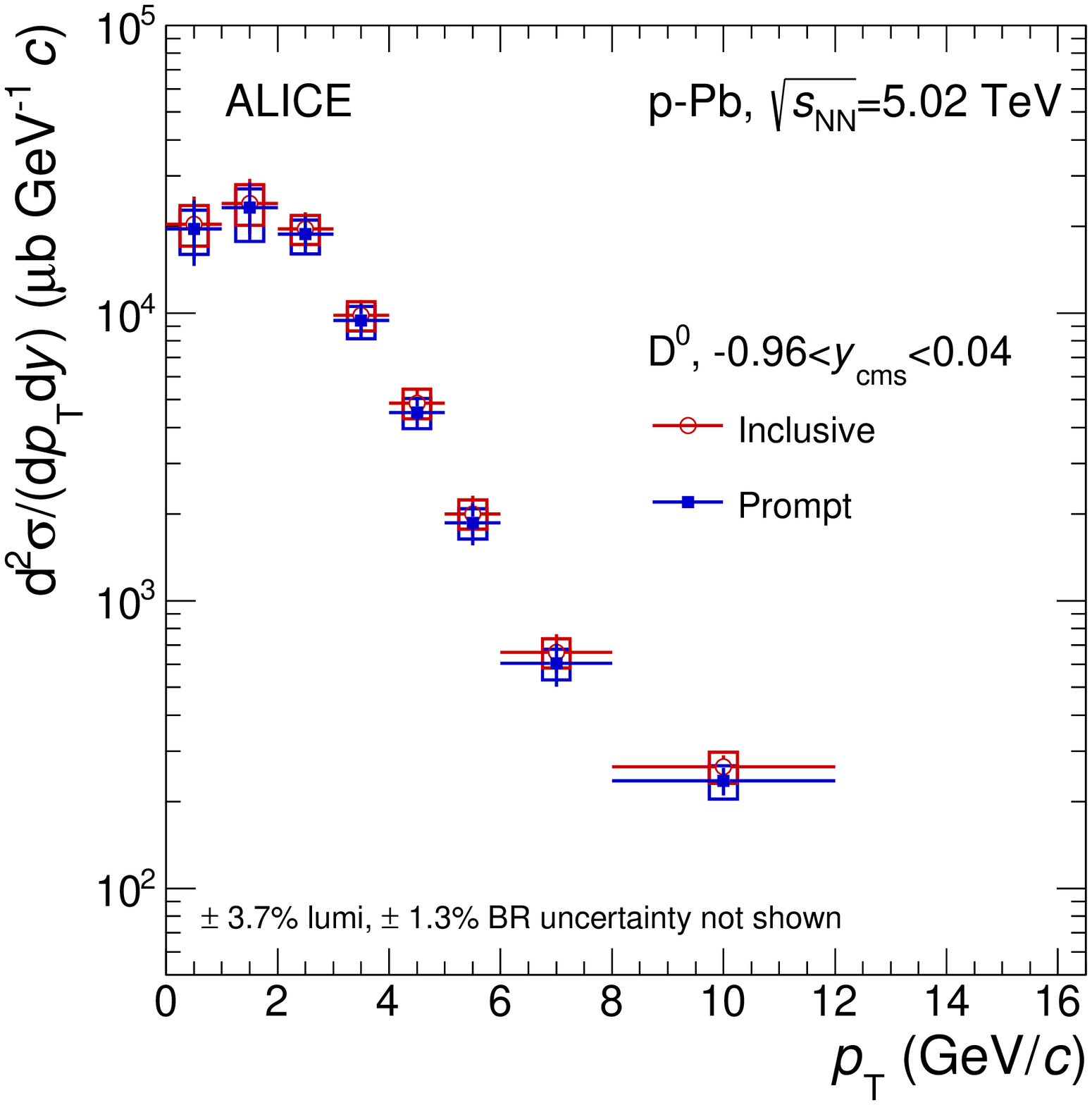}
\includegraphics[width=0.48\textwidth]{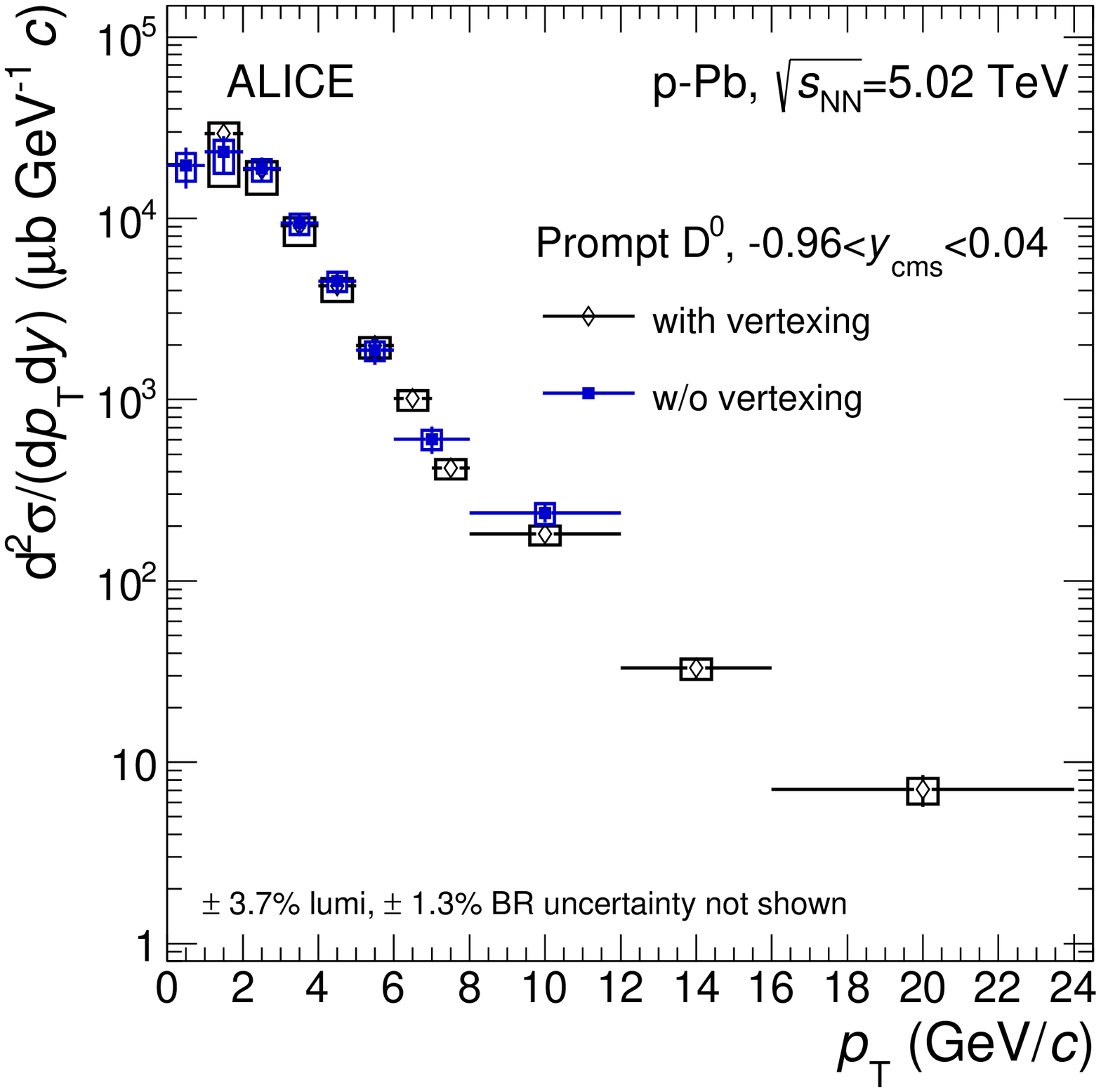}
\caption{$\pt$-differential production cross section of $\Dzero$ mesons with $-0.96<y_{\rm cms}<0.04$ in p--Pb collisions at $\sqrtsNN =5.02~\tev$. 
Left: comparison of prompt and inclusive $\Dzero$ mesons (the latter including also $\Dzero$ mesons from beauty-hadron decays) from the analysis without decay-vertex reconstruction. Right: comparison between the prompt $\Dzero$ cross sections measured with~\cite{Abelev:2014hha} and without decay-vertex reconstruction.}
\label{fig:CrossSecPPB}
\end{center}
\end{figure}

Figure~\ref{fig:CrossSecPPB} shows the $\pt$-differential production cross section for $\Dzero$ mesons with $-0.96<y_{\rm cms}<0.04$ in p--Pb collisions at $\sqrtsNN=5.02~\tev$. 
In the left-hand panel of the figure, the cross section obtained from the analysis without decay-vertex reconstruction is shown for 
inclusive and for prompt $\Dzero$ mesons, while in the right-hand panel the cross section for prompt $\Dzero$ mesons 
is compared with that obtained with decay-vertex reconstruction~\cite{Abelev:2014hha}. The results are consistent within one $\sigma$ of the statistical uncertainties.

\begin{figure}[!t]
\begin{center}
\includegraphics[width=0.48\textwidth]{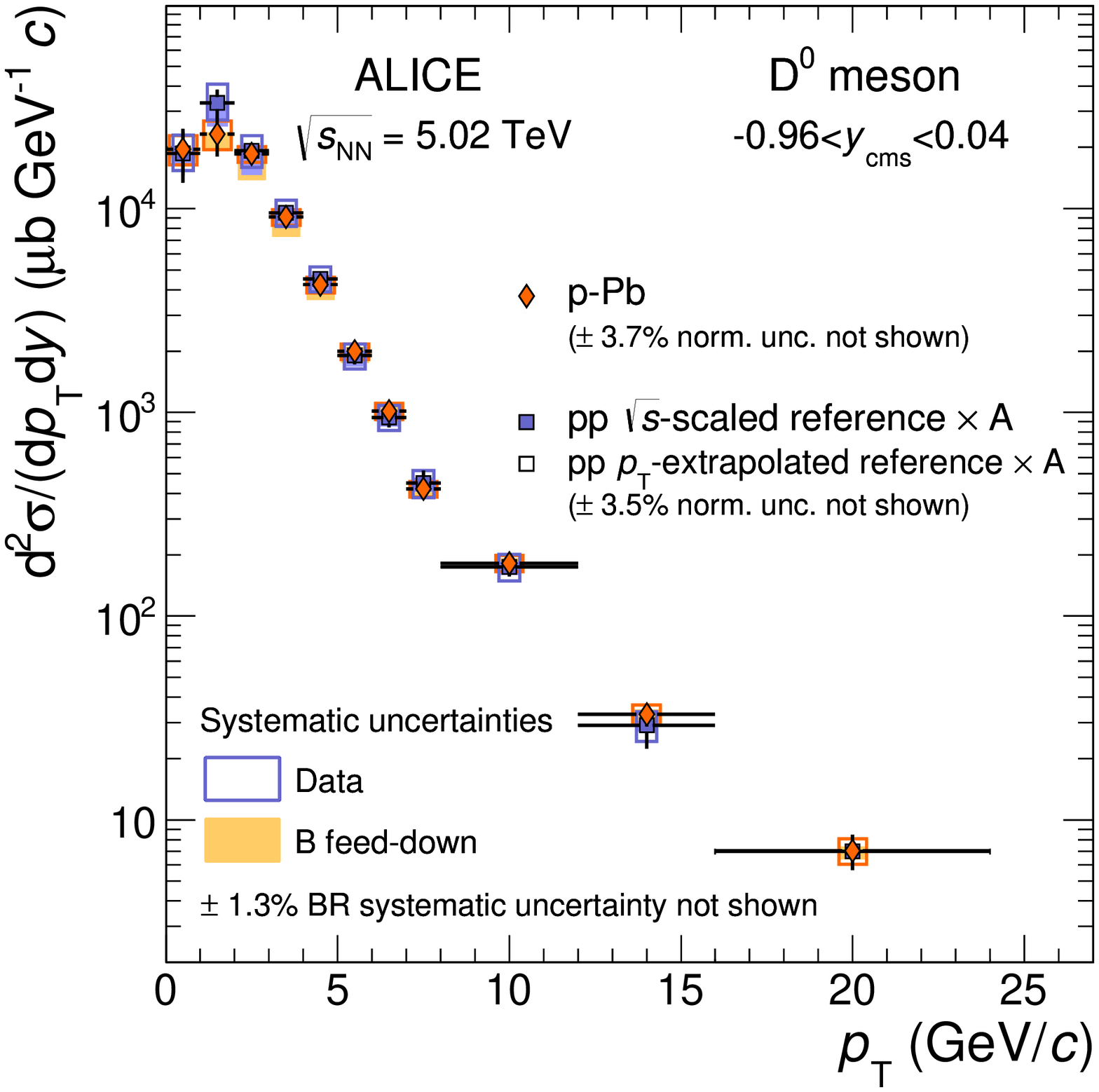}
\includegraphics[width=0.48\textwidth]{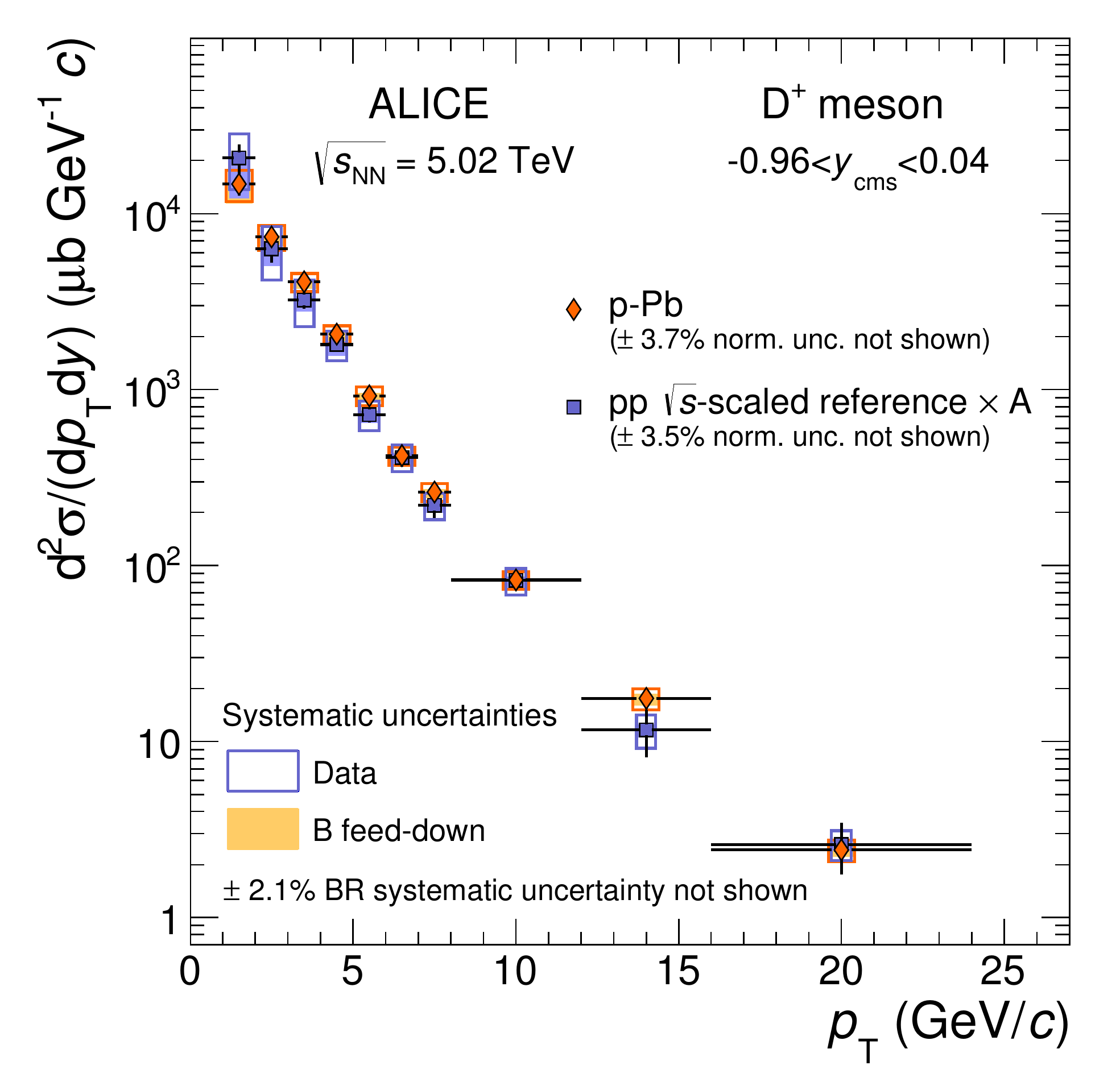}
\includegraphics[width=0.48\textwidth]{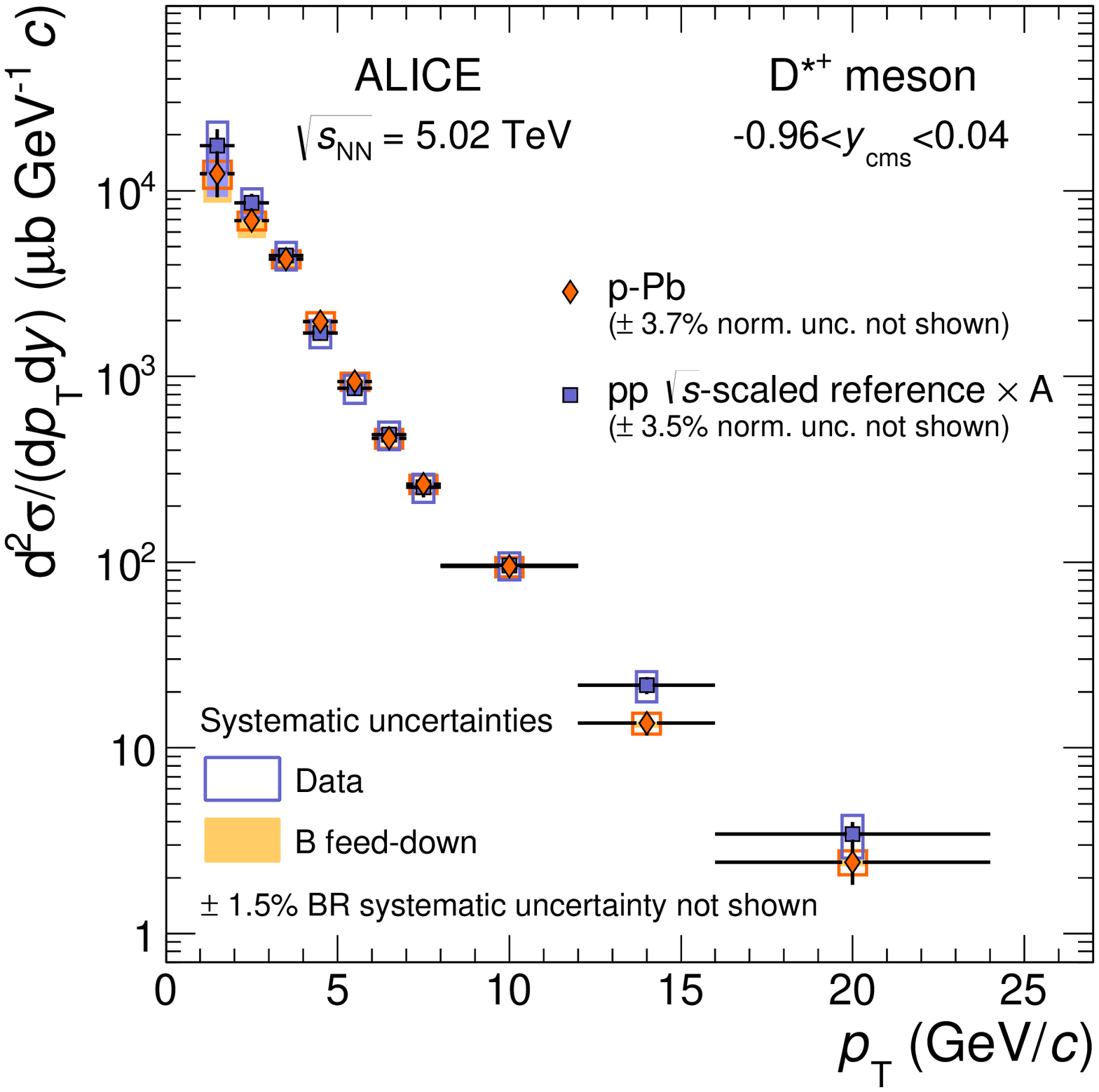}
\includegraphics[width=0.48\textwidth]{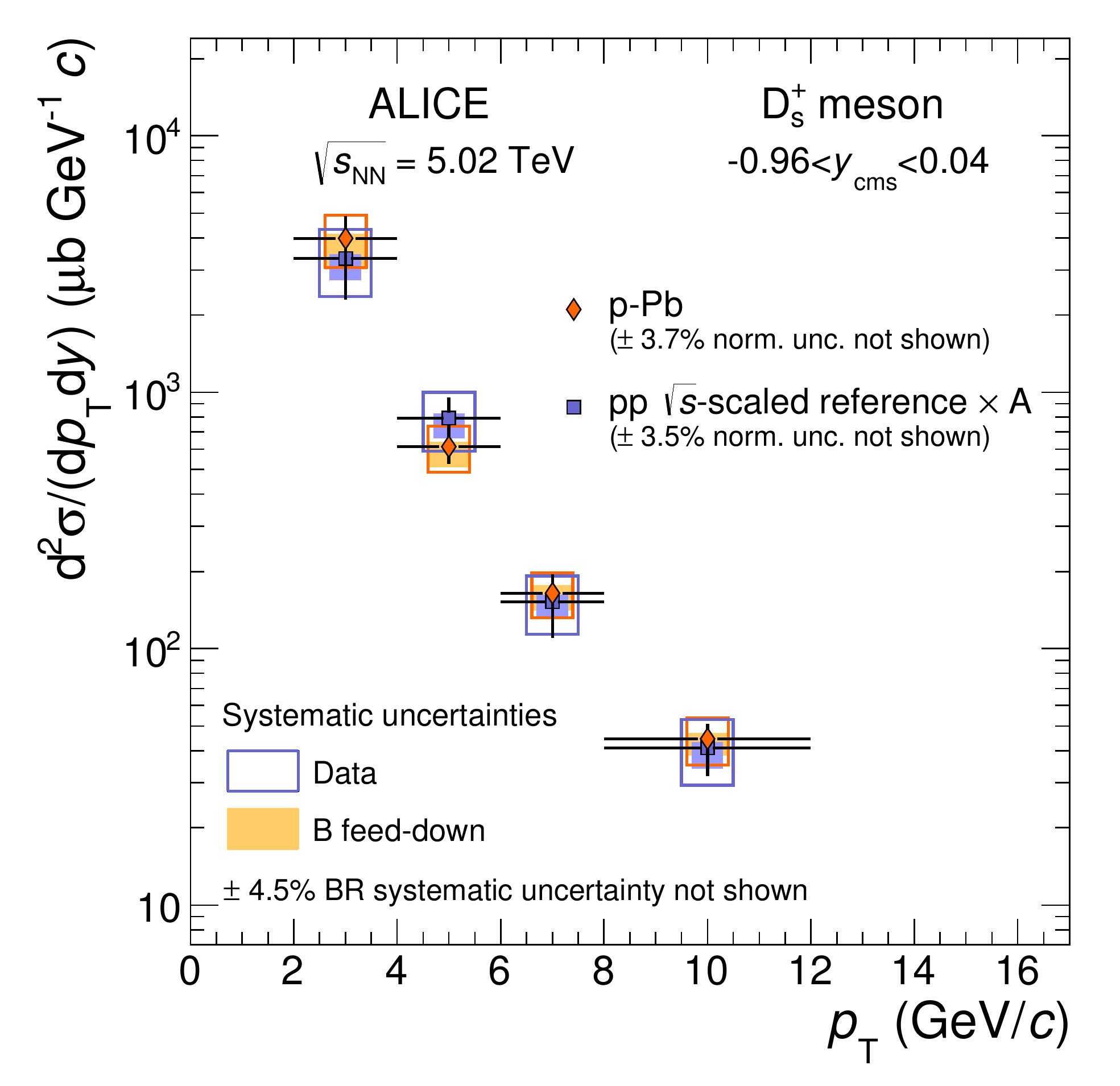}
\caption{$\pt$-differential production cross sections of prompt $\Dzero$ (top-left), $\Dplus$ (top-right), $\Dstar$ (bottom-left) and $\Ds$ (bottom-right) mesons with $-0.96<y_{\rm cms}<0.04$ in p--Pb collisions at $\sqrtsNN=5.02~\tev$, compared with the respective pp reference cross sections scaled by the Pb mass number $A=208$. 
For the $\Dzero$ meson, the results in $0<\pt<2~\gev/c$ are obtained from the analysis without decay-vertex reconstruction, while those in $2<\pt<24~\gev/c$ are taken from the analysis with decay-vertex reconstruction.
The results from the other three D-meson species are the same as in Ref.~\cite{Abelev:2014hha}. The systematic uncertainty of the feed-down correction is displayed separately.}
\label{fig:dsdptppref} 
\end{center}
\end{figure}

As for pp collisions, the most precise measurement of the prompt $\Dzero$ production cross section is obtained 
using the results of the analysis without decay-vertex reconstruction in the interval  $0<\pt<2~\gev/c$ and 
those of the analysis with decay-vertex reconstruction for $\pt>2~\gev/c$~\cite{Abelev:2014hha}. 
The cross section is shown in the top-left panel of Fig.~\ref{fig:dsdptppref}.  
The total cross section for prompt and inclusive $\Dzero$-meson production per unit of rapidity in $-0.96<y_{\rm cms}<0.04$ was calculated
in the same way as for pp collisions. The resulting values are:
\begin{equation}
{\rm d}\sigma^{\rm prompt\,D^0}_{\rm p-Pb,\,5.02\,TeV}/{\rm d}y=79.0\pm 7.3\,({\rm stat.}) ^{+\phantom{0}7.1}_{-13.4}\,({\rm syst.})\pm 2.9\,({\rm lumi.})\pm 1.0\,({\rm BR})~{\rm mb}
\label{eq:sigD0pPb}
\end{equation}
\begin{equation}
{\rm d}\sigma^{\rm inclusive\,D^0}_{\rm p-Pb,\,5.02\,TeV}/{\rm d}y=83.0\pm 7.9\,({\rm stat.}) \pm 7.2\,({\rm syst.})\pm 3.1\,({\rm lumi.})\pm 1.1\,({\rm BR})~{\rm mb}\,.
\end{equation}
The ${\rm c\overline{c}}$ production cross section in $-0.96<y_{\rm cms}<0.04$ 
is:
\begin{equation}
{\rm d}\sigma^{\rm c\overline{c}}_{\rm p-Pb,\,5.02\,TeV}/{\rm d}y =  151\pm 14\,({\rm stat.}) ^{+13}_{-26}\,({\rm syst.})\pm 6\,({\rm lumi.})\pm 7\,({\rm FF})\pm 5\,({\rm rap.\,shape})~{\rm mb}\,.
\end{equation}

The average transverse momentum $\meanpt$ of prompt $\Dzero$ mesons, obtained
with the same procedure described above for pp collisions, is:
\begin{equation}
\meanpt^{\rm prompt\,D^0}_{\rm p-Pb,\,5.02\,TeV} = 2.13 \pm 0.05\,({\rm stat.})\, \pm 0.10\,({\rm syst.})~\gev/c\,.
\end{equation}

The $\pt$-differential cross sections for the other three D-meson species 
($\Dplus$, $\Dstar$ and $\Ds$)~\cite{Abelev:2014hha}\,\footnote{\label{noteDsBR}The cross section for $\Ds$ mesons in p--Pb collisions and the corresponding pp reference were updated with respect to Ref.~\cite{Abelev:2014hha} to account for the change of the 
world-average branching ratio of $\rm \Ds\to\phi \pi^+\to K^-K^+\pi^+$ from 2.28\%~\cite{Beringer:1900zz} 
to 2.24\%~\cite{Agashe:2014kda}.}  
are shown in the other panels of Fig.~\ref{fig:dsdptppref}.

In the same figure, the cross sections in p--Pb collisions are compared with the corresponding pp reference cross sections, scaled by the
Pb mass number $A=208$. The pp reference cross sections at $\sqrt s=5.02~\tev$ were obtained by applying a $\pt$- and 
D-species-dependent scaling factor to the cross sections measured at $\sqrt s= 7~\tev$, namely the cross section shown 
in Fig.~\ref{fig:CrossSecPPFONLL} for $\Dzero$ mesons and those published in Refs.~\cite{ALICE:2011aa,Abelev:2012tca} for the other species.
 The scaling factor was defined as the ratio of the cross sections at 5.02~TeV
(in $-0.96<y_{\rm cms}<0.04$) and 7~TeV (in $|y_{\rm cms}|<0.5$) from the FONLL calculation~\cite{Cacciari:2012ny},
 as described in Ref.~\cite{Averbeck:2011ga}. Its systematic uncertainty was defined by consistently varying the charm-quark mass
 and the values of the factorisation and renormalisation scales at the two energies~\cite{Averbeck:2011ga}. The uncertainty decreases with 
 increasing $\pt$, with values of, for example, $^{+15}_{-\phantom{1}5}\%$ for $0<\pt<1~\gev/c$, $^{+6}_{-3}\%$ for $3<\pt<4~\gev/c$ 
 and $\pm 2\%$ for $\pt>12~\gev/c$.
For $\Dzero$ mesons, the cross section was measured in pp collisions at $\sqrt s=7~\tev$ up to only $\pt=16~\gev/c$; the pp reference for the interval $16<\pt<24~\gev/c$ was defined using the FONLL cross section multiplied by the ratio of data/FONLL in the interval $5<\pt<16~\gev/c$, which has a value of about 1.4 (see Ref.~\cite{Adam:2015sza} for more details).

\begin{figure}[!t]
\begin{center}
\includegraphics[width=\textwidth]{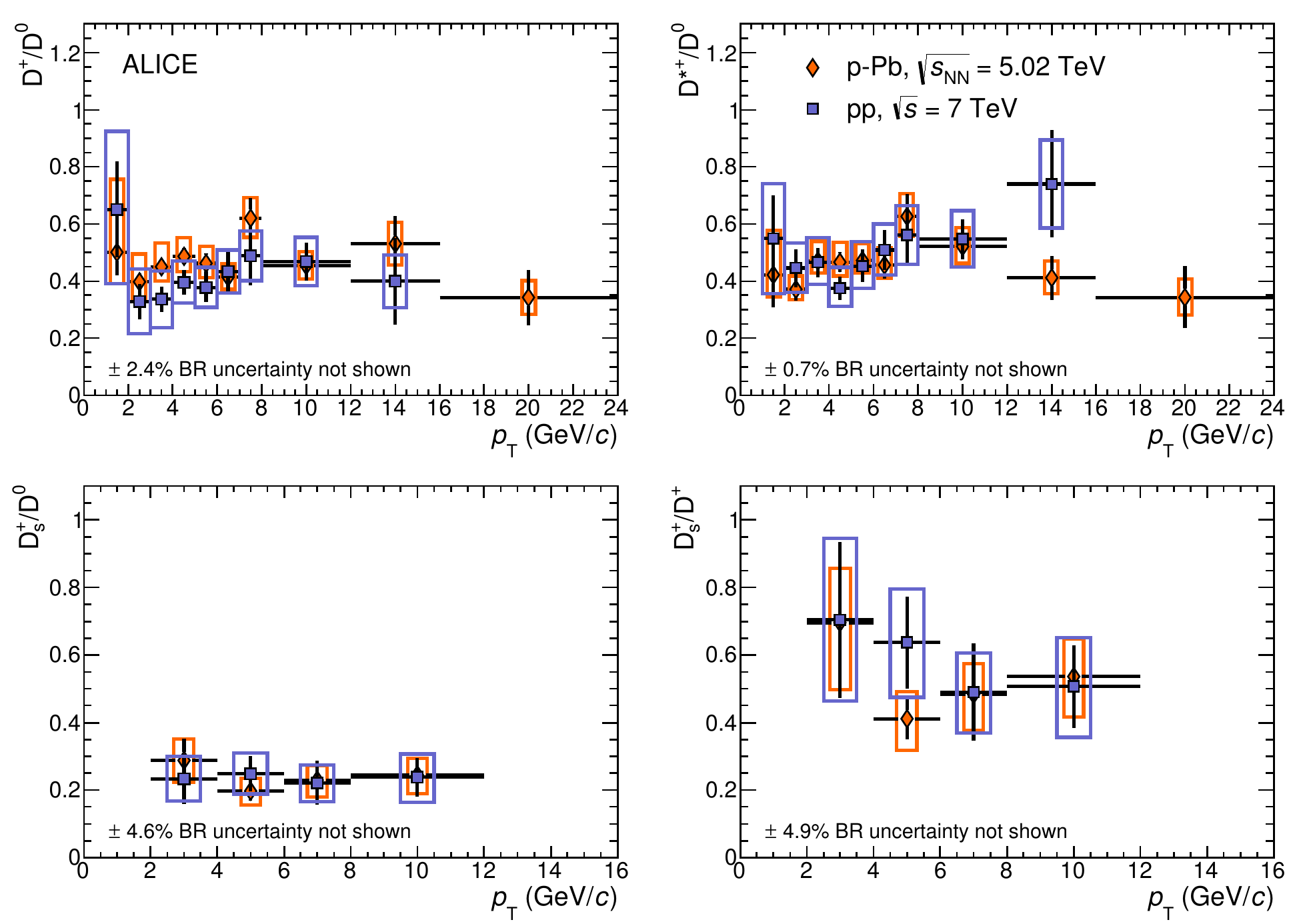}
\caption{Ratios of prompt D-meson production cross sections as a function of $\pt$ in pp collisions at $\sqrt s=7~\tev$ ($|y_{\rm cms}|<0.5$) and p--Pb collisions at $\sqrtsNN=5.02~\tev$ ($-0.96<y_{\rm cms}<0.04$).}
\label{fig:Dratios} 
\end{center}
\end{figure}

The ratios of the $\pt$-differential cross sections of the various D-meson species were calculated taking into account the correlation 
of the systematic uncertainties induced by the corrections for tracking efficiency and feed-down from beauty decays.
In Fig.~\ref{fig:Dratios} these ratios are shown together with those for pp collisions at $\sqrt s=7~\tev$ (from Ref.~\cite{Abelev:2012tca}\,\footnote{The ratios involving the $\Ds$ meson were updated, see footnote~\ref{noteDsBR}.}): within uncertainties, the relative abundances of the four species are not modified in p--Pb with respect to pp collisions. 

\begin{figure}[!t]
\begin{center}
\includegraphics[width=0.45\textwidth]{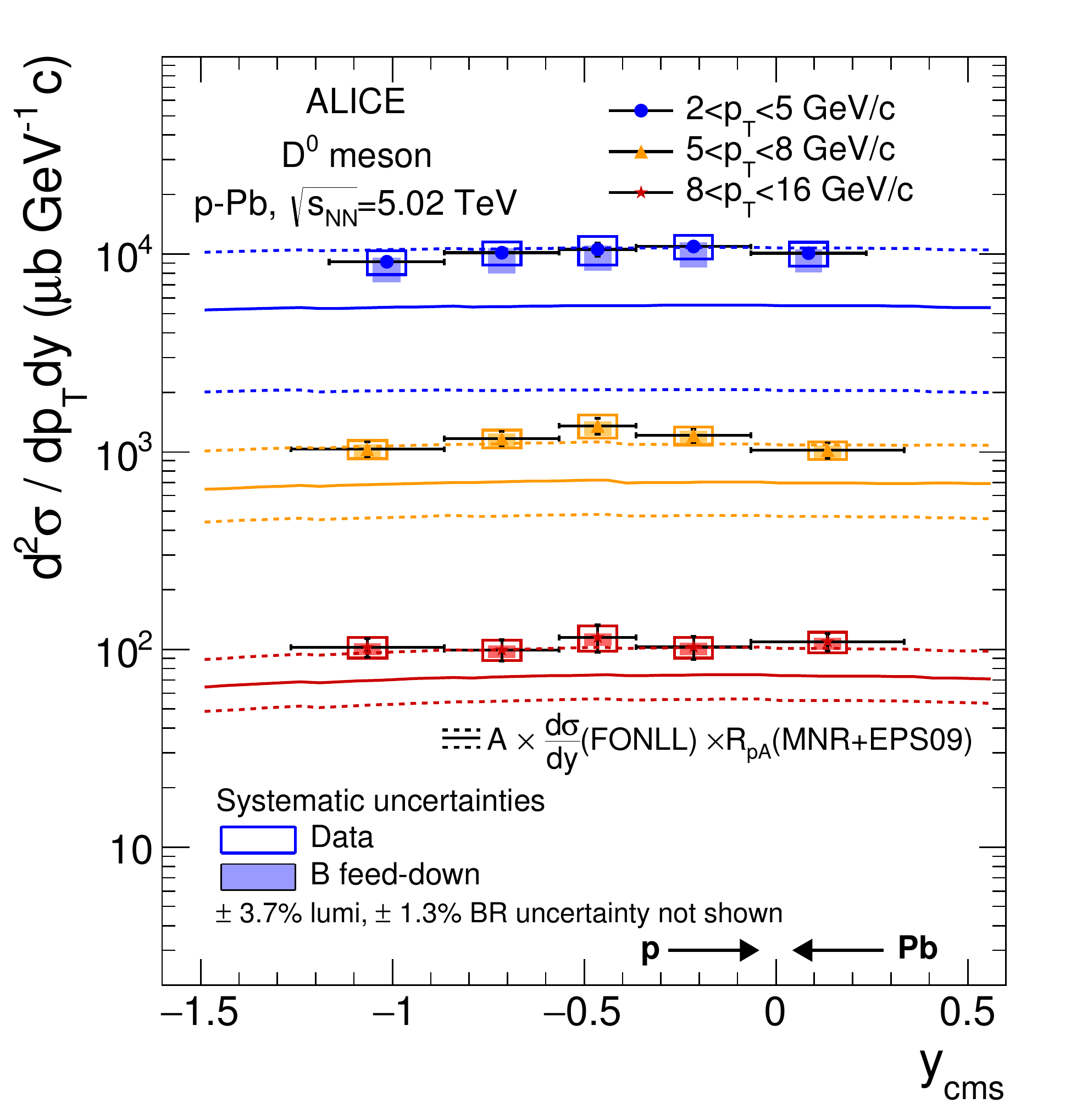}
\includegraphics[width=0.45\textwidth]{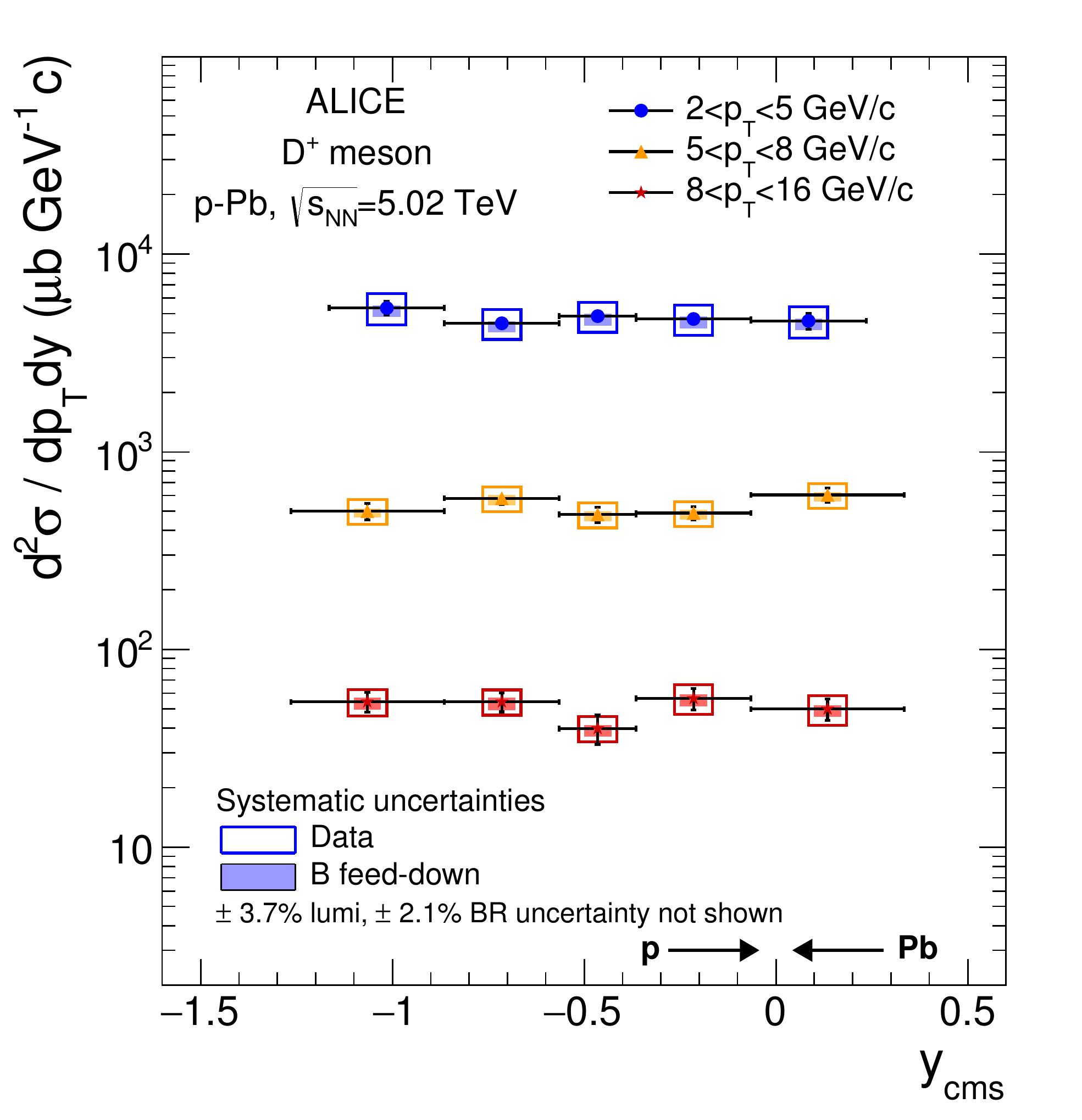}
\includegraphics[width=0.45\textwidth]{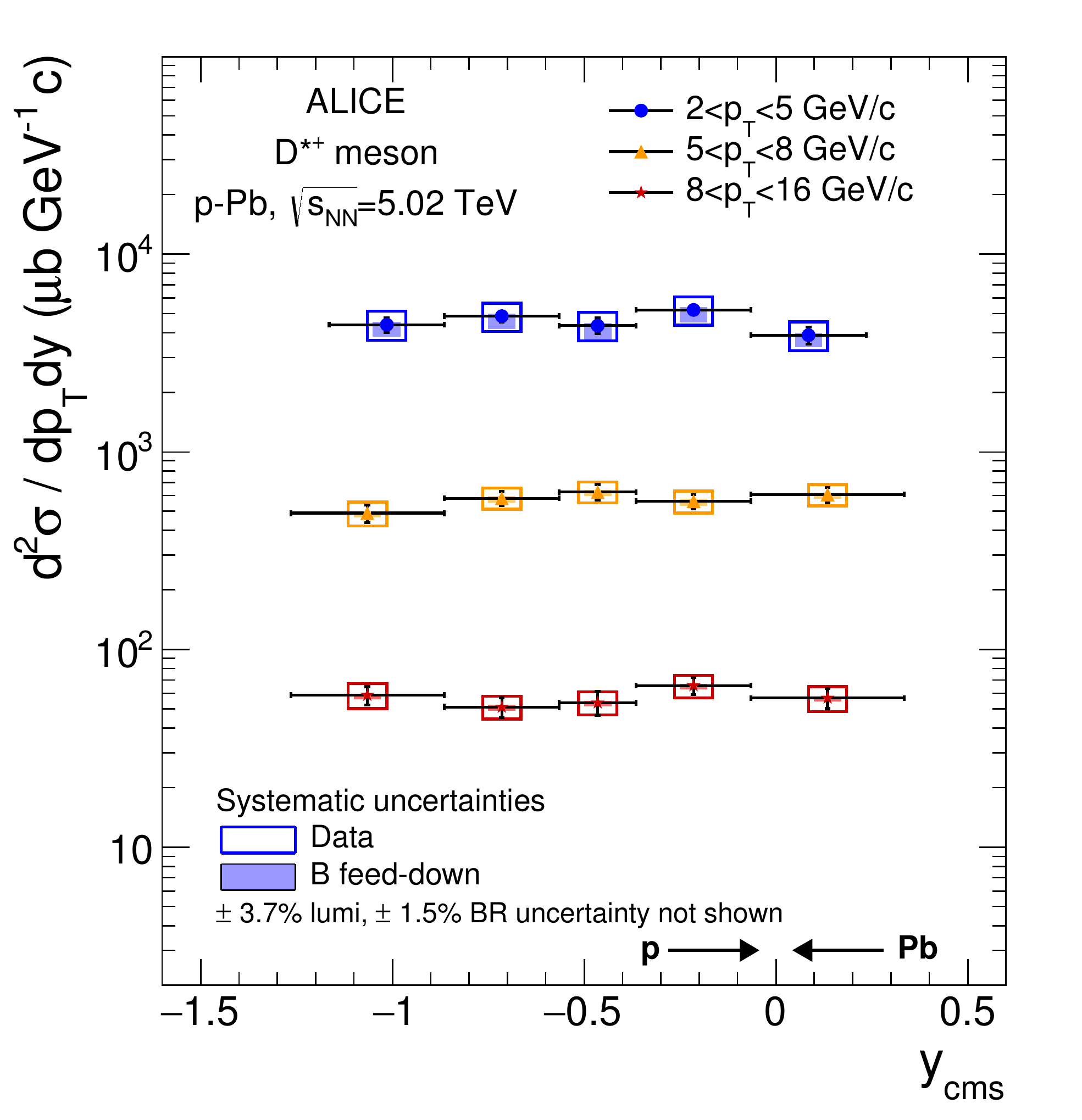}
\caption{Production cross sections as a function of rapidity ($y_{\rm cms}$) for prompt $\Dzero$, $\Dplus$ and $\Dstar$ mesons in p--Pb collisions at $\sqrtsNN=5.02~\tev$, for three $\pt$ intervals.
The $\Dzero$-meson data are compared with a cross section obtained by multiplying the FONLL~\cite{Cacciari:2012ny} calculation by the mass number $A$
and the nuclear modification factor $R_{\rm pA}$ estimated as a function of $y$ with the MNR NLO pQCD calculation~\cite{Mangano:1991jk}, with CTEQ6M PDFs~\cite{Pumplin:2002vw} and the EPS09NLO nuclear PDF parametrisation~\cite{Eskola:2009uj}.}
\label{fig:dsdy} 
\end{center}
\end{figure}

Figure~\ref{fig:dsdy} shows the cross sections as a function of rapidity for prompt $\Dzero$, $\Dplus$ and $\Dstar$ mesons in p--Pb collisions at $\sqrtsNN=5.02~\tev$ in three $\pt$ intervals: 2--5~GeV/$c$ (for $-1.16<y_{\rm cms}<0.24$), 5--8~GeV/$c$ and 8--16~GeV/$c$ (for $-1.26<y_{\rm cms}<0.34$). The cross sections do not vary with $y_{\rm cms}$, within uncertainties, for all three $\pt$ intervals.
The $\Dzero$-meson data are compared with a cross section obtained by multiplying the FONLL~\cite{Cacciari:2012ny} result by the mass number $A$
and the nuclear modification factor $R_{\rm pPb}$ estimated as a function of $y$ with the MNR NLO pQCD calculation~\cite{Mangano:1991jk} with CTEQ6M PDFs~\cite{Pumplin:2002vw} and the EPS09NLO nuclear PDF parametrisation~\cite{Eskola:2009uj}. 
The uncertainty of the calculation is the quadratic sum of the FONLL uncertainty on the cross section and the EPS09NLO uncertainty on $R_{\rm pPb}$. The calculation describes the measurements within uncertainties. As already observed for pp collisions at $\sqrt s=2.76$ and 7~TeV~\cite{ALICE:2011aa,Abelev:2012vra}, the data points lie close to the upper limit of the FONLL uncertainty band.
The absence of a visible rapidity dependence in $-1.26<y_{\rm cms}<0.34$ 
is common to the data and the calculation. For the latter, nuclear shadowing
induces a cross section variation of only about 2--3\% within this interval.

\subsection{$\rm D$-meson nuclear modification factor in p--Pb collisions at $\sqrtsNN=5.02~\tev$}
\label{sec:resultsRpA}

The nuclear modification factor was computed by dividing the $\pt$-differential cross section in p--Pb collisions at 
$\sqrtsNN=5.02~\tev$ by the cross section in pp collisions at the same energy (see Fig.~\ref{fig:dsdptppref}) scaled by the lead mass number $A=208$:
\begin{equation}
R_{\rm pPb}= \frac{1}{A}\,\frac{{\rm d}\sigma_{\rm pPb}^{\rm prompt\,D}/{\rm d}\pt}{{\rm d}\sigma^{\rm prompt\,D}_{\rm pp}/{\rm d}\pt}\,.
\end{equation}
The systematic uncertainties of the p--Pb and pp measurements were considered as independent and propagated quadratically, except for the uncertainty on the feed-down correction, which was recalculated for the ratio of cross sections by consistently varying the FONLL calculation parameters in the numerator and in the denominator.

\begin{figure}[!t]
\begin{center}
\includegraphics[width=0.48\textwidth]{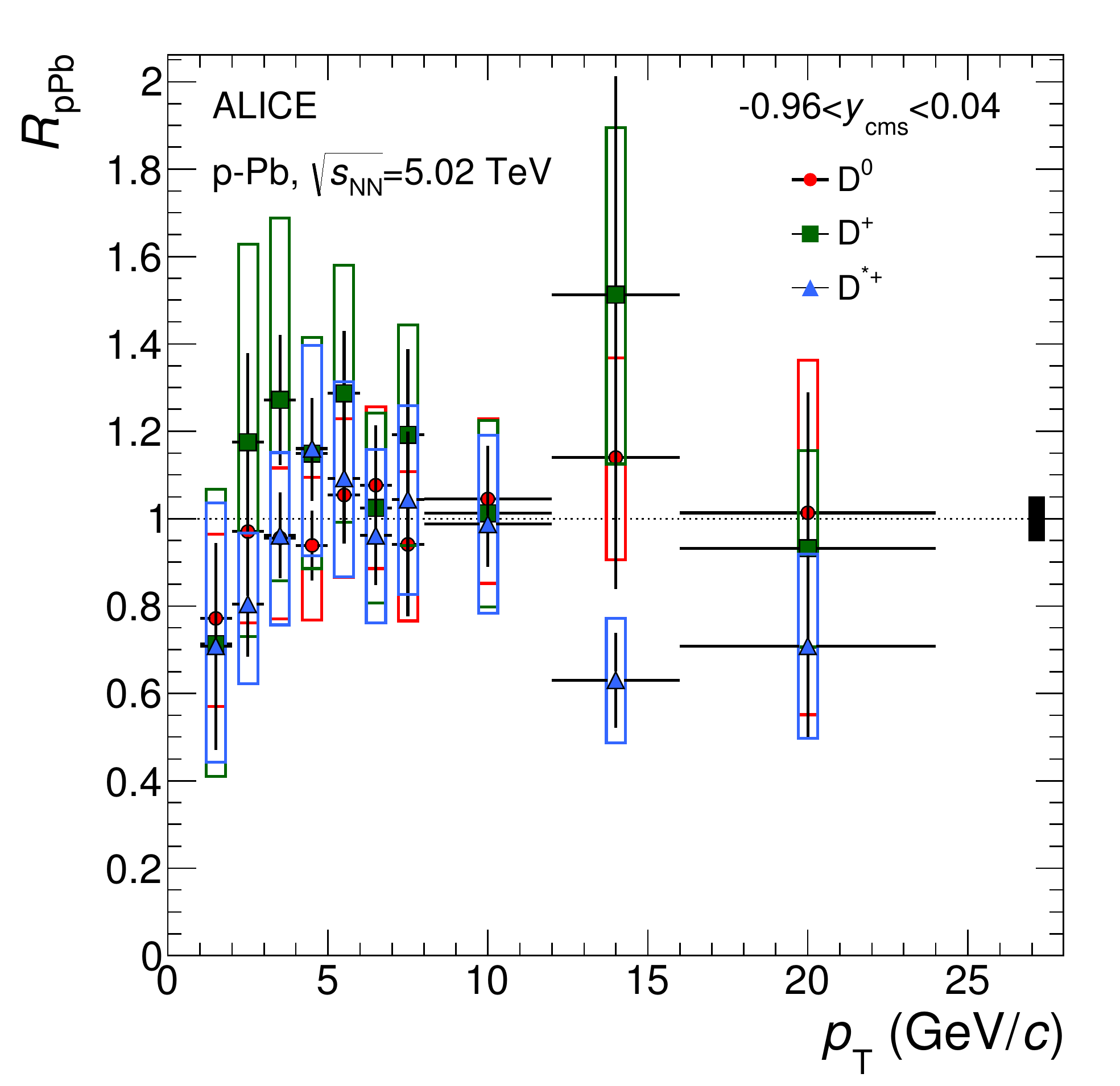}
\includegraphics[width=0.48\textwidth]{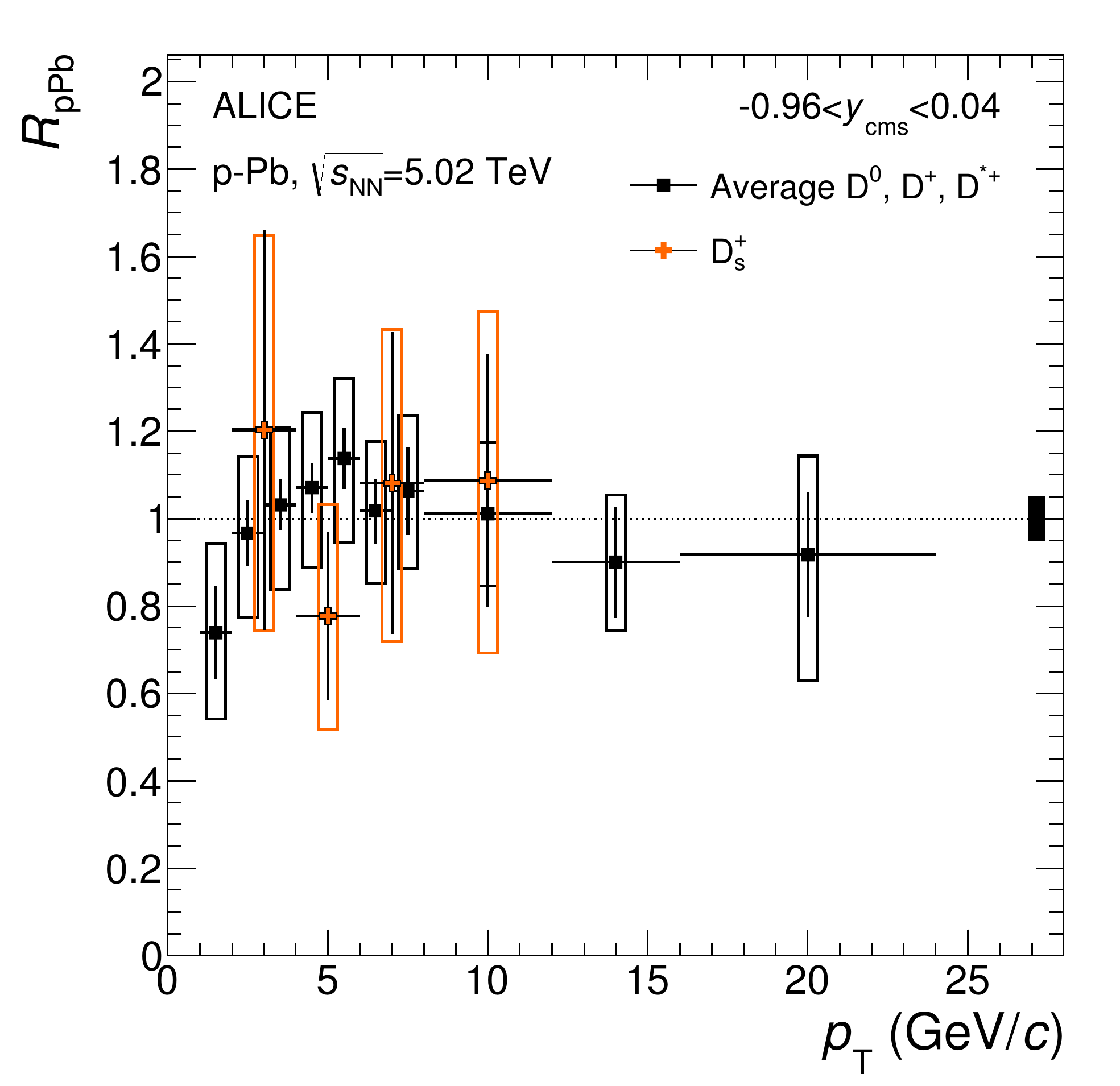}
\caption{Nuclear modification factor $R_{\rm pPb}$ of prompt D mesons in p--Pb collisions at $\sqrtsNN=5.02~\tev$~\cite{Abelev:2014hha}. 
Left: $R_{\rm pPb}$  of $\Dzero$, $\Dplus$ and $\Dstar$ mesons. 
Right: average $R_{\rm pPb}$ of the three non-strange D-meson species and $R_{\rm pPb}$ of $\Ds$ mesons. All results are obtained from the analysis with decay-vertex reconstruction.}
\label{fig:RpAallmesons}
\end{center}
\end{figure}

Figure~\ref{fig:RpAallmesons} shows the nuclear modification factors 
$R_{\rm pPb}$ of prompt $\Dzero$, $\Dplus$ and $\Dstar$ mesons in the left-hand panel and 
their average, along with the $R_{\rm pPb}$ of $\Ds$ mesons, in the right-hand panel.  All the results are obtained with the analysis 
based on decay-vertex reconstruction~\cite{Abelev:2014hha}. 
The average of the nuclear modification factors of the three non-strange D-meson species was calculated 
using the inverse of the relative statistical uncertainties as weights. The systematic error of the average was calculated by propagating the uncertainties through the weighted average, where the contributions from tracking efficiency, beauty feed-down correction, and scaling of the pp reference were taken as fully correlated among the three species. 
$R_{\rm pPb}$ is compatible with unity over the full $\pt$ interval 
covered by the measurements and it is also compatible for non-strange and strange D mesons.

\begin{figure}[!t]
\begin{center}
\includegraphics[width=0.48\textwidth]{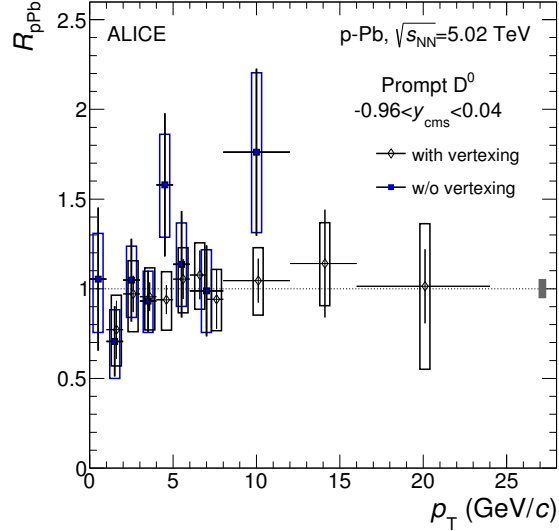}
\caption{Comparison of the nuclear modification factors of prompt $\Dzero$ mesons as obtained in the analysis with decay-vertex reconstruction~\cite{Abelev:2014hha} and in the analysis without decay-vertex reconstruction. }
\label{fig:RpPbVsPt}
\end{center}
\end{figure}

The nuclear modification factor of prompt $\Dzero$ mesons in the interval $0<\pt<12~\gev/c$ was also computed using the cross sections in pp and p--Pb collisions resulting from the analysis without decay-vertex reconstruction. In Fig.~\ref{fig:RpPbVsPt} it is compared
with the result obtained from the analysis with decay-vertex reconstruction, which covers the interval $1<\pt<24~\gev/c$~\cite{Abelev:2014hha}.
The two measurements are consistent within statistical uncertainties. 
In the previous subsections it was shown that the analysis without decay-vertex reconstruction provides the best determination 
of the $\Dzero$ cross section in the interval $1<\pt<2~\gev/c$, where the analysis with decay-vertex reconstruction is affected
by a large uncertainty on the feed-down correction.  
This is not the case for the $R_{\rm pPb}$ measurement, because the feed-down uncertainty cancels to a large extent for this observable.

\begin{figure}[!t]
\begin{center}
\includegraphics[width=0.48\textwidth]{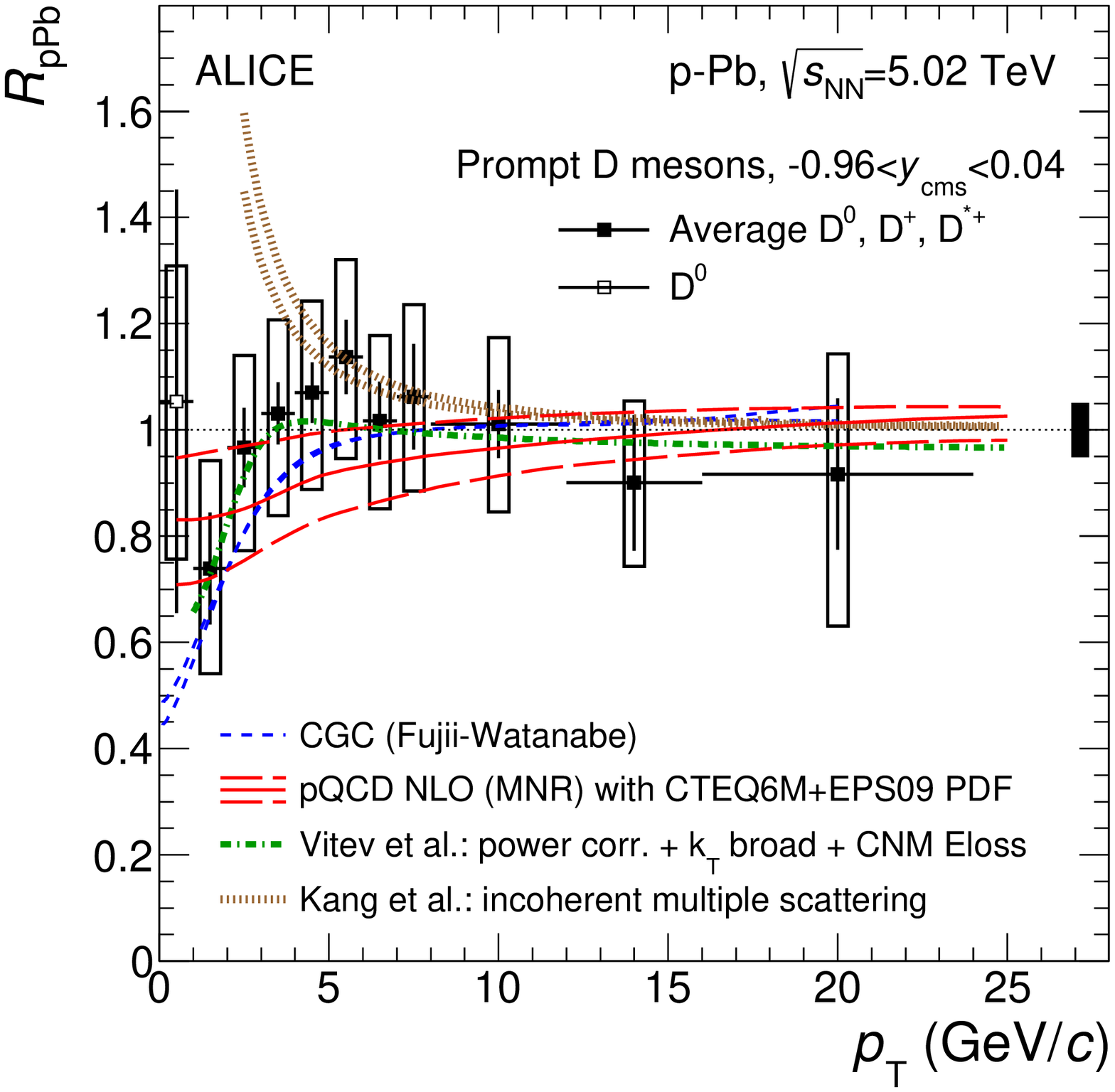}
\includegraphics[width=0.48\textwidth]{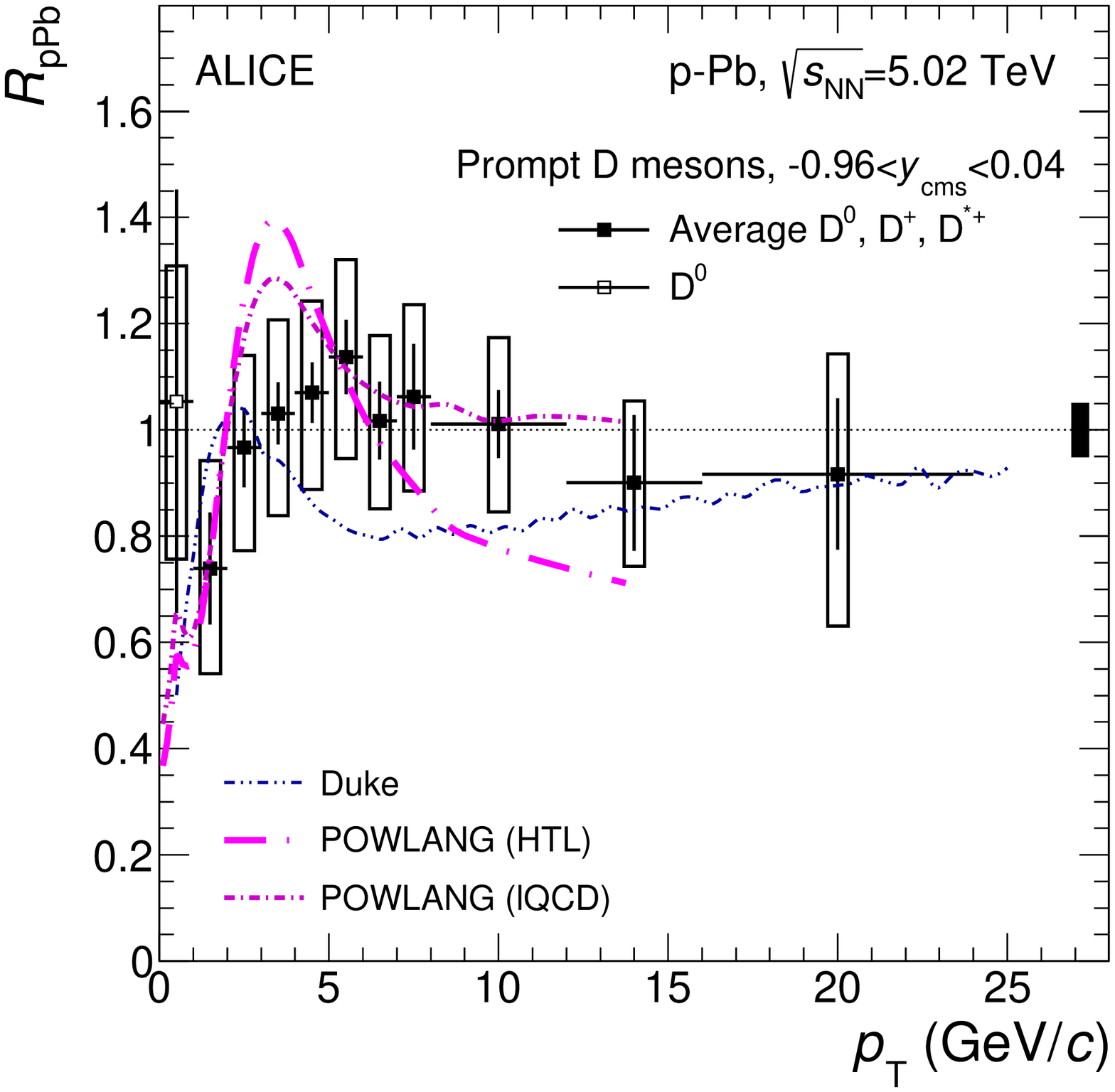}
\caption{Nuclear modification factor $R_{\rm pPb}$ of prompt D mesons in p--Pb collisions at $\sqrtsNN=5.02~\tev$: average $R_{\rm pPb}$ of $\Dzero$, $\Dplus$ and $\Dstar$ mesons in the interval $1<\pt<24~\gev/c$~\cite{Abelev:2014hha}, shown together with the $\Dzero$ $R_{\rm pPb}$ in $0<\pt<1~\GeV/c$. In the left-hand panel, the data are compared with results of theoretical calculations including only CNM effects: CGC~\cite{Fujii:2013yja}, NLO pQCD~\cite{Mangano:1991jk} with EPS09 nPDFs~\cite{Eskola:2009uj}, a LO pQCD calculation with CNM effects (Vitev et al.)~\cite{Sharma:2009hn} and a calculation based on incoherent multiple scatterings (Kang et al.)~\cite{Kang:2014hha}. In the right-hand panel, the results of the Duke~\cite{Xu:2015iha} and POWLANG~\cite{Beraudo:2015wsd} transport models are compared to the measured D-meson $R_{\rm pPb}$.}
\label{fig:RpPbVsPtModels}
\end{center}
\end{figure}

Figure~\ref{fig:RpPbVsPtModels} shows the combined measurement of the nuclear modification factor of prompt (non-strange) D mesons,
as obtained by using the $\Dzero$ measurement without decay-vertex reconstruction for the interval $0<\pt<1~\gev/c$ and the average of the measurements for $\Dzero$, $\Dplus$ and $\Dstar$ mesons in the interval $1<\pt<24~\gev/c$~\cite{Abelev:2014hha}.
The data are compared with theoretical results.
In the left-hand panel of this figure, four models including only CNM effects are displayed: a calculation based on the Color Glass Condensate
formalism~\cite{Fujii:2013yja}, a pQCD calculation based on the MNR formalism~\cite{Mangano:1991jk} with CTEQ6M PDFs~\cite{Pumplin:2002vw} and EPS09NLO nuclear modification~\cite{Eskola:2009uj}, a LO pQCD calculation with intrinsic $k_{\rm T}$ broadening, nuclear shadowing and energy loss of the charm quarks in cold nuclear matter~\cite{Sharma:2009hn}, and a higher-twist calculation based on incoherent multiple scatterings (Kang et al.)~\cite{Kang:2014hha}. 
The three former calculations describe the data within uncertainties in the entire $\pt$ range, while the last one (Kang et al.), which has a different trend with respect to the others, is disfavoured by the data at $\pt<3$--$4~\gev/c$.
CNM effects are expected to be largest for small $\pt$, where, in addition, the predictions of the different theoretical approaches differ. The uncertainty of the present measurement for the lowest $\pt$ interval is about 50\% and does not allow us to 
draw a conclusion. However, the analysis technique without decay-vertex reconstruction, applied on future 
larger data samples, should provide access to the physics-rich range down to $\pt=0$.
In the right-hand panel of Fig.~\ref{fig:RpPbVsPtModels}, the data are 
compared to the results of two transport model 
calculations, Duke~\cite{Xu:2015iha} and 
POWLANG~\cite{Beraudo:2015wsd}, both of them assuming that a 
Quark-Gluon Plasma is formed in p--Pb collisions.
Both models are based on the Langevin approach for the transport of heavy 
quarks through an expanding deconfined medium described by relativistic viscous 
hydrodynamics. 
The Duke model includes both collisional and radiative energy loss.
The POWLANG model considers only collisional processes with two choices for 
the transport coefficients, based on hard-thermal-loop (HTL) and
lattice-QCD (lQCD) calculations, respectively.
In both approaches the D-meson nuclear modification
factor shows  a structure with a 
maximum at $\pt \approx 2.5~\GeV/c$, possibly followed by a moderate 
($<20$--30\%)
suppression at higher $\pt$,
resulting from the interplay of CNM effects and interactions of charm quarks with the
radially expanding medium.
The precision of the measured D-meson $R_{\rm pPb}$ does not allow us 
to discriminate between scenarios with only CNM effects or 
hot medium effects in addition, even though the data seem to disfavour a suppression 
larger than 15--20\% in the interval $5<\pt<10~\GeV/c$.

The $\pt$-integrated nuclear modification factor of prompt 
$\Dzero$ mesons in $-0.96<y_{\rm cms}<0.04$ was computed using the 
${\rm d}\sigma^{\rm prompt\,D^0}/{\rm d}y$ values for pp and p--Pb collisions 
reported in Eqs.~(\ref{eq:sigD0pp}) and (\ref{eq:sigD0pPb}) and using FONLL to 
scale the pp cross section to the centre-of-mass energy and rapidity interval 
of the p--Pb measurement.
The result is:
\begin{equation}
R_{\rm pPb}^{\rm prompt\,D^0}(\pt>0,\,-0.96<y_{\rm cms}<0.04) = 0.89 \pm 0.11\,({\rm stat.}) ^{+0.13}_{-0.18}\,({\rm syst.})\,.
\end{equation}

\section{Summary}
\label{sec:summary}

We have presented a comprehensive set of results on 
charm production in p--Pb and pp collisions, complementing the measurements
reported in Refs.~\cite{Abelev:2014hha} and ~\cite{ALICE:2011aa}.
The production cross sections of the prompt charmed 
mesons $\Dzero$, $\Dplus$, $\Dstar$ and $\Ds$ in p--Pb 
collisions at a centre-of-mass energy per nucleon pair $\sqrtsNN=5.02~\tev$ 
were measured as a function of $\pt$ in the rapidity interval 
$-0.96<y_{\rm cms}<0.04$.
The $\pt$-differential production cross sections, obtained with
an analysis method based on the selection of decay topologies
displaced from the interaction vertex, were reported
in the transverse momentum range $1<\pt<24~\GeV/c$ for $\Dzero$, $\Dplus$ and 
$\Dstar$ mesons and in the range $2<\pt<12~\gev/c$ for $\Ds$ mesons.
The ratios of the cross sections of the four D-meson species were determined 
as a function of $\pt$ and were found to be compatible with those measured in 
pp collisions at $\sqrt{s}=7~\TeV$ in the rapidity interval $|y_{\rm cms}|<0.5$.

The production cross sections of the non-strange D mesons, 
$\Dzero$, $\Dplus$ and $\Dstar$, were also measured in p--Pb collisions
as a function of rapidity in three $\pt$ intervals. 
No significant rapidity dependence was observed in the range 
$-1.26<y_{\rm cms}<0.34$. 

In addition, employing an analysis technique that does not use the 
reconstruction of the $\Dzero$ decay vertex, the prompt $\Dzero$ production 
cross section was measured down to $\pt=0$ in
pp collisions at $\sqrt{s}=7~\TeV$ and p--Pb collisions at
$\sqrtsNN=5.02~\tev$.
The results of the two different analysis techniques, with and without
decay-vertex reconstruction, were found to be compatible
in the common $\pt$ range.
The analysis without decay-vertex reconstruction provides a more precise 
measurement of the $\Dzero$ cross section for $\pt<2~\gev/c$.
This allowed a determination of the total ($\pt$ integrated) $\Dzero$ 
production cross section, d$\sigma$/d$y$, at mid-rapidity, which is
not affected by uncertainties due to the extrapolation to $\pt=0$.
The resulting cross section in pp collisions at $\sqrt{s}=7~\TeV$ is
\begin{equation}
\nonumber
{\rm d}\sigma^{\rm prompt\,D^0}_{\rm pp,\,7\,TeV}/{\rm d}y=518\pm 43\,({\rm stat.}) ^{+\phantom{1}57}_{-102}\,({\rm syst.})\pm 18\,({\rm lumi.})\pm 7\,({\rm BR})~\mu{\rm b}\,.
\end{equation}
The total systematic uncertainty is smaller by a factor of about two on the 
low side and almost three on the high side as compared to our previous
result~\cite{ALICE:2011aa}.
The resulting total ${\rm c\overline{c}}$ production cross section in pp collisions at $\sqrt{s}=7~\TeV$ is:
\begin{equation}
\nonumber
{\rm \sigma^{\rm c\overline{c}}_{\rm pp\,7\,TeV}= 8.18 \pm 0.67 \,({\rm stat.}) \,^{+0.90}_{-1.62}\,({\rm syst.})  \,^{+2.40}_{-0.36} ({\rm extr.}) \pm 0.29\,({\rm lumi.})\, \pm 0.36\,({\rm FF})~{\rm mb}\,.}
\end{equation}

In p--Pb collisions at $\sqrtsNN=5.02~\tev$, the $\pt$-integrated 
prompt-$\Dzero$ production cross section at mid-rapidity 
($-0.96<y_{\rm cms}<0.04$) is
\begin{equation}
\nonumber
{\rm d}\sigma^{\rm prompt\,D^0}_{\rm p-Pb,\,5.02\,TeV}/{\rm d}y=79.0\pm 7.3\,({\rm stat.}) ^{+\phantom{0}7.1}_{-13.4}\,({\rm syst.})\pm 2.9\,({\rm lumi.})\pm 1.0\,({\rm BR})~{\rm mb}\,.
\end{equation}

The $\pt$-differential nuclear modification factor $\RpPb$ was 
found to be compatible with unity in the transverse-momentum interval 
$0<\pt<24~\gev/c$.
This result provides clear experimental 
evidence~\cite{Abelev:2014hha,Adam:2015sza} 
that the modification of the D-meson transverse momentum distributions 
observed in Pb--Pb collisions as compared to pp interactions is due to 
final-state effects induced by the interactions of the charm quarks with 
the hot and dense partonic medium created in ultra-relativistic heavy-ion 
collisions.
The uncertainties of the present measurement are about 20--30\% for 
$\pt>1~\GeV/c$, considering the average of $\Dzero$, $\Dplus$ and $\Dstar$
 $\RpPb$, and about 50\% in the interval $0<\pt<1~\GeV/c$, where
the $\Dzero$ could be reconstructed with the analysis technique without 
decay-vertex reconstruction.
The results are described within uncertainties by theoretical calculations 
that include initial-state effects, which are expected to be small
for $\pt>2~\gev/c$ but significant for $\pt$ close to 0, where 
the predictions of the different theoretical approaches differ.
The observed $R_{\rm pPb}$ is also described by transport calculations assuming 
the formation of a deconfined medium in p--Pb collisions,
even though the data seem to disfavour a suppression 
larger than 15--20\% in the interval $5<\pt<10~\GeV/c$.
The current precision of the measurement does not allow us to draw conclusions 
on the role of the different CNM effects and on the possible presence of 
additional hot-medium effects.
However, the analysis technique without decay-vertex reconstruction, 
applied on future larger data samples, should provide access to the 
physics-rich range down to $\pt=0$.

\newenvironment{acknowledgement}{\relax}{\relax}
\begin{acknowledgement}
\section*{Acknowledgements}
\input{acknowledgements.tex}    
\end{acknowledgement}

\bibliographystyle{utphys}
\bibliography{biblio}
%

%

\newpage
\appendix
\section{The ALICE Collaboration}
\label{app:collab}
\input{Alice_Authorlist_2016-05-16_mod.tex}  
\end{document}

%% file: acknowledgements.tex

The ALICE Collaboration would like to thank all its engineers and technicians for their invaluable contributions to the construction of the experiment and the CERN accelerator teams for the outstanding performance of the LHC complex.
The ALICE Collaboration gratefully acknowledges the resources and support provided by all Grid centres and the Worldwide LHC Computing Grid (WLCG) collaboration.
The ALICE Collaboration acknowledges the following funding agencies for their support in building and
running the ALICE detector:
State Committee of Science,  World Federation of Scientists (WFS)
and Swiss Fonds Kidagan, Armenia;
Conselho Nacional de Desenvolvimento Cient\'{\i}fico e Tecnol\'{o}gico (CNPq), Financiadora de Estudos e Projetos (FINEP),
Funda\c{c}\~{a}o de Amparo \`{a} Pesquisa do Estado de S\~{a}o Paulo (FAPESP);
Ministry of Science \& Technology of China (MSTC), National Natural Science Foundation of China (NSFC) and Ministry of Education of China (MOEC);
Ministry of Science, Education and Sports of Croatia and  Unity through Knowledge Fund, Croatia;
Ministry of Education and Youth of the Czech Republic;
Danish Natural Science Research Council, the Carlsberg Foundation and the Danish National Research Foundation;
The European Research Council under the European Community's Seventh Framework Programme;
Helsinki Institute of Physics and the Academy of Finland;
French CNRS-IN2P3, the `R\'{e}gion Pays de Loire', `R\'{e}gion Alsace', `R\'{e}gion Auvergne' and CEA, France;
German Bundesministerium f\"ur Bildung, Wissenschaft, Forschung und Technologie (BMBF) and the Helmholtz Association;
General Secretariat for Research and Technology, Ministry of Development, Greece;
National Research, Development and Innovation Office (NKFIH), Hungary;
Council of Scientific and Industrial Research (CSIR), New Delhi;
Department of Atomic Energy and Department of Science and Technology of the Government of India;
Istituto Nazionale di Fisica Nucleare (INFN) and Centro Fermi - Museo Storico della Fisica e Centro Studi e Ricerche ``Enrico Fermi'', Italy;
Japan Society for the Promotion of Science (JSPS) KAKENHI and MEXT, Japan;
National Research Foundation of Korea (NRF);
Consejo Nacional de Cienca y Tecnologia (CONACYT), Direcci\'{o}n General de Asuntos del Personal Acad\'{e}mico (DGAPA), M\'{e}xico, Am\'{e}rique Latine Formation acad\'{e}mique - 
European Commission~(ALFA-EC) and the EPLANET Program~(European Particle Physics Latin American Network);
Stichting voor Fundamenteel Onderzoek der Materie (FOM) and the Nederlandse Organisatie voor Wetenschappelijk Onderzoek (NWO), Netherlands;
Research Council of Norway (NFR);
Pontificia Universidad Cat\'{o}lica del Per\'{u};
National Science Centre, Poland;
Ministry of National Education/Institute for Atomic Physics and National Council of Scientific Research in Higher Education~(CNCSI-UEFISCDI), Romania;
Joint Institute for Nuclear Research, Dubna;
Ministry of Education and Science of Russian Federation, Russian Academy of Sciences, Russian Federal Agency of Atomic Energy, Russian Federal Agency for Science and Innovations and The Russian Foundation for Basic Research;
Ministry of Education of Slovakia;
Department of Science and Technology, South Africa;
Centro de Investigaciones Energ\'{e}ticas, Medioambientales y Tecnol\'{o}gicas (CIEMAT), E-Infrastructure shared between Europe and Latin America (EELA), 
Ministerio de Econom\'{i}a y Competitividad (MINECO) of Spain, Xunta de Galicia (Conseller\'{\i}a de Educaci\'{o}n),
Centro de Aplicaciones Tecnol\'{o}gicas y Desarrollo Nuclear (CEA\-DEN), Cubaenerg\'{\i}a, Cuba, and IAEA (International Atomic Energy Agency);
Swedish Research Council (VR) and Knut $\&$ Alice Wallenberg Foundation (KAW);
National Science and Technology Development Agency (NSDTA), Suranaree University of Technology (SUT) and Office of the Higher Education Commission under NRU project of Thailand;
Ukraine Ministry of Education and Science;
United Kingdom Science and Technology Facilities Council (STFC);
The United States Department of Energy, the United States National Science Foundation, the State of Texas, and the State of Ohio.

%% file: Alice_Authorlist_2016-05-16_mod.tex


\begingroup
\small
\begin{flushleft}
J.~Adam$^\textrm{\scriptsize 39}$,
D.~Adamov\'{a}$^\textrm{\scriptsize 85}$,
M.M.~Aggarwal$^\textrm{\scriptsize 89}$,
G.~Aglieri Rinella$^\textrm{\scriptsize 35}$,
M.~Agnello$^\textrm{\scriptsize 112}$\textsuperscript{,}$^\textrm{\scriptsize 31}$,
N.~Agrawal$^\textrm{\scriptsize 48}$,
Z.~Ahammed$^\textrm{\scriptsize 136}$,
S.~Ahmad$^\textrm{\scriptsize 18}$,
S.U.~Ahn$^\textrm{\scriptsize 69}$,
S.~Aiola$^\textrm{\scriptsize 140}$
% ...
\renewcommand\labelenumi{\textsuperscript{\theenumi}~}

\section*{Affiliation notes}
\renewcommand\theenumi{\roman{enumi}}
\begin{Authlist}
\item \Adef{0}Deceased
\item \Adef{idp1819072}{Also at: Georgia State University, Atlanta, Georgia, United States}
\item \Adef{idp3242880}{Also at: Also at Department of Applied Physics, Aligarh Muslim University, Aligarh, India}
\item \Adef{idp3975120}{Also at: M.V. Lomonosov Moscow State University, D.V. Skobeltsyn Institute of Nuclear, Physics, Moscow, Russia}
\end{Authlist}

\section*{Collaboration Institutes}
\renewcommand\theenumi{\arabic{enumi}~}

$^{1}$A.I. Alikhanyan National Science Laboratory (Yerevan Physics Institute) Foundation, Yerevan, Armenia
\\
$^{2}$Benem\'{e}rita Universidad Aut\'{o}noma de Puebla, Puebla, Mexico
\\
$^{3}$Bogolyubov Institute for Theoretical Physics, Kiev, Ukraine
\\
$^{4}$Bose Institute, Department of Physics 
and Centre for Astroparticle Physics and Space Science (CAPSS), Kolkata, India
\\
$^{5}$Budker Institute for Nuclear Physics, Novosibirsk, Russia
\\
$^{6}$California Polytechnic State University, San Luis Obispo, California, United States
\\
$^{7}$Central China Normal University, Wuhan, China
\\
$^{8}$Centre de Calcul de l'IN2P3, Villeurbanne, Lyon, France
\\
$^{9}$Centro de Aplicaciones Tecnol\'{o}gicas y Desarrollo Nuclear (CEADEN), Havana, Cuba
\\
$^{10}$Centro de Investigaciones Energ\'{e}ticas Medioambientales y Tecnol\'{o}gicas (CIEMAT), Madrid, Spain
\\
$^{11}$Centro de Investigaci\'{o}n y de Estudios Avanzados (CINVESTAV), Mexico City and M\'{e}rida, Mexico
\\
$^{12}$Centro Fermi - Museo Storico della Fisica e Centro Studi e Ricerche ``Enrico Fermi', Rome, Italy
\\
$^{13}$Chicago State University, Chicago, Illinois, United States
\\
$^{14}$China Institute of Atomic Energy, Beijing, China
\\
$^{15}$Commissariat \`{a} l'Energie Atomique, IRFU, Saclay, France
\\
$^{16}$COMSATS Institute of Information Technology (CIIT), Islamabad, Pakistan
\\
$^{17}$Departamento de F\'{\i}sica de Part\'{\i}culas and IGFAE, Universidad de Santiago de Compostela, Santiago de Compostela, Spain
\\
$^{18}$Department of Physics, Aligarh Muslim University, Aligarh, India
\\
$^{19}$Department of Physics, Ohio State University, Columbus, Ohio, United States
\\
$^{20}$Department of Physics, Sejong University, Seoul, South Korea
\\
$^{21}$Department of Physics, University of Oslo, Oslo, Norway
\\
$^{22}$Department of Physics and Technology, University of Bergen, Bergen, Norway
\\
$^{23}$Dipartimento di Fisica dell'Universit\`{a} 'La Sapienza'
and Sezione INFN, Rome, Italy
\\
$^{24}$Dipartimento di Fisica dell'Universit\`{a}
and Sezione INFN, Cagliari, Italy
\\
$^{25}$Dipartimento di Fisica dell'Universit\`{a}
and Sezione INFN, Trieste, Italy
\\
$^{26}$Dipartimento di Fisica dell'Universit\`{a}
and Sezione INFN, Turin, Italy
\\
$^{27}$Dipartimento di Fisica e Astronomia dell'Universit\`{a}
and Sezione INFN, Bologna, Italy
\\
$^{28}$Dipartimento di Fisica e Astronomia dell'Universit\`{a}
and Sezione INFN, Catania, Italy
\\
$^{29}$Dipartimento di Fisica e Astronomia dell'Universit\`{a}
and Sezione INFN, Padova, Italy
\\
$^{30}$Dipartimento di Fisica `E.R.~Caianiello' dell'Universit\`{a}
and Gruppo Collegato INFN, Salerno, Italy
\\
$^{31}$Dipartimento DISAT del Politecnico and Sezione INFN, Turin, Italy
\\
$^{32}$Dipartimento di Scienze e Innovazione Tecnologica dell'Universit\`{a} del Piemonte Orientale and INFN Sezione di Torino, Alessandria, Italy
\\
$^{33}$Dipartimento Interateneo di Fisica `M.~Merlin'
and Sezione INFN, Bari, Italy
\\
$^{34}$Division of Experimental High Energy Physics, University of Lund, Lund, Sweden
\\
$^{35}$European Organization for Nuclear Research (CERN), Geneva, Switzerland
\\
$^{36}$Excellence Cluster Universe, Technische Universit\"{a}t M\"{u}nchen, Munich, Germany
\\
$^{37}$Faculty of Engineering, Bergen University College, Bergen, Norway
\\
$^{38}$Faculty of Mathematics, Physics and Informatics, Comenius University, Bratislava, Slovakia
\\
$^{39}$Faculty of Nuclear Sciences and Physical Engineering, Czech Technical University in Prague, Prague, Czech Republic
\\
$^{40}$Faculty of Science, P.J.~\v{S}af\'{a}rik University, Ko\v{s}ice, Slovakia
\\
$^{41}$Faculty of Technology, Buskerud and Vestfold University College, Tonsberg, Norway
\\
$^{42}$Frankfurt Institute for Advanced Studies, Johann Wolfgang Goethe-Universit\"{a}t Frankfurt, Frankfurt, Germany
\\
$^{43}$Gangneung-Wonju National University, Gangneung, South Korea
\\
$^{44}$Gauhati University, Department of Physics, Guwahati, India
\\
$^{45}$Helmholtz-Institut f\"{u}r Strahlen- und Kernphysik, Rheinische Friedrich-Wilhelms-Universit\"{a}t Bonn, Bonn, Germany
\\
$^{46}$Helsinki Institute of Physics (HIP), Helsinki, Finland
\\
$^{47}$Hiroshima University, Hiroshima, Japan
\\
$^{48}$Indian Institute of Technology Bombay (IIT), Mumbai, India
\\
$^{49}$Indian Institute of Technology Indore, Indore, India
\\
$^{50}$Indonesian Institute of Sciences, Jakarta, Indonesia
\\
$^{51}$Inha University, Incheon, South Korea
\\
$^{52}$Institut de Physique Nucl\'eaire d'Orsay (IPNO), Universit\'e Paris-Sud, CNRS-IN2P3, Orsay, France
\\
$^{53}$Institute for Nuclear Research, Academy of Sciences, Moscow, Russia
\\
$^{54}$Institute for Subatomic Physics of Utrecht University, Utrecht, Netherlands
\\
$^{55}$Institute for Theoretical and Experimental Physics, Moscow, Russia
\\
$^{56}$Institute of Experimental Physics, Slovak Academy of Sciences, Ko\v{s}ice, Slovakia
\\
$^{57}$Institute of Physics, Academy of Sciences of the Czech Republic, Prague, Czech Republic
\\
$^{58}$Institute of Physics, Bhubaneswar, India
\\
$^{59}$Institute of Space Science (ISS), Bucharest, Romania
\\
$^{60}$Institut f\"{u}r Informatik, Johann Wolfgang Goethe-Universit\"{a}t Frankfurt, Frankfurt, Germany
\\
$^{61}$Institut f\"{u}r Kernphysik, Johann Wolfgang Goethe-Universit\"{a}t Frankfurt, Frankfurt, Germany
\\
$^{62}$Institut f\"{u}r Kernphysik, Westf\"{a}lische Wilhelms-Universit\"{a}t M\"{u}nster, M\"{u}nster, Germany
\\
$^{63}$Instituto de Ciencias Nucleares, Universidad Nacional Aut\'{o}noma de M\'{e}xico, Mexico City, Mexico
\\
$^{64}$Instituto de F\'{\i}sica, Universidad Nacional Aut\'{o}noma de M\'{e}xico, Mexico City, Mexico
\\
$^{65}$Institut Pluridisciplinaire Hubert Curien (IPHC), Universit\'{e} de Strasbourg, CNRS-IN2P3, Strasbourg, France
\\
$^{66}$iThemba LABS, National Research Foundation, Somerset West, South Africa
\\
$^{67}$Joint Institute for Nuclear Research (JINR), Dubna, Russia
\\
$^{68}$Konkuk University, Seoul, South Korea
\\
$^{69}$Korea Institute of Science and Technology Information, Daejeon, South Korea
\\
$^{70}$KTO Karatay University, Konya, Turkey
\\
$^{71}$Laboratoire de Physique Corpusculaire (LPC), Clermont Universit\'{e}, Universit\'{e} Blaise Pascal, CNRS--IN2P3, Clermont-Ferrand, France
\\
$^{72}$Laboratoire de Physique Subatomique et de Cosmologie, Universit\'{e} Grenoble-Alpes, CNRS-IN2P3, Grenoble, France
\\
$^{73}$Laboratori Nazionali di Frascati, INFN, Frascati, Italy
\\
$^{74}$Laboratori Nazionali di Legnaro, INFN, Legnaro, Italy
\\
$^{75}$Lawrence Berkeley National Laboratory, Berkeley, California, United States
\\
$^{76}$Moscow Engineering Physics Institute, Moscow, Russia
\\
$^{77}$Nagasaki Institute of Applied Science, Nagasaki, Japan
\\
$^{78}$National Centre for Nuclear Studies, Warsaw, Poland
\\
$^{79}$National Institute for Physics and Nuclear Engineering, Bucharest, Romania
\\
$^{80}$National Institute of Science Education and Research, Bhubaneswar, India
\\
$^{81}$National Research Centre Kurchatov Institute, Moscow, Russia
\\
$^{82}$Niels Bohr Institute, University of Copenhagen, Copenhagen, Denmark
\\
$^{83}$Nikhef, Nationaal instituut voor subatomaire fysica, Amsterdam, Netherlands
\\
$^{84}$Nuclear Physics Group, STFC Daresbury Laboratory, Daresbury, United Kingdom
\\
$^{85}$Nuclear Physics Institute, Academy of Sciences of the Czech Republic, \v{R}e\v{z} u Prahy, Czech Republic
\\
$^{86}$Oak Ridge National Laboratory, Oak Ridge, Tennessee, United States
\\
$^{87}$Petersburg Nuclear Physics Institute, Gatchina, Russia
\\
$^{88}$Physics Department, Creighton University, Omaha, Nebraska, United States
\\
$^{89}$Physics Department, Panjab University, Chandigarh, India
\\
$^{90}$Physics Department, University of Athens, Athens, Greece
\\
$^{91}$Physics Department, University of Cape Town, Cape Town, South Africa
\\
$^{92}$Physics Department, University of Jammu, Jammu, India
\\
$^{93}$Physics Department, University of Rajasthan, Jaipur, India
\\
$^{94}$Physikalisches Institut, Eberhard Karls Universit\"{a}t T\"{u}bingen, T\"{u}bingen, Germany
\\
$^{95}$Physikalisches Institut, Ruprecht-Karls-Universit\"{a}t Heidelberg, Heidelberg, Germany
\\
$^{96}$Physik Department, Technische Universit\"{a}t M\"{u}nchen, Munich, Germany
\\
$^{97}$Purdue University, West Lafayette, Indiana, United States
\\
$^{98}$Pusan National University, Pusan, South Korea
\\
$^{99}$Research Division and ExtreMe Matter Institute EMMI, GSI Helmholtzzentrum f\"ur Schwerionenforschung, Darmstadt, Germany
\\
$^{100}$Rudjer Bo\v{s}kovi\'{c} Institute, Zagreb, Croatia
\\
$^{101}$Russian Federal Nuclear Center (VNIIEF), Sarov, Russia
\\
$^{102}$Saha Institute of Nuclear Physics, Kolkata, India
\\
$^{103}$School of Physics and Astronomy, University of Birmingham, Birmingham, United Kingdom
\\
$^{104}$Secci\'{o}n F\'{\i}sica, Departamento de Ciencias, Pontificia Universidad Cat\'{o}lica del Per\'{u}, Lima, Peru
\\
$^{105}$Sezione INFN, Bari, Italy
\\
$^{106}$Sezione INFN, Bologna, Italy
\\
$^{107}$Sezione INFN, Cagliari, Italy
\\
$^{108}$Sezione INFN, Catania, Italy
\\
$^{109}$Sezione INFN, Padova, Italy
\\
$^{110}$Sezione INFN, Rome, Italy
\\
$^{111}$Sezione INFN, Trieste, Italy
\\
$^{112}$Sezione INFN, Turin, Italy
\\
$^{113}$SSC IHEP of NRC Kurchatov institute, Protvino, Russia
\\
$^{114}$Stefan Meyer Institut f\"{u}r Subatomare Physik (SMI), Vienna, Austria
\\
$^{115}$SUBATECH, Ecole des Mines de Nantes, Universit\'{e} de Nantes, CNRS-IN2P3, Nantes, France
\\
$^{116}$Suranaree University of Technology, Nakhon Ratchasima, Thailand
\\
$^{117}$Technical University of Ko\v{s}ice, Ko\v{s}ice, Slovakia
\\
$^{118}$Technical University of Split FESB, Split, Croatia
\\
$^{119}$The Henryk Niewodniczanski Institute of Nuclear Physics, Polish Academy of Sciences, Cracow, Poland
\\
$^{120}$The University of Texas at Austin, Physics Department, Austin, Texas, United States
\\
$^{121}$Universidad Aut\'{o}noma de Sinaloa, Culiac\'{a}n, Mexico
\\
$^{122}$Universidade de S\~{a}o Paulo (USP), S\~{a}o Paulo, Brazil
\\
$^{123}$Universidade Estadual de Campinas (UNICAMP), Campinas, Brazil
\\
$^{124}$Universidade Federal do ABC, Santo Andre, Brazil
\\
$^{125}$University of Houston, Houston, Texas, United States
\\
$^{126}$University of Jyv\"{a}skyl\"{a}, Jyv\"{a}skyl\"{a}, Finland
\\
$^{127}$University of Liverpool, Liverpool, United Kingdom
\\
$^{128}$University of Tennessee, Knoxville, Tennessee, United States
\\
$^{129}$University of the Witwatersrand, Johannesburg, South Africa
\\
$^{130}$University of Tokyo, Tokyo, Japan
\\
$^{131}$University of Tsukuba, Tsukuba, Japan
\\
$^{132}$University of Zagreb, Zagreb, Croatia
\\
$^{133}$Universit\'{e} de Lyon, Universit\'{e} Lyon 1, CNRS/IN2P3, IPN-Lyon, Villeurbanne, Lyon, France
\\
$^{134}$Universit\`{a} di Brescia, Brescia, Italy
\\
$^{135}$V.~Fock Institute for Physics, St. Petersburg State University, St. Petersburg, Russia
\\
$^{136}$Variable Energy Cyclotron Centre, Kolkata, India
\\
$^{137}$Warsaw University of Technology, Warsaw, Poland
\\
$^{138}$Wayne State University, Detroit, Michigan, United States
\\
$^{139}$Wigner Research Centre for Physics, Hungarian Academy of Sciences, Budapest, Hungary
\\
$^{140}$Yale University, New Haven, Connecticut, United States
\\
$^{141}$Yonsei University, Seoul, South Korea
\\
$^{142}$Zentrum f\"{u}r Technologietransfer und Telekommunikation (ZTT), Fachhochschule Worms, Worms, Germany
\endgroup